\documentclass[useAMS,usenatbib]{mnras}

\usepackage{graphicx}
\usepackage{txfonts}

\title[A study of 102 star forming regions]
      {A spectral and photometric study of 102 star forming regions in 
       seven spiral galaxies}

\author[A.~S.~Gusev et al.]
       {A.~S.~Gusev,$^{1}$
        F.~Sakhibov,$^{2}$
        A.~E.~Piskunov,$^{3}$
        N.~V.~Kharchenko,$^{4}$
        V.~V.~Bruevich,$^{1}$
\newauthor
        O.~V.~Ezhkova,$^{1}$
        S.~A.~Guslyakova,$^{5}$
        V.~Lang,$^{2}$
        E.~V.~Shimanovskaya,$^{1}$ 
\newauthor
        and Y.~N.~Efremov$^{1}$ \\
 $^{1}$ Sternberg Astronomical Institute, Lomonosov Moscow State University, 
        Universitetsky pr. 13, 119992 Moscow, Russia \\
 $^{2}$ University of Applied Sciences of Mittelhessen, Campus Friedberg,
        Department of Mathematics, Natural Sciences and Data Processing, \\
        Wilhelm-Leuschner-Strasse 13, 61169 Friedberg, Germany \\
 $^{3}$ Institute of Astronomy, Russian Academy of Sciences,
        ul. Pyatnitskaya 48, 119017 Moscow, Russia \\       
 $^{4}$ Main Astronomical Observatory, National Academy of Sciences of 
        Ukraine, ul. Zabolotnogo 27, 03680 Kiev, Ukraine \\
 $^{5}$ Institute for Space Research, Russian Academy of Sciences,
        ul. Profsoyuznaya 84/32, 117997 Moscow, Russia \\
             }

\date{Accepted 2016 January 25. Received 2016 January 25;
in original form 2015 April 1}

\begin{document}

\maketitle

\begin{abstract}
We present a study of complexes of young massive star clusters 
(YMCs), embedded in extragalactic giant H\,{\sc ii} regions, based on the 
coupling of spectroscopic with photometric and spectrophotometric 
observations of about 100 star forming regions in seven spiral galaxies 
(NGC~628, NGC~783, NGC~2336, NGC~6217, NGC~6946, NGC~7331, and NGC~7678). The 
complete observational database has been observed and accumulated within the 
framework of 
our comprehensive study of extragalactic star forming regions. The 
current paper presents the last part of either unpublished or refreshed 
photometric and spectrophotometric observations of the galaxies NGC~6217, 
NGC~6946, NGC~7331, and NGC~7678. We derive extinctions, chemical abundances, 
continuum and line emissions of ionised gas, ages and masses for 
cluster complexes. We find the young massive cluster complexes to 
have ages no greater than $10$~Myr and masses between $10^4 M_{\odot}$ and 
$10^7 M_{\odot}$, and the extinctions $A_V$ vary between 
$\sim 0$ and 3~mag, while the impact of the nebular emission on integrated 
broadband photometry mainly is not greater than $40\%$ of the total flux and 
is comparable with accuracies of dereddened photometric quantities. We also 
find evidence of differential extinction of stellar and gas emissions in some 
clusters, which hinders the photometric determination of ages and masses in 
these cases. Finally, we show that young massive cluster complexes in 
the studied galaxies and open clusters in the Milky Way form a continuous 
sequence of luminosities/masses and colour/ages.
\end{abstract}


\section{Introduction}
\label{sect:intro}

This paper presents the second part of our comprehensive study of star 
forming (SF) regions in spiral galaxies. The results of spectroscopic 
observations of 102 H\,{\sc ii} regions in seven galaxies were presented in 
the previous papers \citep*{gusev2012,gusev2013}. Here we present the 
results of $UBVRI$ photometry and H$\alpha$ spectrophotometry for the same 
SF~regions in the galaxies. We obtained and analyzed $UBVRI$ photometric 
data for 101 out of the 102 studied H\,{\sc ii} regions in seven galaxies 
(except object No.~1 in NGC~7331), and H$\alpha$ spectrophotometric 
data for 53 H\,{\sc ii} regions in galaxies NGC~628, NGC~6946, and NGC~7331. 
A combination of the multicolour photometry and spectroscopic observations of 
SF~regions provides a clue to the properties of young massive cluster 
complexes embedded in giant H\,{\sc ii} regions in seven spiral galaxies 
NGC~628 (10), NGC~783 (8), NGC~2336 (28), NGC~6217 (3), NGC~6946 (39), 
NGC~7331 (3), and NGC~7678 (10). Since we observed both giant 
H\,{\sc ii} regions with embedded cluster complexes, and the embedded 
cluster complexes themselves, we call the studied objects `star 
forming regions' (SF~regions).

\begin{table*}
\caption[]{\label{table:sample}
The galaxy sample.
}
\begin{center}
\begin{tabular}{cccccccccccc} \hline \hline
Galaxy & Type & $B_t$ & $M_B^a$ & Inclination & PA       & $R_{25}^b$ & 
$R_{25}^b$ & $D$   & $A(B)_{\rm Gal}$ & $A(B)_{\rm in}$ & Notes$^c$ \\
       &      & (mag) & (mag)   & (degree)    & (degree) & (arcmin)   & 
(kpc)      & (Mpc) & (mag)        & (mag)       &           \\
1 & 2 & 3 & 4 & 5 & 6 & 7 & 8 & 9 & 10 & 11 & 12 \\
\hline
NGC~628   & Sc & 9.70 & $-20.72$ & 7 & 25 & 5.23 & 10.96 & 7.2 & 0.254 & 0.04 & 1, 2 \\
NGC~783   & Sc & 13.18 & $-22.01$ & 43 & 57 & 0.71 & 14.56 & 70.5 & 0.222 & 0.45 & 3 \\
NGC~2336  & SB(R)bc & 11.19 & $-22.14$ & 55 & 175 & 2.51 & 23.51 & 32.2 & 0.120 & 0.41 & 4 \\
NGC~6217  & SB(R)bc & 11.89 & $-20.45$ & 33 & 162 & 1.15 & 6.89 & 20.6 & 0.158 & 0.22 & 5, 6 \\
NGC~6946  & SABc & 9.75 & $-20.68$ & 31 &  62 & 7.74 & 13.28 & 5.9 & 1.241 & 0.04 & 6 \\
NGC~7331  & Sbc & 10.20 & $-21.68$ & 75 & 169 & 4.89 & 20.06 & 14.1 & 0.331 & 0.61 & 6 \\
NGC~7678  & SBc & 12.50 & $-21.55$ & 44 &  21 & 1.04 & 14.46 & 47.8 & 0.178 & 0.23 & 6, 7 \\
\hline
\end{tabular}\\
\end{center}
\begin{flushleft}
$^a$ Absolute magnitude of a galaxy corrected for Galactic extinction and 
inclination effects. \\
$^b$ Radius of a galaxy at the isophotal level 25 mag/arcsec$^2$ in the 
$B$ band corrected for Galactic extinction and inclination effects. \\
$^c$ References: 1 -- \citet{bruevich2007}, 2 -- \citet{gusev2013b}, 
3 -- \citet{gusev2006a,gusev2006b}, 4 -- \citet{gusev2003}, 
5 -- \citet{artamonov1999}, 6 -- \citet{gusev2015}, 
7 -- \citet*{artamonov1997}. \\
\end{flushleft}
\end{table*}

A star forming region in a galaxy is a single conglomerate of newly 
formed star clusters, dust clouds, and ionised gas. Star forming regions 
have a hierarchical structure over a large range of scales. On small scales, 
there are young star clusters with diameters of a few parsecs and OB 
associations with diameters of several tens of parsecs. The largest coherent 
star formation regions are star complexes with diameters of several hundred 
parsecs \citep{elmegreen1996,efremov1998}. The sizes of the largest complexes 
can reach 2 kpc \citep{elmegreen1996b}. Common star complexes are huge groups 
of relatively young stars, associations, and clusters. Younger clusters form 
within the larger and older ones. Complexes are the largest and oldest 
`clusters', the final step in the hierarchical sequence of embedded young 
star groups. Their sizes are limited by the effective thickness of the 
galactic gaseous discs. The flocculent spiral arms and some spurs from the 
grand design spirals are, plausibly, sheared older complexes. The sizes 
of the SF~regions studied here are within the range from several 
tens to $\sim1000$~pc, with a mean value $\sim300$~pc.

Being bright, SF~regions can be observed in nearby galaxies as objects of 
magnitude  $16-20$ with emission spectra, but they cannot be resolved into 
separate stars. Our linear resolution, $\approx40$~pc in the nearest 
galaxies NGC~628 and NGC~6946, does not allow us to separate young star 
clusters and OB associations: smaller star clusters will be observed as 
star-like objects with sizes $\approx40$~pc. In more distant galaxies, 
we can observe SF~regions with sizes of 200--300~pc, i.e. star complexes 
are conglomerates of young star clusters and/or associations. 
Young massive star clusters (YMCs), embedded 
in SF~regions, are dense aggregates of young stars, formed at essentially 
the same time in the same region of space \citep*{zwart2010}. 
Most of the objects studied here are complexes of 
young massive clusters. From now on, we will use the common term 
`YMC complex' for the studied stellar populations in SF~regions. 
It should be realized that this term covers young objects of different 
types: star complexes, cluster complexes, stellar aggregates, OB associations, 
and star clusters.

The typical gas velocities in an individual H\,{\sc ii}~region 
are $15-30$ km\,s$^{-1}$ -- the velocity of hydrogen atoms at a 
temperature of $10^4$~K, as well as the velocity of the stellar wind. 
At such velocities, the lifetime  of an H\,{\sc ii} region with a typical 
size of about 100~pc is about $3-7$~Myr. After a few Myrs, when 
the surrounding gas is blown up, the SF~region exists in the form of bounded 
stellar agglomerates, which are referred to as star clusters by 
\citet{gieles2011}. Ionised gas is not observed around star clusters older 
than 10~Myr. Thus, we study in the present paper only star clusters/cluster 
complexes younger than 10~Myr.

\begin{figure}
\resizebox{1.00\hsize}{!}{\includegraphics[angle=000]{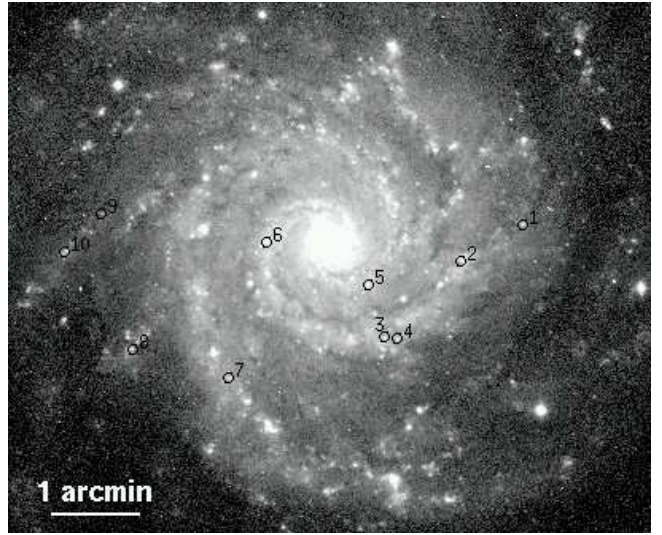}}
\caption{$B$ image of NGC~628 and positions of the galaxy's star 
forming regions. The numbers of the star forming regions from 
Table~\ref{table:phot} are indicated. North is upward and East is to the 
left.
}
\label{figure:fig628}
\end{figure}

A study of the earliest stages of these clusters and their complexes is a 
difficult task because of the impact of gas and dust on the observations of 
young massive clusters. Accounting for the effects of gas and dust on 
observations of YMCs and their complexes is very important for the 
interpretation of multicolour photometry in terms of the initial mass 
function (IMF) and the star formation rate (SFR) history. Despite a huge 
amount of both spectroscopic and photometric observations of 
the extragalactic giant H\,{\sc ii} regions, SF~regions, the overlaps of 
the spectroscopic observations of SF~regions with the photometric ones are 
very poor. Using both techniques, we can eliminate the 
degeneracy between age and extinction, and between age and metallicity. These 
degeneracies  present a hurdle to the analysis of photometric data 
\citep{scalo1986}. In addition, continuum and line emissions from ionised 
gas rival the stellar luminosity at optical wavelengths 
\citep[see compilations in][]{reines2010}.

In the case of massive star forming regions, the nebula emission is strong 
enough to affect the integrated broadband photometry. \citet{reines2010} 
found that nebular line emission is significant in many commonly used 
broad-band {\it Hubble Space Telescope (HST)} filters, including the F555W 
$V$ band and the F435W $B$ band. Emission lines detected via spectroscopy 
can be used for disentangling the effects of extinction, accounting for the 
impact of the nebula emission on integrated broadband photometry, and serve 
as valuable diagnostics of gas abundances \citep[see][and references therein] 
{dinerstein1990,sakhibov1990,reines2010}. In other words, the 
combination of spectroscopic and multicolour photometric observations of 
SF~regions provides the true colours and metallicities of young massive 
cluster complexes. This is necessary to account for the nebular emission 
impact and to eliminate `age--extinction' and `age--metallicity' 
degenerations in the comparative analysis with the theoretical evolutionary 
models of star clusters. Hereafter, we call the colours and luminosities, 
which are corrected for both the extinction and the nebula emission in broad 
bands, the `true' colours and luminosities.

With these ideas in mind, several years ago we started simultaneous 
spectroscopic and photometric observations of SF~regions in nearby galaxies. 
We should outline the progress in the solution of the above mentioned problem 
of degeneracies achieved by coupling the spectroscopy with $UBVI$ photometry 
of star clusters in M82 \citep{konstantopoulos2009}. Having combined 
information from spectroscopy and imaging, \citet{konstantopoulos2009} 
found the disc clusters to form a uniform population, displaying no 
positional dependencies with respect to age. Unlike current 
research, \citet{konstantopoulos2009} studied relatively older objects using 
absorption spectra. The spectroscopic data itself plays an important role 
in the study of the chemical evolution of galaxies, where oxygen and nitrogen 
are key elements. These aspects of the application of our spectral 
observations results, as well as the data reduction, are summarised in our 
previous papers \citep{gusev2012,gusev2013}. 

\begin{figure}
\resizebox{1.00\hsize}{!}{\includegraphics[angle=000]{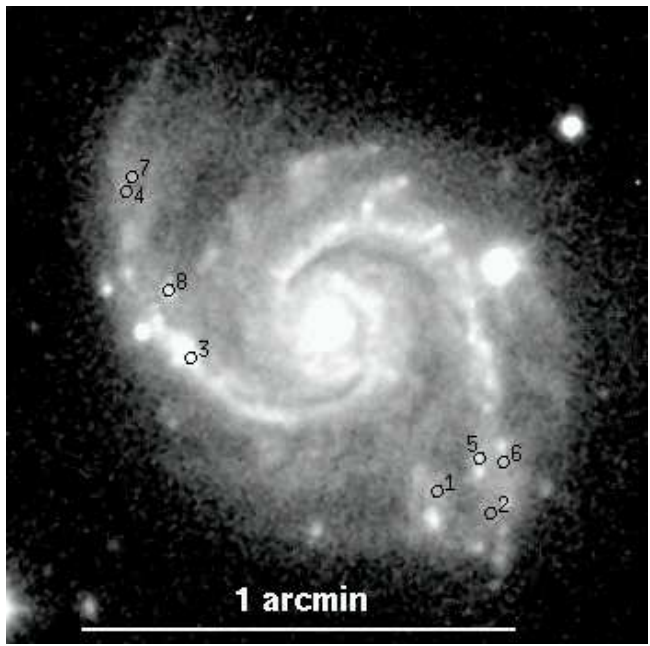}}
\caption{Same as Fig.~\ref{figure:fig628}, but for NGC~783.
}
\label{figure:fig783}
\end{figure}

This paper is organized as follows. The observations and reduction are
described in Section~\ref{sect:observ}. Section~\ref{sect:sfrs} presents 
the observed photometric parameters of the SF~regions. The reduction for the 
light absorption, and the analysis of the observational effects caused by 
observation constraints, are described in Section~\ref{sect:obs_eff}. 
In Section~\ref{sect:n2336} we compare the photometric properties of the 
cluster complexes in the studied SF~regions with the synthetic colours of 
evolution models of star clusters. Section~\ref{sect:phys} describes the 
estimation of the physical parameters of the studied SF~regions. In 
Section~\ref{sect:ocs} we also present a comparative analysis of the derived 
true colours of the YMC complexes with the intrinsic colours of 650 Galactic 
open clusters derived by \citet{kharchenko2005} and \citet{piskunov2005} and 
the synthetic colours of the evolution models. We discuss our results in 
Section~\ref{sect:discus}. Some conclusions are presented in 
Section~\ref{sect:concl}.

\section{Observations and data reduction}
\label{sect:observ}

\begin{figure}
\resizebox{1.00\hsize}{!}{\includegraphics[angle=000]{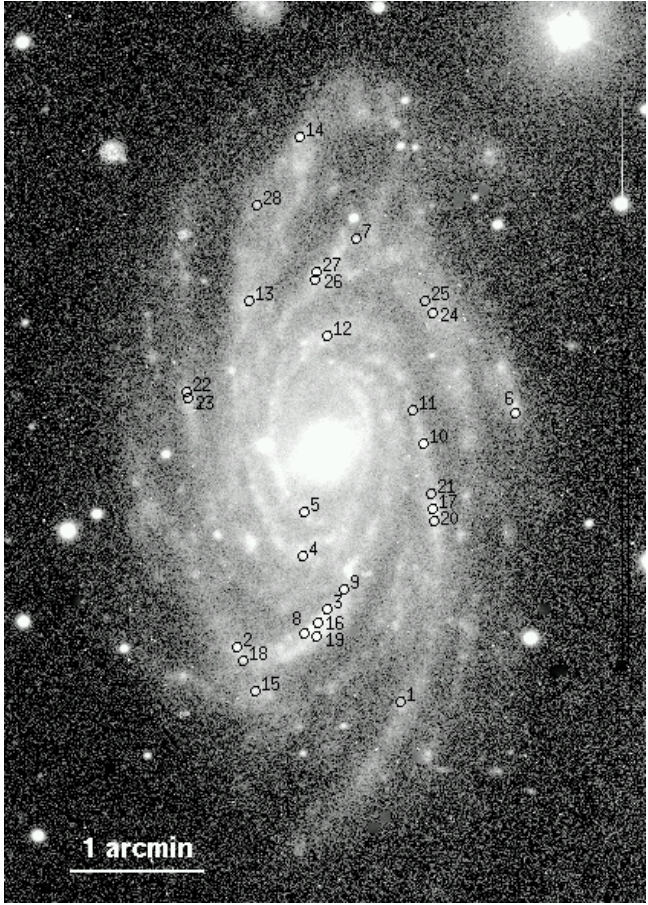}}
\caption{Same as Fig.~\ref{figure:fig628}, but for NGC~2336.
}
\label{figure:fig2336}
\end{figure}

The galaxy sample is presented in Table~\ref{table:sample}, 
where fundamental parameters from the 
LEDA\footnote{http://leda.univ-lyon1.fr/} database \citep{paturel2003} 
are provided.  The morphological type of the galaxy is given in column (2), 
the apparent and absolute magnitudes are listed in columns (3) and (4), 
the inclination and position angles are listed in columns (5) and (6), and the 
isophotal radius in arcmin and kpc are listed in columns (7) and (8). The 
adopted distances are given in column (9). Finally, the Galactic absorption 
and the dust absorption due to the inclination of a galaxy are presented 
in columns (10) and (11). The Galactic absorptions, $A(B)_{\rm Gal}$,
are taken from the NED\footnote{http://ned.ipac.caltech.edu/} database. 
The other parameters are taken from the LEDA database \citep{paturel2003}. 
Maps of the galaxies are given in 
Figs.~\ref{figure:fig628}--\ref{figure:fig7678}. The adopted value of the 
Hubble constant is equal to $H_0 = 75$ km\,s$^{-1}$Mpc$^{-1}$.

The photometric ($UBVRI$) and spectrophotometric (H$\alpha$ line) 
results for the galaxies from our sample were published earlier (see 
the notes in Table~\ref{table:sample}), with the exception of the H$\alpha$ 
spectrophotometric observations for NGC~6946 and NGC~7331. In 
Section~\ref{sect:observ1} we give a brief compilation of our earlier 
published results.

\subsection{Observations}
\label{sect:observ1}

$UBVRI$ CCD observations of NGC~628, NGC~783, NGC~6217, NGC~6946, 
NGC~7331, and NGC~7678 were obtained in 2002--2006 with the 1.5~m telescope 
of the Maidanak Observatory (Institute of Astronomy of the Academy of 
Sciences of Uzbekistan) using a SITe-2000 CCD array. The focal length of 
the telescope is 12~m. For a detailed description of the telescope and the 
CCD camera, see \citet{artamonov2010}. With broadband $U$, $B$, $V$, $R$, 
and $I$ filters, the CCD array realizes a photometric system close to the 
standard Johnson--Cousins $UBVRI$ system. The camera is cooled with liquid 
nitrogen. The size of the array is $2000\times800$ pixels. It provides a 
$8.9\times3.6$ arcmin$^2$ field of view with an image scale of 
0.267~arcsec\,pixel$^{-1}$. The seeing during our observation sessions was 
about $0.7-1.1$~arcsec.

Since the angular sizes of NGC~6946 and NGC~7331 are larger than the field of 
view, we acquired two separate images for the northern and southern parts of 
NGC~6946 and one image for the central part of NGC~7331.

The $UBVRI$ observations of NGC~2336 were carried out in 2001 with the 
1.8~m telescope of the Bohyunsan Optical Astronomy Observatory (Korea 
Astronomy and Space Science Institute) equipped with a CCD camera at the 
Ritchey--Chretien f/8 focus. The camera was a SITe AR-coated 
$2048\times2048$ pixel array, with a plate scale of 0.34 arcsec\,pixel$^{-1}$. 
The field of view was about $11.7\times11.7$ arcmin$^2$. The seeing 
was $1.7-2.3$~arcsec \citep{gusev2003}.

Spectrophotometric H$\alpha$ observations of three galaxies from our 
sample (NGC~628, NGC~6946, and NGC~7731) were made on 
26 September 2006 with the 1.5-m telescope of the Mt.~Maidanak Observatory 
with the SI-4000 CCD camera. The chip size, $4096\times4096$ pixels, provides 
a field of view of $18.1\times18.1$ arcmin$^2$, with an image scale of 
0.267 arcsec\,pixel$^{-1}$. The total exposure time was 1200~s 
($4\times300$~s) for NGC~628 and 900~s ($3\times300$~s) for NGC~6946 and 
NGC~7331. The seeing was $0.9-1.2$~arcsec.

A wide-band interference H$\alpha$ filter ($\lambda_{eff}$ =
6569\AA, FWHM = 44\AA) was used for the observations. 
The  filter parameters provide H$\alpha$+[N\,{\sc
ii}]$\lambda$6548+$\lambda$6584 imaging for the nearby galaxies 
NGC~628, NGC~6946, and NGC~7331.

\subsection{Data reduction}

\begin{figure}
\resizebox{1.00\hsize}{!}{\includegraphics[angle=000]{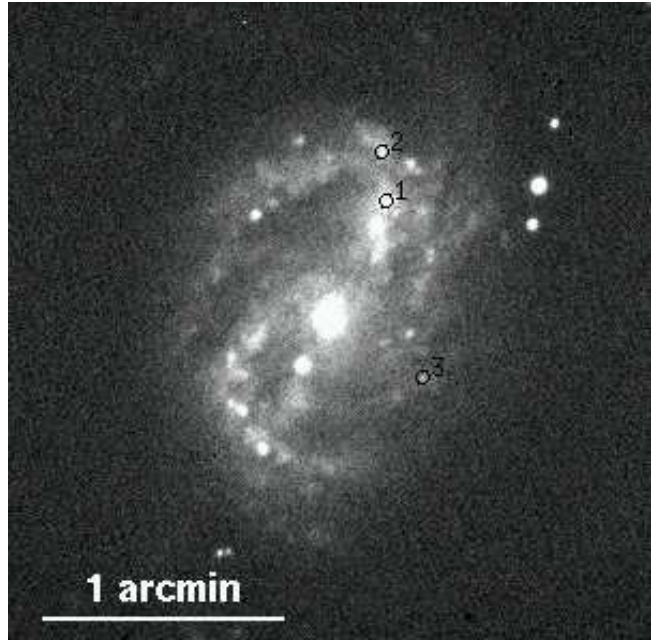}}
\caption{Same as Fig.~\ref{figure:fig628}, but for NGC~6217.
}
\label{figure:fig6217}
\end{figure}

The reduction of the photometric and spectrophotometric data was carried 
out using standard techniques, with the European Southern Observatory 
Munich Image Data Analysis 
System\footnote{http://www.eso.org/sci/software/esomidas/} ({\sc eso-midas}). 
The main image reduction stages were as follows: \\
(a) Correction for bias and flat field; \\
(b) Removal of cosmic ray traces; \\
(c) Determining the sky background, then subtracting it from each
    image frame; \\
(d) Aligning the images; \\
(e) Normalizing and joining the northern and southern parts of the
    galaxy (for NGC~628 and NGC~6946 in the $U$, $B$, $V$, $R$, and $I$ 
    passbands); \\
(f) Adding up the galaxy images taken with the same filter; \\
(g) Absolute calibration.

The absolute calibration of the photometric data involved reducing the data 
from the instrumental photometric system to the standard Johnson--Cousins 
system and correcting for the air mass using the derived colour 
equations and the results of the aperture photometry of the galaxy.

We derived the colour equations and corrected for atmospheric extinction 
using observations of stars from the fields of \citet{landolt1992} 
PG~0231+051, PG~2213--006, PG~2331+055, SA~92, SA~110, SA~111, SA~113, 
and SA~114, acquired on the same nights in the $U$, $B$, $V$, $R$, $I$, 
and H$\alpha$ filters in the wide air mass interval. A detailed study of 
the instrumental photometric system and the atmospheric extinction at the
Maidanak Observatory was presented in \citet{artamonov2010}.

In addition, the aperture photometric data for the galaxies 
from the LEDA database \citep{paturel2003} are used for the absolute 
calibration of the galaxies. The uncertainties of our photometric data and 
zero-point error for the galaxies are presented in 
Table~\ref{table:errors}.

\begin{table*}
\caption[]{\label{table:errors}
The photometric accuracies and image scales of the galaxies.
}
\begin{center}
\begin{tabular}{ccccccc} \hline \hline
Galaxy & $\Delta B$ (zero-point) & $\Delta (U-B)$ & $\Delta (B-V)$ & 
         $\Delta (V-R)$          & $\Delta (V-I)$ & scale \\
       & (mag)                   & (mag)          & (mag)          & 
         (mag)                   & (mag)          & (pc/arcsec) \\
\hline
NGC~628   & 0.03 & 0.10 & 0.03 & 0.02 & 0.03 &  34.9 \\
NGC~783   & 0.15 & 0.04 & 0.02 & 0.03 & 0.02 & 342   \\
NGC~2336  & 0.02 & 0.03 & 0.04 & 0.06 & 0.07 & 156   \\
NGC~6217  & 0.13 & 0.11 & 0.02 & 0.09 & 0.09 &  99.9 \\
NGC~6946  & 0.06 & 0.06 & 0.02 & 0.04 & 0.04 &  28.6 \\
NGC~7331  & 0.02 & 0.11 & 0.03 & 0.03 & 0.03 &  68.4 \\
NGC~7678  & 0.04 & 0.04 & 0.02 & 0.01 & 0.03 & 232   \\
\hline
\end{tabular}\\
\end{center}
\end{table*}

\begin{figure*}
\resizebox{1.00\hsize}{!}{\includegraphics[angle=000]{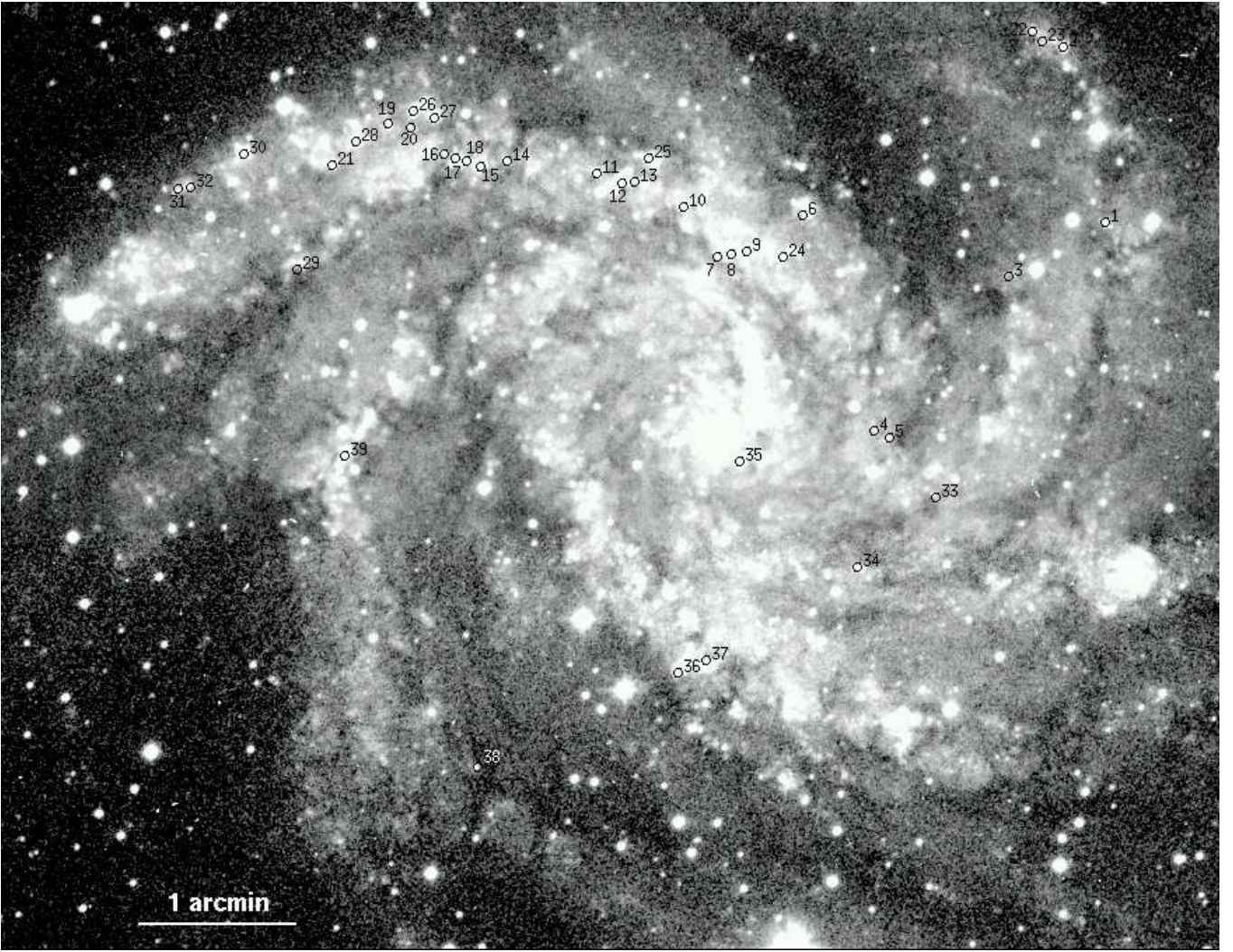}}
\caption{Same as Fig.~\ref{figure:fig628}, but for NGC~6946.
}
\label{figure:fig6946}
\end{figure*}

The spectrophotometric data reduction and the calibration of H$\alpha$ fluxes 
were described in detail in \citet{gusev2013b}. The calibration of 
H$\alpha$ fluxes was made using the results of the spectrophotometric 
observations of \citet{belley1992} and \citet{kennicutt1980} for NGC~628 and 
NGC~6946. We adopted the value 
$3.70\times10^{-18}$~erg\,s$^{-1}$\,cm$^{-2}$/ADU for 
the coefficient of reduction from our instrumental H$\alpha$ flux to the flux 
in physical units for all three galaxies. Corrections for the air mass and 
the radial velocity  were also taken into account.

The derived spectrophotometric H$\alpha$+[N\,{\sc ii}] fluxes from 
the H\,{\sc ii}~regions are presented in Table~\ref{table:phot}. The data are 
not corrected for interstellar reddening.

\begin{figure}
\resizebox{1.00\hsize}{!}{\includegraphics[angle=000]{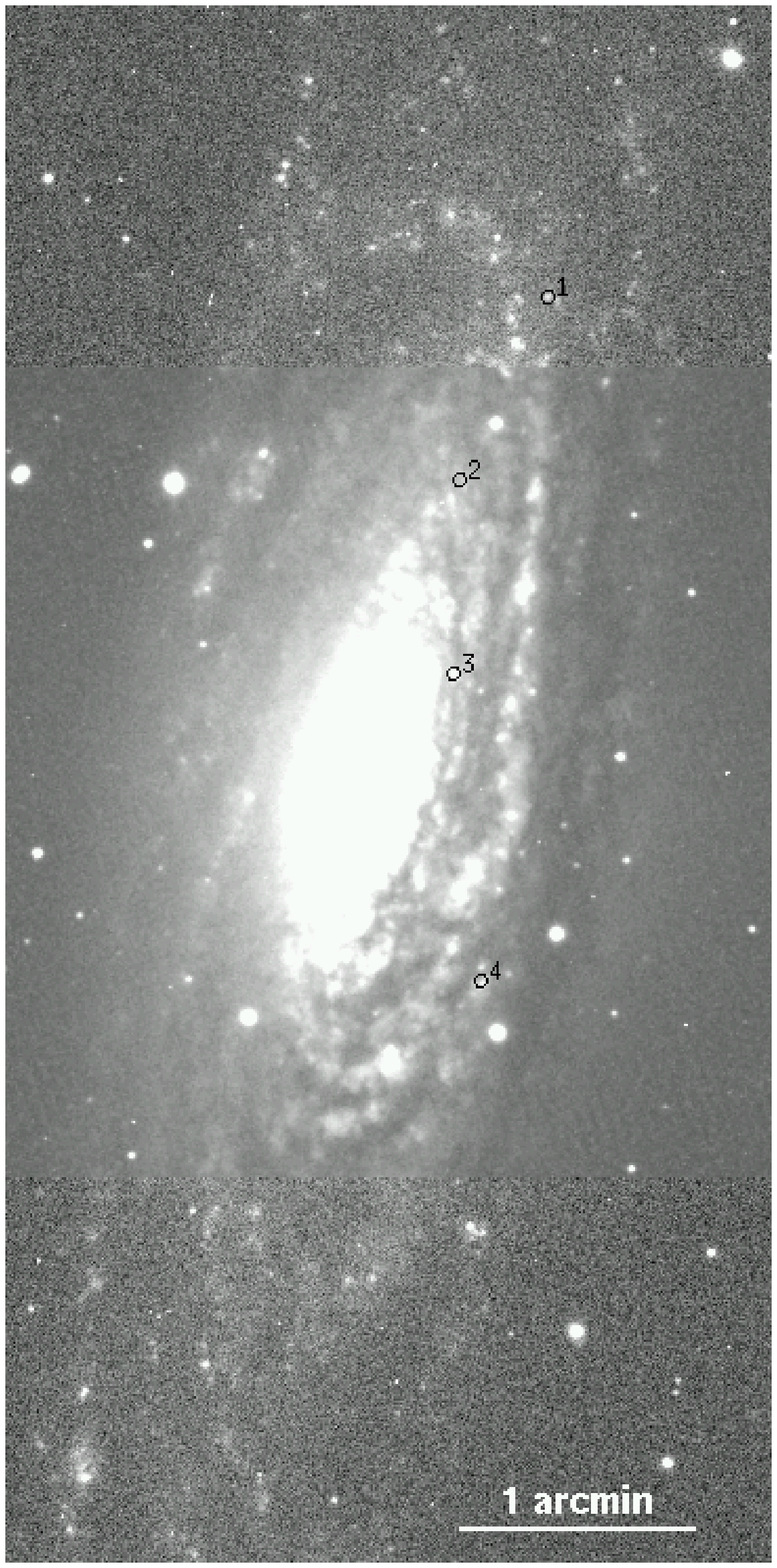}}
\caption{$B$ (central part) and H$\alpha$ (northern and southern parts) 
image of NGC~7331. Numbering of star forming regions and 
the orientation are as in Fig.~\ref{figure:fig628}.
}
\label{figure:fig7331}
\end{figure}

\begin{figure}
\resizebox{1.00\hsize}{!}{\includegraphics[angle=000]{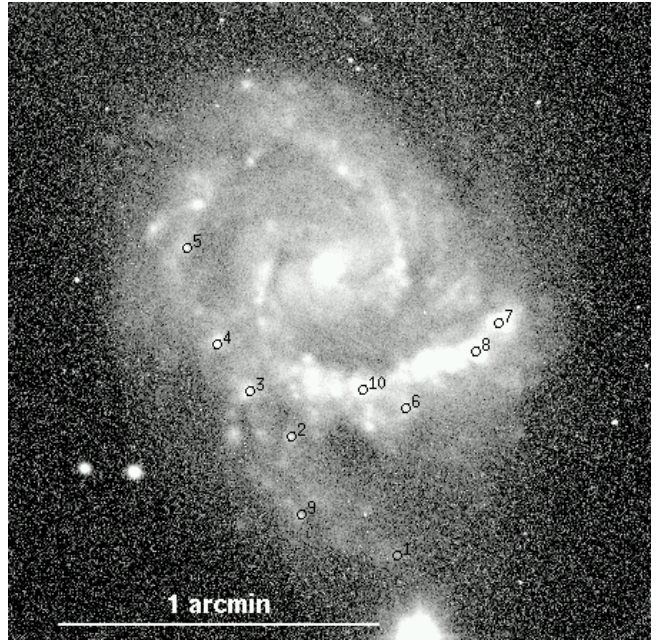}}
\caption{Same as Fig.~\ref{figure:fig628}, but for NGC~7678.
}
\label{figure:fig7678}
\end{figure}

\begin{figure}
\vspace{2.6mm}
\resizebox{1.00\hsize}{!}{\includegraphics[angle=000]{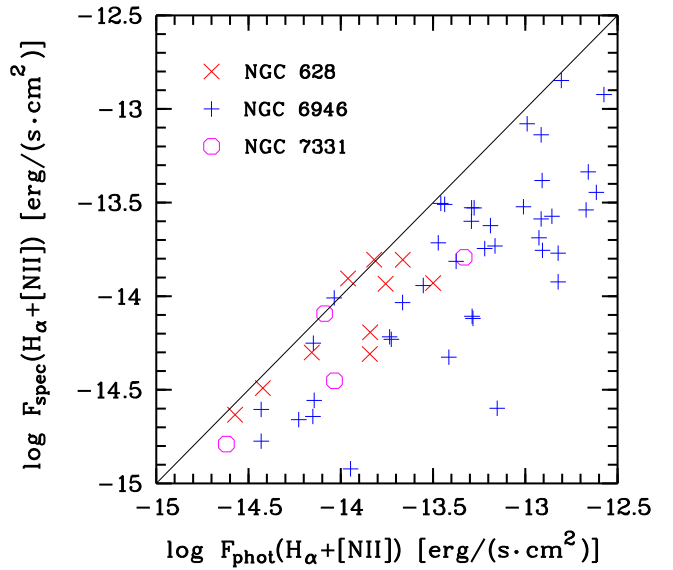}}
\caption{Comparison between the spectroscopic (spec) and spectrophotometric 
(phot) H$\alpha$+[N\,{\sc ii}] fluxes for H\,{\sc ii} regions based on our 
data.
}
\label{figure:ha}
\end{figure}

In Fig.~\ref{figure:ha} we compare our spectrophotometric fluxes (see 
Table~\ref{table:phot}) in three galaxies with the spectroscopic 
H$\alpha$+[N\,{\sc ii}] fluxes \citep{gusev2012,gusev2013}. 
Fig.~\ref{figure:ha} shows that the spectroscopic flux value is $\sim40\%$ 
of the total flux. The spectroscopic flux values are lower limits for the 
absolute fluxes for slit spectroscopy measurements, because not all 
radiation from the H\,{\sc ii} regions falls into the slit. The lack of 
H\,{\sc ii} regions with $F_{\rm spec} > F_{\rm phot}$ testifies to the 
correctness of the mutual calibration of the spectral fluxes with the 
spectrophotometric ones.

The linear scales of our images calculated for the adopted distances to the 
galaxies are given in the last column of Table~\ref{table:errors}.

\section{Photometric parameters of star forming regions}
\label{sect:sfrs}

We identified the SF~regions in NGC~628 and NGC~6946 using 
the list of H\,{\sc ii} regions of \citet{belley1992}. The identification 
of the SF~regions in the other galaxies was made by eye. Maps of the galaxies 
with identified star forming complexes are presented in 
Figs.~\ref{figure:fig628}--\ref{figure:fig7678}.

We took the geometric mean of the major and minor axes of the star forming 
complex for the SF~region's characteristic diameter $d$: 
$d = \sqrt{d_{max} \times d_{min}}$. We measured $d_{max}$ and $d_{min}$ 
from the radial $V$ profiles at the half-maximum brightness level (FWHM) 
for regions having a starlike profile, or by the distance between 
the points of maximum flux gradient for regions having extended (diffuse) 
profiles. We adopted the seeing as the error of the size measurements, which 
definitely exceeds all other errors.

\begin{figure}
\vspace{0.5cm}
\resizebox{0.90\hsize}{!}{\includegraphics[angle=000]{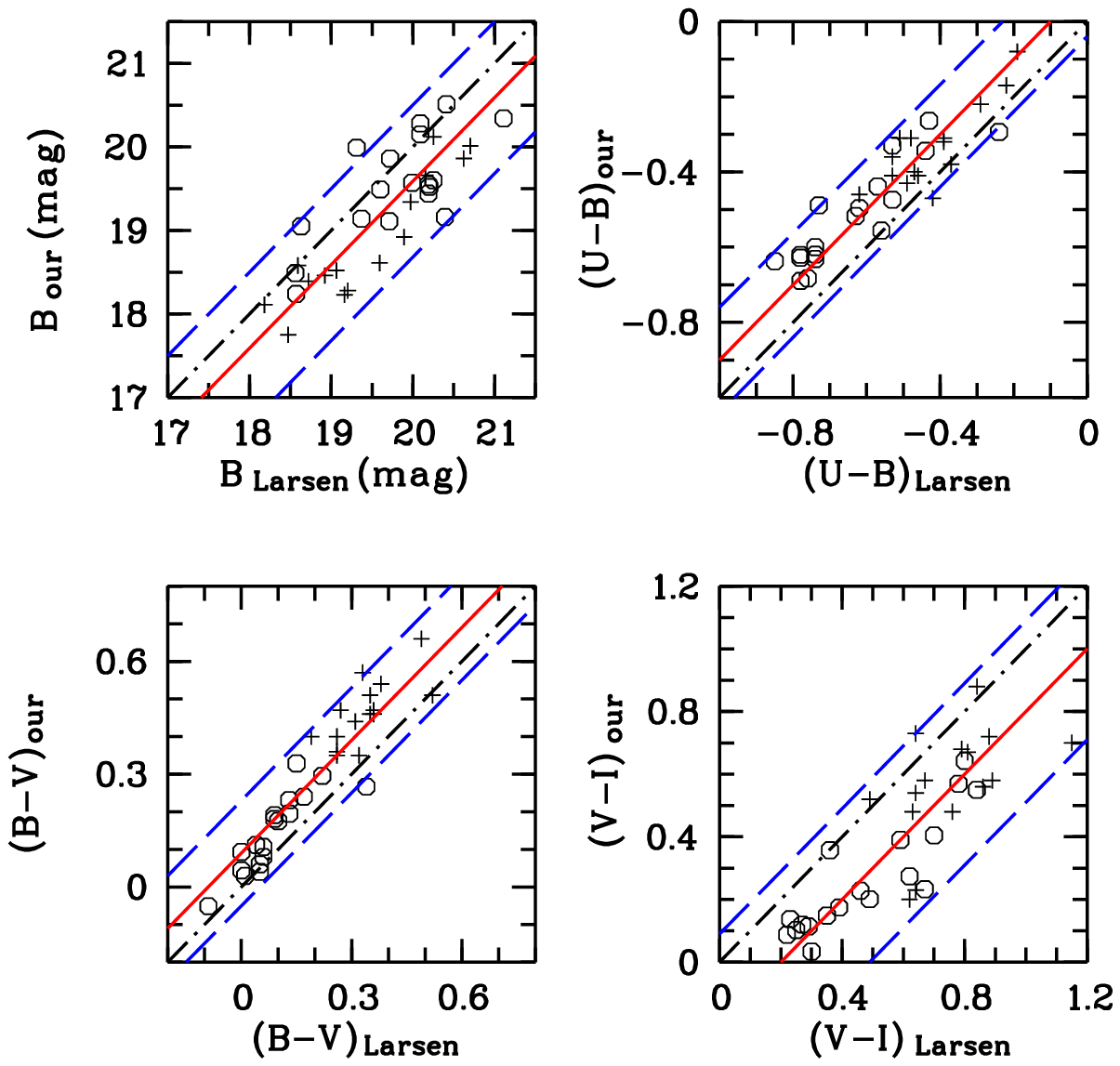}}
\caption{Comparison between our and the reference 
\citep{larsen1999,larsen2004} $B$ magnitudes (top-left), $U-B$ 
(top-right), $B-V$ (bottom-left), and $V-I$ (bottom-right) colour indices 
for star forming regions in NGC~628 and NGC~6946. The pluses are objects in 
NGC~6946, the circles are objects in NGC~628. The solid lines are 
linear fits, computed for SF~regions (pluses and circles). Dashed lines are 
upper and lower $95\%$ prediction limits of the linear fit. Dot dashed lines 
are one-to-one correlation. Magnitudes and colour indices are not corrected 
for extinction.
}
\label{figure:lofig}
\end{figure}

The measurements of the apparent total $B$ magnitude and colour indices 
$U-B$, $B-V$, $V-R$, and $V-I$ were made within a round 
aperture. To measure $B$ magnitudes, we used an aperture size equal to 
the sum of the $d_{max}$ and the seeing in the $B$ band; in the case of 
measurements of colour indices, the aperture size is equal to $d_{max}$. 

\begin{table*}
\caption[]{\label{table:phot}
Photometric parameters of the star forming regions.
}
\begin{center}
\begin{tabular}{ccccccccccc} \hline \hline
H\,{\sc ii} & Ref.$^a$ & N--S$^b$ & E--W$^b$ & $B$ & $M_B$ & $U-B$ & 
$B-V$ & $V-R$ & $V-I$ & $\log F$(H$\alpha$) \\
region      &          & (arcsec) & (arcsec) & (mag) & (mag) & (mag) &  
(mag) & (mag) & (mag) & ([erg \,s$^{-1}$\,cm$^{-2}$]) \\
1 & 2 & 3 & 4 & 5 & 6 & 7 & 8 & 9 & 10 & 11 \\
\hline
\multicolumn{11}{c}{NGC~628} \\
 1 & 94 & +12.5 & +129.0 & 19.91 &  $-9.38\pm0.01$ & $-0.70\pm0.01$ &
$0.11\pm0.01$ & $0.23\pm0.01$ &  $0.51\pm0.01$ & $-14.16\pm0.01$ \\
 2 & 92 & $-12.0$ &  +87.7 & 19.16 & $-10.13\pm0.01$ & $-0.33\pm0.02$ &
$0.24\pm0.01$ & $0.36\pm0.01$ &  $0.76\pm0.01$ & $-13.84\pm0.01$ \\
 3 & 77 & $-64.3$ &  +36.5 & 20.54 &  $-8.75\pm0.02$ & $-0.78\pm0.02$ &
$0.17\pm0.04$ & $0.87\pm0.04$ &  $0.52\pm0.06$ & $-13.96\pm0.01$ \\
 4 & 75 & $-65.4$ &  +45.0 & 20.15 &  $-9.14\pm0.03$ & $-0.49\pm0.03$ &
$0.10\pm0.03$ & $0.24\pm0.04$ &  $0.04\pm0.06$ & $-14.57\pm0.01$ \\
 5 & 73 & $-28.8$ &  +25.3 & 19.55 &  $-9.74\pm0.01$ & $-0.69\pm0.02$ &
$0.04\pm0.03$ & $0.41\pm0.04$ &  $0.60\pm0.05$ & $-13.84\pm0.01$ \\
 6 & 25 &  +0.2 &  $-43.0$ & 19.44 &  $-9.85\pm0.01$ & $-0.62\pm0.02$ &
$0.08\pm0.03$ & $0.09\pm0.04$ &  $0.32\pm0.06$ & $-14.42\pm0.01$ \\
 7 & 62 & $-91.8$ &  $-68.1$ & 20.11 &  $-9.18\pm0.02$ & $-0.72\pm0.02$ &
$0.18\pm0.02$ & $0.76\pm0.02$ &  $0.55\pm0.03$ & $-13.76\pm0.01$ \\
 8 & 58 & $-72.0$ & $-132.6$ & 19.76 &  $-9.53\pm0.01$ & $-0.89\pm0.01$ &
$0.11\pm0.01$ & $0.64\pm0.01$ &  $0.33\pm0.02$ & $-13.66\pm0.01$ \\
 9 & 35 & +19.7 & $-153.1$ & 21.50 &  $-7.79\pm0.02$ & $-0.99\pm0.02$ &
$0.27\pm0.02$ & $0.70\pm0.02$ & $-0.16\pm0.06$ & $-13.82\pm0.01$ \\
10 & 38 &  $-6.4$ & $-177.9$ & 18.82 & $-10.47\pm0.01$ & $-0.51\pm0.01$ &
$0.15\pm0.01$ & $0.38\pm0.01$ &  $0.55\pm0.01$ & $-13.50\pm0.01$ \\
\multicolumn{11}{c}{NGC~783} \\
1   & --- & $-24.3$ & +16.0 & 22.23 & $-12.01\pm0.12$ & $-0.30\pm0.16$ &
$0.20\pm0.18$ & $0.54\pm0.16$ &  $0.74\pm0.20$ & --- \\
2   & --- & $-27.5$ & +23.7 & 20.85 & $-13.39\pm0.02$ &  $0.19\pm0.04$ & 
$0.54\pm0.03$ & $0.19\pm0.05$ &  $0.80\pm0.05$ & --- \\
3   & --- &  $-4.8$ & $-19.7$ & 20.67 & $-13.57\pm0.02$ & $-0.24\pm0.02$ & 
$0.31\pm0.04$ & $0.30\pm0.06$ &  $0.61\pm0.08$ & --- \\
4   & --- & +19.5 & $-29.1$ & 21.71 & $-12.53\pm0.04$ & $-0.15\pm0.05$ & 
$0.54\pm0.07$ & $0.44\pm0.07$ &  $1.07\pm0.07$ & --- \\
5   & --- & $-19.5$ & +22.1 & 21.81 & $-12.43\pm0.02$ & $-0.37\pm0.03$ & 
$0.37\pm0.03$ & $0.20\pm0.05$ &  $0.45\pm0.06$ & --- \\
6   & --- & $-20.0$ & +25.6 & 22.12 & $-12.12\pm0.05$ & $-0.52\pm0.05$ & 
$0.53\pm0.07$ & $0.40\pm0.08$ &  $0.81\pm0.10$ & --- \\
7   & --- & +21.6 & $-28.3$ & 21.83 & $-12.41\pm0.04$ & $-0.25\pm0.05$ & 
$0.44\pm0.07$ & $0.41\pm0.07$ &  $0.53\pm0.11$ & --- \\
8   & --- &  +5.1 & $-22.9$ & 20.92 & $-13.32\pm0.04$ & $-0.62\pm0.04$ & 
$0.28\pm0.07$ & $0.17\pm0.10$ &  $0.10\pm0.15$ & --- \\
\multicolumn{11}{c}{NGC~2336} \\
  1   & --- & $-113.8$ & +33.5 & 19.83 & $-12.71\pm0.03$ & $-0.69\pm0.05$ & 
$0.42\pm0.03$ & $0.34\pm0.04$ & $0.51\pm0.04$ & --- \\
  2   & 28  &  $-89.0$ & $-40.0$ & 19.84 & $-12.70\pm0.03$ & $-0.30\pm0.05$ & 
$0.39\pm0.03$ & $0.52\pm0.02$ & $0.50\pm0.03$ & --- \\
  3   & 27a &  $-71.8$ &  +0.2 & 20.19 & $-12.35\pm0.05$ & $-0.43\pm0.06$ & 
$0.27\pm0.07$ & $0.44\pm0.08$ & $0.69\pm0.08$ & --- \\
  4   & 26  &  $-47.8$ & $-10.5$ & 19.51 & $-13.03\pm0.05$ & $-0.58\pm0.07$ & 
$0.52\pm0.06$ & $0.54\pm0.07$ & $0.95\pm0.08$ & --- \\
  5   & 22  &  $-27.5$ &  $-9.8$ & 20.55 & $-11.99\pm0.03$ & $-0.51\pm0.07$ & 
$0.54\pm0.05$ & $0.54\pm0.05$ & $0.94\pm0.07$ & --- \\
  6   & 17  &  +17.5 & +85.4 & 19.80 & $-12.74\pm0.01$ & $-0.50\pm0.03$ & 
$0.32\pm0.02$ & $0.45\pm0.02$ & $0.52\pm0.03$ & --- \\
  7   &  5  &  +95.9 & +13.6 & 21.07 & $-11.47\pm0.03$ & $-0.53\pm0.08$ & 
$0.44\pm0.04$ & $0.73\pm0.04$ & $0.76\pm0.05$ & --- \\
  8   & 27d &  $-82.5$ &  $-9.5$ & 19.67 & $-12.87\pm0.01$ & $-0.42\pm0.03$ & 
$0.42\pm0.02$ & $0.41\pm0.03$ & $0.59\pm0.04$ & --- \\
  9   & --- &  $-62.6$ &  +8.1 & 20.66 & $-11.88\pm0.03$ & $-0.36\pm0.07$ & 
$0.48\pm0.04$ & $0.48\pm0.06$ & $0.48\pm0.07$ & --- \\
 10   & 19  &   +3.4 & +43.8 & 19.42 & $-13.12\pm0.02$ & $-0.46\pm0.03$ & 
$0.27\pm0.02$ & $0.36\pm0.03$ & $0.53\pm0.04$ & --- \\
 11   & 16  &  +18.2 & +38.7 & 20.90 & $-11.64\pm0.05$ & $-0.70\pm0.09$ & 
$0.52\pm0.06$ & $0.48\pm0.05$ & $0.83\pm0.07$ & --- \\
 12   & 10  &  +52.3 &  +0.2 & 20.38 & $-12.16\pm0.05$ & $-0.25\pm0.10$ & 
$0.55\pm0.06$ & $0.52\pm0.06$ & $0.82\pm0.08$ & --- \\
 13   &  8  &  +68.1 & $-34.5$ & 19.33 & $-13.21\pm0.03$ & $-0.66\pm0.04$ & 
$0.48\pm0.03$ & $0.51\pm0.03$ & $0.71\pm0.03$ & --- \\
 14   &  2  & +142.0 & $-11.5$ & 20.01 & $-12.53\pm0.03$ & $-0.65\pm0.04$ & 
$0.32\pm0.03$ & $0.37\pm0.04$ & $0.44\pm0.05$ & --- \\
 15   & 30  & $-109.0$ & $-32.1$ & 19.93 & $-12.61\pm0.03$ & $-0.58\pm0.05$ & 
$0.39\pm0.03$ & $0.32\pm0.03$ & $0.47\pm0.04$ & --- \\
 16   & 27b &  $-77.3$ &  $-3.6$ & 20.52 & $-12.02\pm0.08$ & $-0.26\pm0.13$ & 
$0.40\pm0.12$ & $0.43\pm0.13$ & $0.61\pm0.13$ & --- \\
 17   & 21  &  $-26.1$ & +48.3 & 18.26 & $-14.28\pm0.01$ & $-0.33\pm0.01$ & 
$0.29\pm0.01$ & $0.43\pm0.01$ & $0.56\pm0.02$ & --- \\
 18   & 29  &  $-94.5$ & $-37.3$ & 21.20 & $-11.34\pm0.04$ & $-0.50\pm0.10$ & 
$0.38\pm0.06$ & $0.66\pm0.07$ & $0.57\pm0.09$ & --- \\
 19   & 27e &  $-83.9$ &  $-4.3$ & 21.12 & $-11.42\pm0.10$ & $-0.19\pm0.15$ & 
$0.39\pm0.13$ & $0.37\pm0.15$ & $0.49\pm0.17$ & --- \\
 20   & --- &  $-31.6$ & +48.6 & 21.14 & $-11.40\pm0.03$ & $-0.45\pm0.05$ & 
$0.71\pm0.04$ & $0.34\pm0.05$ & $0.70\pm0.06$ & --- \\
 21   & --- &  $-19.3$ & +47.6 & 20.87 & $-11.67\pm0.05$ & $-0.24\pm0.10$ & 
$0.59\pm0.09$ & $0.39\pm0.12$ & $0.43\pm0.17$ & --- \\
 22   & --- &  +27.2 & $-62.7$ & 21.76 & $-10.78\pm0.02$ & $-0.91\pm0.04$ & 
$0.68\pm0.03$ & $0.61\pm0.03$ & $0.65\pm0.04$ & --- \\
 23   & 23  &  +24.4 & $-61.7$ & 20.27 & $-12.27\pm0.01$ & $-0.97\pm0.02$ & 
$0.57\pm0.02$ & $0.52\pm0.02$ & $0.68\pm0.03$ & --- \\
 24   &  9  &  +62.6 & +48.3 & 20.84 & $-11.70\pm0.05$ & $-0.57\pm0.09$ & 
$0.45\pm0.05$ & $0.25\pm0.05$ & $0.46\pm0.06$ & --- \\
 25   & --- &  +67.7 & +44.5 & 20.27 & $-12.27\pm0.03$ & $-0.60\pm0.05$ & 
$0.55\pm0.04$ & $0.22\pm0.04$ & $0.38\pm0.06$ & --- \\
 26   & --- &  +77.7 &  $-5.3$ & 21.18 & $-11.36\pm0.04$ & $-0.64\pm0.06$ & 
$0.57\pm0.06$ & $0.35\pm0.07$ & $0.55\pm0.07$ & --- \\
 27   &  7  &  +80.8 &  $-4.6$ & 20.07 & $-12.47\pm0.02$ & $-0.31\pm0.05$ & 
$0.32\pm0.05$ & $0.25\pm0.06$ & $0.63\pm0.07$ & --- \\
 28   &  4  & +111.7 & $-31.1$ & 20.73 & $-11.81\pm0.04$ & $-0.60\pm0.09$ & 
$0.49\pm0.06$ & $0.56\pm0.06$ & $0.42\pm0.07$ & --- \\
\multicolumn{11}{c}{NGC~6217} \\
1 & --- & +29.8 & +13.9 & 17.19 & $-14.35\pm0.01$ & $-0.09\pm0.01$ & 
$0.58\pm0.01$ & $0.50\pm0.01$ & $1.13\pm0.01$ & --- \\
2 & --- & +42.2 & +13.0 & 17.66 & $-13.44\pm0.04$ & $-0.56\pm0.05$ & 
$0.17\pm0.07$ & $0.45\pm0.09$ & $0.99\pm0.10$ & --- \\
3 & --- & $-14.6$ & +23.2 & 19.83 & $-11.67\pm0.08$ & $-0.83\pm0.05$ & 
$0.06\pm0.06$ & $0.53\pm0.07$ & $0.65\pm0.16$ & --- \\
\multicolumn{11}{c}{NGC~6946} \\
 1 & 116 &  +79.7 & +146.5 & 20.09 &  $-8.76\pm0.01$ & $-0.33\pm0.01$ & 
$0.41\pm0.02$ & $0.64\pm0.02$ & $0.61\pm0.03$ & $-13.73\pm0.01$ \\
 2 & 143 & +146.9 & +130.5 & 18.76 & $-10.09\pm0.01$ & $-0.37\pm0.01$ & 
$0.50\pm0.01$ & $0.85\pm0.01$ & $0.79\pm0.02$ & $-12.80\pm0.01$ \\
 3    & --- &  +58.9 & +110.0 & 20.57 &  $-8.28\pm0.07$ & $-0.64\pm0.05$ & 
$0.53\pm0.07$ & $0.23\pm0.07$ & $0.67\pm0.11$ & $-13.99\pm0.01$ \\
 4    & --- &   $-0.3$ &  +58.8 & 21.52 &  $-7.33\pm0.08$ & $-0.45\pm0.08$ & 
$0.62\pm0.12$ & $0.60\pm0.13$ & $1.05\pm0.17$ & $-14.14\pm0.01$ \\
 5    & --- &   $-2.7$ &  +64.6 & 20.73 &  $-8.12\pm0.06$ & $-0.02\pm0.10$ & 
$0.55\pm0.15$ & $0.36\pm0.18$ & $0.93\pm0.21$ & $-14.43\pm0.02$ \\
\hline
\end{tabular}\\
\end{center}
\begin{flushleft}
$^a$ ID number of H\,{\sc ii} region by \citet{belley1992} for NGC~628
and NGC~6946, and by \citet{gusev2003} for NGC~2336. \\
$^b$ Galactocentric coordinates. Positive values correspond to the 
northern and western positions. \\
\end{flushleft}
\end{table*}

\setcounter{table}{2}
\begin{table*}
\caption[]{
Continued}
\begin{center}
\begin{tabular}{ccccccccccc} \hline \hline
H\,{\sc ii} & Ref.$^a$ & N--S$^b$ & E--W$^b$ & $B$ & $M_B$ & $U-B$ &
$B-V$ & $V-R$ & $V-I$ & $\log F$(H$\alpha$) \\
region      &          & (arcsec) & (arcsec) & (mag) & (mag) & (mag) &
(mag) & (mag) & (mag) & ([erg \,s$^{-1}$\,cm$^{-2}$]) \\
1 & 2 & 3 & 4 & 5 & 6 & 7 & 8 & 9 & 10 & 11 \\
\hline
\multicolumn{11}{c}{NGC~6946} \\
 6    & 149 &  +82.6 &  +31.6 & 18.11 & $-10.74\pm0.01$ & $-0.41\pm0.01$ & 
$0.40\pm0.01$ & $0.32\pm0.01$ & $0.69\pm0.01$ & $-13.22\pm0.01$ \\
 7    & --- &  +66.3 &   $-1.2$ & 21.29 &  $-7.56\pm0.07$ & $-0.50\pm0.08$ & 
$0.32\pm0.14$ & $0.67\pm0.18$ & $0.44\pm0.41$ & $-13.95\pm0.01$ \\
 8    & --- &  +67.7 &   +4.1 & 19.36 &  $-9.49\pm0.02$ & $-0.08\pm0.02$ & 
$0.51\pm0.03$ & $0.39\pm0.05$ & $0.91\pm0.06$ & $-14.15\pm0.01$ \\
 9    & 150 &  +68.5 &  +10.2 & 18.39 & $-10.46\pm0.01$ & $-0.36\pm0.01$ & 
$0.47\pm0.02$ & $0.65\pm0.02$ & $0.89\pm0.03$ & $-12.91\pm0.01$ \\
10    &  10 &  +85.5 &  $-14.0$ & 18.58 & $-10.27\pm0.02$ & $-0.32\pm0.02$ & 
$0.54\pm0.02$ & $0.44\pm0.02$ & $0.69\pm0.02$ & $-13.19\pm0.01$ \\
11    & --- &  +98.6 &  $-46.8$ & 18.92 &  $-9.93\pm0.01$ & $-0.31\pm0.01$ & 
$0.57\pm0.02$ & $0.72\pm0.02$ & $0.79\pm0.05$ & $-13.01\pm0.01$ \\
12    &   8 &  +94.9 &  $-37.5$ & 18.28 & $-10.57\pm0.01$ & $-0.31\pm0.01$ & 
$0.47\pm0.01$ & $0.43\pm0.01$ & $0.79\pm0.02$ & $-13.16\pm0.01$ \\
13    &   6 &  +95.4 &  $-32.4$ & 19.82 &  $-9.03\pm0.02$ & $-0.33\pm0.03$ & 
$0.69\pm0.04$ & $0.73\pm0.04$ & $1.22\pm0.05$ & $-13.47\pm0.01$ \\
14    &  24 & +103.4 &  $-81.0$ & 20.12 &  $-8.73\pm0.02$ & $-0.22\pm0.03$ & 
$0.66\pm0.03$ & $0.78\pm0.02$ & $0.77\pm0.04$ & $-13.44\pm0.01$ \\
15    &  25 & +101.0 &  $-91.4$ & 18.46 & $-10.39\pm0.01$ & $-0.43\pm0.01$ & 
$0.40\pm0.01$ & $0.65\pm0.02$ & $0.73\pm0.02$ & $-12.82\pm0.01$ \\
16    &  29 & +106.1 & $-105.0$ & 20.32 &  $-8.53\pm0.01$ & $-0.30\pm0.02$ & 
$0.70\pm0.02$ & $1.16\pm0.02$ & $1.53\pm0.03$ & $-13.29\pm0.01$ \\
17    &  28 & +104.2 & $-101.0$ & 18.21 & $-10.64\pm0.01$ & $-0.21\pm0.01$ & 
$0.61\pm0.01$ & $0.87\pm0.01$ & $1.23\pm0.01$ & $-12.67\pm0.01$ \\
18    &  26 & +103.1 &  $-96.7$ & 19.52 &  $-9.33\pm0.01$ & $-0.33\pm0.01$ & 
$0.49\pm0.01$ & $0.52\pm0.01$ & $0.58\pm0.02$ & $-13.37\pm0.01$ \\
19    &  16 & +117.8 & $-126.3$ & 19.66 &  $-9.19\pm0.02$ & $-0.47\pm0.03$ & 
$0.44\pm0.04$ & $0.69\pm0.03$ & $1.09\pm0.04$ & $-13.29\pm0.01$ \\
20    & --- & +116.2 & $-118.0$ & 21.33 &  $-7.52\pm0.05$ & $-0.37\pm0.07$ & 
$0.25\pm0.09$ & $0.57\pm0.11$ &      ---      & $-14.04\pm0.01$ \\
21    &  19 & +101.8 & $-147.6$ & 18.61 & $-10.24\pm0.01$ & $-0.46\pm0.01$ & 
$0.35\pm0.01$ & $0.74\pm0.01$ & $0.94\pm0.02$ & $-12.91\pm0.01$ \\
22    & 146 & +153.0 & +118.8 & 19.61 &  $-9.24\pm0.01$ & $-0.19\pm0.02$ & 
$0.54\pm0.01$ & $0.77\pm0.01$ & $1.10\pm0.02$ & $-13.28\pm0.01$ \\
23    & 145 & +149.3 & +122.5 & 19.28 &  $-9.57\pm0.01$ & $-0.35\pm0.01$ & 
$0.53\pm0.01$ & $0.97\pm0.01$ & $1.13\pm0.02$ & $-12.93\pm0.01$ \\
24    & 151 &  +66.3 &  +23.8 & 19.74 &  $-9.11\pm0.03$ &  $0.33\pm0.04$ & 
$0.98\pm0.04$ & $0.50\pm0.04$ & $1.05\pm0.04$ & $-14.15\pm0.01$ \\
25    &   5 & +104.5 &  $-27.1$ & 17.96 & $-10.89\pm0.01$ & $-0.51\pm0.02$ & 
$0.38\pm0.02$ & $0.63\pm0.03$ & $0.63\pm0.04$ & $-12.61\pm0.01$ \\
26    &  15 & +122.3 & $-116.7$ & 18.23 & $-10.62\pm0.01$ & $-0.41\pm0.01$ & 
$0.36\pm0.02$ & $0.55\pm0.02$ & $0.43\pm0.04$ & $-12.91\pm0.01$ \\
27    &  14 & +119.9 & $-108.7$ & 17.75 & $-11.10\pm0.01$ & $-0.40\pm0.01$ & 
$0.35\pm0.01$ & $0.42\pm0.02$ & $0.40\pm0.03$ & $-12.66\pm0.01$ \\
28    &  17 & +110.6 & $-138.6$ & 18.52 & $-10.33\pm0.01$ & $-0.17\pm0.01$ & 
$0.46\pm0.01$ & $0.34\pm0.01$ & $0.88\pm0.01$ & $-13.41\pm0.01$ \\
29    & --- &  +61.8 & $-161.2$ & 18.96 &  $-9.89\pm0.01$ & $-0.24\pm0.02$ & 
$0.50\pm0.02$ & $0.40\pm0.02$ & $0.89\pm0.03$ & $-13.29\pm0.01$ \\
30    &  39 & +105.8 & $-181.2$ & 17.93 & $-10.92\pm0.01$ & $-0.44\pm0.01$ & 
$0.59\pm0.02$ & $0.62\pm0.01$ & $0.74\pm0.02$ & $-12.57\pm0.01$ \\
31    & --- &  +92.5 & $-206.3$ & 19.73 &  $-9.12\pm0.02$ & $-0.42\pm0.02$ & 
$0.31\pm0.03$ & $0.33\pm0.04$ & $0.71\pm0.05$ & $-13.73\pm0.01$ \\
32    & --- &  +93.3 & $-201.5$ & 19.66 &  $-9.19\pm0.01$ & $-0.40\pm0.02$ & 
$0.46\pm0.02$ & $0.31\pm0.02$ & $0.52\pm0.03$ & $-13.66\pm0.01$ \\
33    & --- &  $-25.7$ &  +82.0 & 21.30 &  $-7.55\pm0.15$ &  $0.03\pm0.21$ & 
$0.75\pm0.20$ &      ---      & $1.53\pm0.14$ & $-13.28\pm0.01$ \\
34    & --- &  $-52.6$ &  +52.4 & 19.34 &  $-9.51\pm0.01$ & $-0.38\pm0.02$ & 
$0.46\pm0.03$ & $0.49\pm0.03$ & $0.75\pm0.05$ & $-13.46\pm0.01$ \\
35    & 163 &  $-12.1$ &   +7.6 & 19.00 &  $-9.85\pm0.01$ &  $0.43\pm0.02$ & 
$1.12\pm0.02$ & $0.53\pm0.02$ & $1.43\pm0.02$ & $-13.55\pm0.01$ \\
36    &  78 &  $-92.9$ &  $-15.9$ & 18.58 & $-10.27\pm0.02$ & $-0.43\pm0.03$ & 
$0.42\pm0.04$ & $0.71\pm0.04$ & $0.91\pm0.06$ & $-12.86\pm0.01$ \\
37    & --- &  $-88.1$ &   $-5.5$ & 18.21 & $-10.64\pm0.02$ & $-0.31\pm0.02$ & 
$0.51\pm0.03$ & $0.33\pm0.04$ & $0.93\pm0.04$ & $-13.15\pm0.01$ \\
38    & --- & $-129.1$ &  $-92.4$ & 20.41 &  $-8.44\pm0.01$ &  $0.28\pm0.03$ & 
$1.03\pm0.01$ & $0.34\pm0.01$ & $1.02\pm0.01$ & $-14.23\pm0.01$ \\
39    &  49 &   $-9.7$ & $-142.8$ & 18.81 & $-10.04\pm0.01$ & $-0.44\pm0.01$ & 
$0.39\pm0.01$ & $0.65\pm0.01$ & $0.72\pm0.02$ & $-12.99\pm0.01$ \\
\multicolumn{11}{c}{NGC~7331} \\
1 & --- & +123.2 & +40.4 & --- & --- & --- & 
--- & --- & --- & $-14.62\pm0.01$ \\
2 & --- &  +76.8 & +18.0 & 19.66 & $-11.09\pm0.04$ & $-0.06\pm0.05$ & 
$0.81\pm0.05$ & $0.68\pm0.03$ & $1.56\pm0.03$ &  $-14.03\pm0.01$ \\
3 & --- &  +27.4 & +16.2 & 20.05 & $-10.70\pm0.04$ & $-0.09\pm0.06$ & 
$0.76\pm0.07$ & $0.49\pm0.10$ & $1.26\pm0.12$ &  $-14.09\pm0.02$ \\
4 & --- &  $-50.2$ & +23.1 & 18.34 & $-12.41\pm0.02$ & $-0.75\pm0.02$ & 
$0.22\pm0.02$ & $0.41\pm0.02$ & $0.81\pm0.03$ & $-13.33\pm0.01$ \\
\multicolumn{11}{c}{NGC~7678} \\
 1 & --- & $-54.2$ & +12.6 & 21.50 & $-11.90\pm0.05$ & $-0.57\pm0.07$ & 
$0.55\pm0.10$ & $0.47\pm0.10$ & $0.20\pm0.21$ & --- \\
 2 & --- & $-31.8$ &  $-7.1$ & 20.73 & $-12.67\pm0.03$ & $-0.52\pm0.05$ & 
$0.41\pm0.06$ & $0.19\pm0.07$ & $0.44\pm0.10$ & --- \\
 3 & --- & $-23.3$ & $-14.8$ & 19.30 & $-14.10\pm0.02$ & $-0.43\pm0.03$ & 
$0.33\pm0.04$ & $0.35\pm0.04$ & $0.59\pm0.06$ & --- \\
 4 & --- & $-14.5$ & $-21.0$ & 19.47 & $-13.93\pm0.02$ & $-0.71\pm0.03$ & 
$0.51\pm0.04$ & $0.58\pm0.04$ & $0.56\pm0.06$ & --- \\
 5 & --- &  +3.7 & $-26.6$ & 20.18 & $-13.22\pm0.03$ & $-0.43\pm0.05$ & 
$0.31\pm0.06$ & $0.16\pm0.08$ & $0.33\pm0.09$ & --- \\
 6 & --- & $-26.5$ & +14.2 & 19.44 & $-13.96\pm0.02$ & $-0.23\pm0.03$ & 
$0.43\pm0.04$ & $0.35\pm0.05$ & $0.79\pm0.06$ & --- \\
 7 & --- & $-10.5$ & +31.6 & 17.47 & $-15.93\pm0.01$ & $-0.35\pm0.01$ & 
$0.48\pm0.01$ & $0.39\pm0.01$ & $0.93\pm0.01$ & --- \\
 8 & --- & $-15.8$ & +27.3 & 17.14 & $-16.26\pm0.01$ & $-0.56\pm0.01$ & 
$0.36\pm0.01$ & $0.41\pm0.01$ & $0.65\pm0.02$ & --- \\
 9 & --- & $-46.5$ &  $-5.2$ & 20.66 & $-12.74\pm0.03$ & $-0.67\pm0.04$ & 
$0.34\pm0.06$ & $0.59\pm0.07$ & $0.43\pm0.12$ & --- \\
10 & --- & $-23.0$ &  +6.2 & 18.29 & $-15.11\pm0.01$ & $-0.38\pm0.02$ & 
$0.40\pm0.02$ & $0.26\pm0.02$ & $0.52\pm0.03$ & --- \\
\hline
\end{tabular}\\
\end{center}
\begin{flushleft}
$^a$ ID number of H\,{\sc ii} region by \citet{belley1992} for NGC~628
and NGC~6946, and by \citet{gusev2003} for NGC~2336. \\
$^b$ Galactocentric coordinates. Positive values correspond to the
northern and western positions. \\
\end{flushleft}
\end{table*}

The galactic background was subtracted from the flux, measured inside 
the round aperture. For the galactic background, we took the average flux 
from the area close to the SF~region which has no bright objects and has 
a minimal inner diameter greater than the sum of $d_{max}$ and the 
seeing in the $B$ band.

We derived the magnitudes and colour indices for 101 of the 102 SF~regions 
studied in \citet{gusev2012} and \citet{gusev2013}. Region No.~1 in NGC~7331 
is out of the image in the $UBVRI$ bands (Fig.~\ref{figure:fig7331}). 

The photometric parameters of the star formation regions are 
presented in Table~\ref{table:phot}. The columns of the table 
present the following properties: (1) assigned sequence number by 
\citet{gusev2012} and \citet{gusev2013}, (2) cross-reference ID number, 
(3,~4) apparent coordinates, in arcseconds, relative to the galaxy centre 
in the plane of the sky, (5) apparent total $B$ magnitude, 
(6) absolute magnitude $M_B$, $M_B = B - 5\log D - 25$, where $D$ 
is an adopted distance in units of Mpc, (7--10) the colour indices 
$U-B$, $B-V$, $V-R$, and $V-I$ with their uncertainties, 
(11) logarithm of spectrophotometric H$\alpha$+[N\,{\sc ii}] flux.

The photometric errors in Table~\ref{table:phot} correspond to the 
uncertainties in the aperture photometry. A main source of 
photometric errors is the uncertainty of the galactic background. 
Table~\ref{table:phot} presents the pure observed data.

\citet{larsen1999,larsen2004} studied young star clusters in several nearby 
galaxies. We can only compare our data with the data of Larsen qualitatively, 
because of the different apertures, seeings, and photometric systems used. 
We identified 15 objects in NGC~6946 and 18 objects in NGC~628 
from Larsen's list, which are in common with the objects studied here and in 
our previous paper \citet{bruevich2007}. In Fig.~\ref{figure:lofig} we 
compare our $UBVI$ photometry with the results published by \citet{larsen1999} 
and \citet{larsen2004}. Fig.~\ref{figure:lofig} shows a good correlation 
between Larsen's and our observations, 
\begin{eqnarray}
\label{equation:lar_col}
B_{\rm our}=B_{\rm Larsen}-(0.41\pm0.08) \nonumber \\
(U-B)_{\rm our}=(U-B)_{\rm Larsen}+(0.10\pm0.01) \\
(B-V)_{\rm our}=(B-V)_{\rm Larsen}+(0.09\pm0.01) \nonumber \\
(V-I)_{\rm our}=(V-I)_{\rm Larsen}-(0.20\pm0.02) \nonumber
\end{eqnarray}
with the following correlation coefficients for different 
magnitudes and colour indices: $r_B=0.98$, $r_{U-B}=0.92$, $r_{B-V}=0.94$, 
and $r_{V-I}=0.83$.

Our $B$ magnitudes are systematically lower than the $B$ magnitudes measured 
by Larsen. This is a result of the different apertures used by 
\citet{larsen2004} (1.0~arcsec) and by our team (several arcseconds, 
depending on the size of the SF~region). It is notable that the seeing of 
{\it HST} images ($\sim0.1$~arcsec) of \citet{larsen2004} is 
ten times better than the seeing of our images ($\sim1.0$~arcsec). 
The differences and deviations between our data and the results of 
\citet{larsen2004}: $B_{\rm our} = B_{\rm Larsen}-(0.41\pm0.08)$~mag 
(the standard deviation around the mean is 0.44~mag) and 
$V_{\rm our} = V_{\rm Larsen}-(0.50\pm0.07)$~mag (the standard deviation 
around the mean is 0.45~mag) are in good agreement with the differences and 
deviations between the space (2004) and the ground (1999) observations of 
Larsen: $\langle\Delta V_{{\rm ground}-HST}\rangle=-0.76\pm0.61$~mag for 
the ground observations with an aperture of 6.4~arcsec and 
$\langle\Delta V_{{\rm ground}-HST}\rangle=-0.29\pm0.50$~mag for the ground 
observations with an aperture of 3.0~arcsec (here 0.61 and 0.50~mag are 
the standard deviations around the mean). Unfortunately, we can not compare 
the ground observations of Larsen with ours because of the small number of 
objects in common.

Fig.~\ref{figure:lofig} and Eq.~(\ref{equation:lar_col}) show that the 
systematic deviations between the linear fits (solid line) and a one-to-one 
correlation (dot dashed line), in the cases of $U-B$ and $B-V$, lie within the 
accuracy limits $\le0.1$~mag. Note that the deviations for the $B-V$ colour 
index between the standard Johnson--Cousins and the {\it HST} photometric 
systems for young star clusters are $\pm0.15$~mag by \citet{larsen2004} and 
$\pm0.17$~mag by \citet{holtzman1995}.

In the case of the $V-I$ colour index, there is a negative systematic 
deviation $\approx0.20$~mag at the lower correlation coefficient. A lower 
correlation and a quite great systematic deviation $0.20$~mag 
in case of the $V-I$ colours can be caused by a smaller signal-to-noise 
ratio and a larger background fluctuation in the $I$ passband. In comparison 
with evolutionary models, we accounted for this systematic deviation 
using a linear regression, shown in Fig.~\ref{figure:lofig} 
(bottom-right panel), $V-I=(V-I)_{\rm obs}+0.20$.

These $V-I$ colour indices, corrected according to 
Eq.~(\ref{equation:lar_col}), are given in column~(10) of 
Table~\ref{table:phot}. Obviously, the $V-I$ data can be used only for 
qualitative estimations.

\section{Observational effects}
\label{sect:obs_eff}

\subsection{Light absorption}

As mentioned above, combined spectroscopic and multi photometric 
observations of SF~regions provide dereddened colours of the SF~regions. The 
star forming regions studied here constitute a single conglomerate of 
clouds of interstellar dust, ionised hydrogen, and newly formed star 
clusters. Much of the light emitted by the stars in the SF~region's clusters 
undergoes absorption inside rather than outside the region itself. 
Therefore, if a SF~region contains a significant amount of interstellar 
dust, then the light from the stars of that SF~region is strongly attenuated 
even if the host galaxy is seen face-on. 

We assume that the emission from stars embedded in the SF~region is absorbed 
in the same way as the emission in lines of ionised hydrogen surrounding 
the star clusters in the SF~region. In other words, the light extinction for 
the stars is equal to the light extinction for the emission of ionised gas 
$A_V{\rm (stars)} = A_V{\rm (Balmer)}$. For a detailed discussion of this 
topic, see our previous paper \citep{gusev2009}.
 
Using this assumption, the measured $UBVRI$ fluxes have been 
corrected for interstellar reddening using the following expression 
$$A_V=\frac{2.5}{k_{{\rm H}\beta}- k_{{\rm H}\alpha}}
\cdot \log \frac{{(I_{{\rm H}\alpha}}/{{ I_{{\rm H}\beta}})}_{{\rm obs}}}
{{{(I_{{\rm H}\alpha}/I_{{\rm H}\beta})}_{{\rm dereddened}}}}.$$
Here, $(I_{{\rm H}\alpha}/I_{{\rm H}\beta})_{\rm obs}$ 
is the ratio of the intensities $I$ of the Balmer hydrogen emission lines 
derived from the spectra of the H\,{\sc ii} regions of the studied SF~regions, 
while $k_{{\rm H}\alpha}=A_{{\rm H}\alpha}/A_V$ and 
$k_{{\rm H}\beta}=A_{{\rm H}\beta}/A_V$ are the absorption coefficients in 
the emission lines H$\alpha$ and H$\beta$.

We used the theoretical H$\alpha$ to H$\beta$ ratio from 
\citet{osterbrock1989}, assuming case B recombination and an electron 
temperature of $10^4$~K and the analytical approximation to the Whitford 
interstellar reddening law by \citet*{izotov1994}. Note that differences 
between the interstellar extinction laws in different galaxies are negligible 
for the part of the optical spectra where both lines H$\alpha$ and H$\beta$ 
belong. The uncertainty of the extinction value is calculated from the 
measurement errors of the H$\alpha$ and H$\beta$ lines and propagated to 
the dereddened fluxes. The derived estimates of the extinction $A_V$ are 
summarised in Table~\ref{table:phys}.

\begin{figure}
\vspace{0.6cm}
\resizebox{0.90\hsize}{!}{\includegraphics[angle=000]{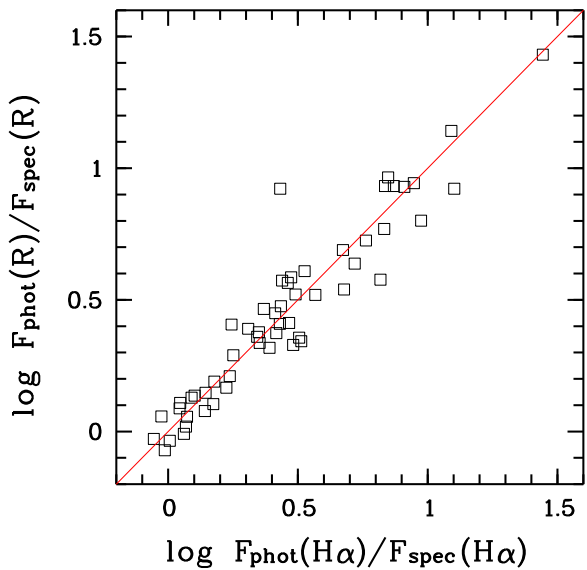}}
\caption{Ratio of fluxes within the round aperture and the area of slit 
in the $R$ passband versus ratio of fluxes in H$\alpha$ filter. Correlation 
coefficient is $98\%$.
}
\label{figure:spectr_photom}
\end{figure}

\subsection{Nebular continuum and line emission}
\label{sect:Nebular}

A number of authors have discussed the impact of nebular emission on 
the broadband photometry of young stellar populations \citep[see references 
therein][]{dinerstein1990,sakhibov1990,reines2010,hollyhead2015}. Here we 
determine the relative contributions of the stellar continuum $F_{\rm stars}$, 
nebular continuum $F_{\rm nebula\,continuum}$, and emission lines 
$F_{\rm lines}$ to the total observed flux $F_{\rm total}$ in the broadband 
filter as follows \citep{sakhibov1990}: 

\begin{equation}
F_{\rm total} = F_{\rm stars}+ F_{\rm nebular\,continuum} + F_{\rm lines}.
\label{equation:f2_eqn}
\end{equation}

Having obtained the emission line ratios for every SF~region in our 
spectroscopic sample, we derived the characteristic values of the electron 
temperatures $T_e$ and metallicities $Z$ in the H\,{\sc ii} regions in the 
studied SF~regions \citep{gusev2012,gusev2013}.

One can see in Fig.~\ref{figure:ha} that the slit in the majority of 
the cases never transmits the full amount of ionised flux in H$\alpha$. 
This leads to underestimating the contribution of the nebular emission 
to the photometric bands. For the three galaxies NGC~628, NGC~6946, 
and NGC~7331, we used the available spectrophotometric H$\alpha$ fluxes 
(see Fig.~\ref{figure:ha} and Table~\ref{table:phot}) for the absolute 
calibration of the emission line intensities derived through spectroscopy. 
To account for this underestimation for the other four 
galaxies with unknown H$\alpha$ photometry, we multiplied the absolute fluxes 
derived through the slit by a factor calculated as 
the ratio of the flux in the $R$ passband within the round aperture for every 
H\,{\sc ii}~region with the flux within the area of the slit: 
$F_{\rm photometry}(R)/F_{\rm spectroscopy}(R)$.

Fig.~\ref{figure:spectr_photom} shows a good correlation ($98\%$) between 
the used factor $F_{\rm photometry}(R)/F_{\rm spectroscopy}(R)$ and the 
ratio of H$\alpha$ intensities 
$F_{\rm photometry}({\rm H}\alpha)/F_{\rm spectroscopy}({\rm H}\alpha)$ 
within the same round aperture and the same area of the slit.
The correlation in Fig.~\ref{figure:spectr_photom} was computed using 
the available spectrophotometric H$\alpha$ fluxes from H\,{\sc ii}~regions 
in the three galaxies NGC~628, NGC~6946, and NGC~7331.

This enables us to estimate the relative contributions of the nebular 
continuum $F_{\rm nebular\,continuum}$, using the equations for 
the continuum emission near the limits of the hydrogen series 
emission, free-free emission, and two-photon emission, given in 
\citet{lang1978}, \citet{kaplan1979}, \citet{brown1970}, and 
\citet{osterbrock1989}. 

\begin{figure}
\vspace{2.8mm}
\resizebox{1.00\hsize}{!}{\includegraphics[angle=000]{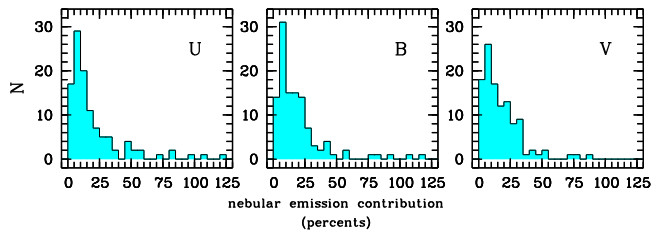}}
\caption{Distributions of studied SF~regions over nebular emission 
contribution in the $U$ band (left), in the $B$ band (centre), and in 
the $V$ band (right).
}
\label{figure:sakh5}
\end{figure}

The contribution from the nebular line emission was computed through the 
summation of the emission line intensities that appear in a given photometric 
band. The fluxes for the non-measured emission lines were computed from the 
derived estimations of the emission measures $EM$, using the equations given 
in \citet{kaplan1979} and \citet{osterbrock1989}. A total of 18 main lines of 
interstellar medium were taken into account.

Fig.~\ref{figure:sakh5} shows the distributions of the studied SF regions over 
the relative contribution of the nebular emission to the $U$, $B$, and $V$ 
fluxes. 

The number of objects decreases with an increase of the relative contributions 
of the nebular emission. Approximately $50\%$ of the objects show relative 
contributions of less than $15\%$ from the total fluxes. This indicates that 
$50\%$ of the star clusters are older than 4~Myr, according to the evolution 
of the ratio of the nebular continuum to the total continuum computed in 
\citet{reines2010} as a function of wavelength for a Starburst99 SSP. 
Another $50\%$ of the objects  show relative contributions of more than $15\%$ 
from the total fluxes and are younger than 4~Myr. Note that these 
estimates are qualitative. Objects with relative contributions 
of nebular emission to the $UBVR$ broadbands greater than $1$ are 
excluded from further consideration. Such not sensible cases could be 
observational effects, caused by overestimated absolute H$\alpha$ fluxes, 
or a wrong computed correction factor 
$F_{\rm photometry}(R)/F_{\rm spectroscopy}(R)$, or by underestimated 
absolute fluxes in the photometric bands because of differential extinction 
in the stellar continuum and nebular emission. The estimated relative 
contributions of the nebular emission to broadband $B$, 
$I_B({\rm gas})/I_B({\rm total})$, are displayed in Table~\ref{table:phys}.

\begin{figure}
\vspace{0.3cm}
\resizebox{1.00\hsize}{!}{\includegraphics[angle=000]{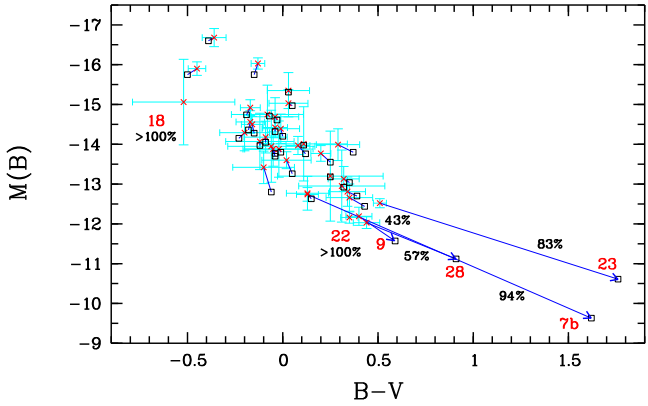}}
\caption{Comparison of the true colours and luminosities of star
clusters (open squares) embedded in SF~regions with the dereddened 
ones (crosses) of star forming regions in NGC~2336 in the 
colour--magnitude diagram. 
}
\label{figure:sakh4}
\end{figure}

In Fig.~\ref{figure:sakh4} we compare the dereddened  colours and 
luminosities (crosses) of the star forming regions in NGC~2336 with the true 
colours and luminosities of the stellar clusters (open squares) embedded in 
the SF~regions, in a colour--magnitude diagram. We computed the true colours 
and luminosities from the dereddened ones using Eq.~(\ref{equation:f2_eqn}). 
The dereddened fluxes are adopted as the total broadband fluxes $F_{\rm total}$ 
in Eq.~(\ref{equation:f2_eqn}). The deviations between the crosses (colours and 
luminosities computed from the dereddened broadband total fluxes from the 
SF~regions) and the open squares (the true colours and luminosities of cluster 
complexes embedded in the SF~regions) in most cases are smaller than the 
absolute errors. In all these cases, the contribution of nebular emission is 
less than $40\%$ of the total fluxes of the SF~regions.

\begin{figure*}
\resizebox{1.00\hsize}{!}{\includegraphics[angle=000]{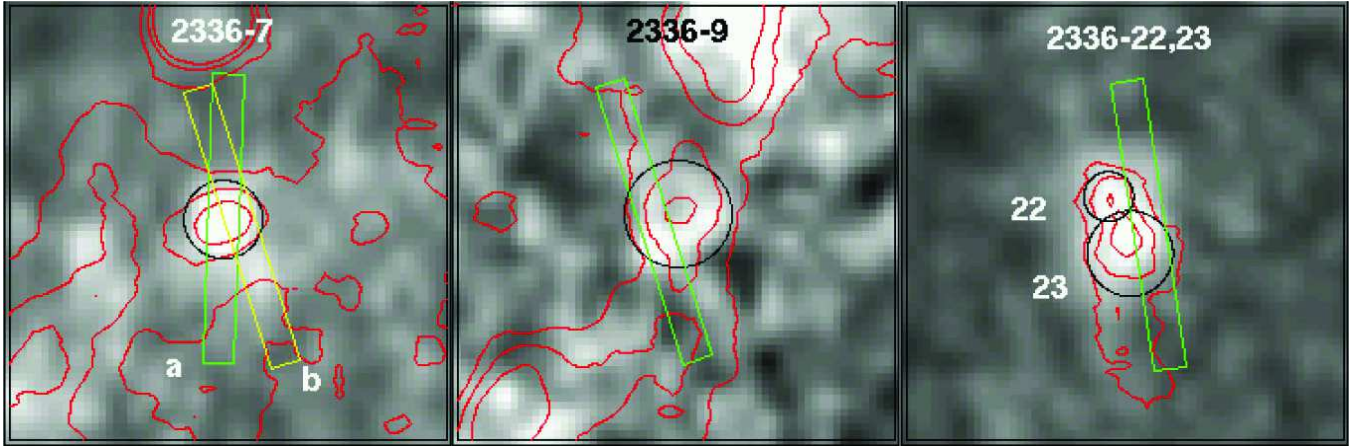}}
\caption{H$\alpha$ images of SF~regions Nos.~7, 9, 22, and 23 in NGC~2336 
with overlaid contours of isophotes in the $B$ band. H$\alpha$ FITS image of 
NGC~2336 was taken from the NED 
\citep*[http://ned.ipac.caltech.edu;][]{epinat2008}. Circles show the 
position and size of apertures for the measurement of $B$ and H$\alpha$ 
fluxes. Bars show the position of the slits during spectroscopic observations. 
The size of the images is $21.3\times21.3$~arcsec$^2$. North is upward and 
East is to the left. See the text for details.
}
\label{figure:map2336}
\end{figure*}

Some objects (No.~7b\footnote{Object No.~7 is corrected for the gas 
contribution using the spectrum `b' for this region.}, No.~18, No.~19, 
No.~22, No.~23,  No.~28) show extremely high nebular emission fluxes, 
which are comparable with or even exceed the total fluxes of the SF~regions at 
least in one of the $UBVR$ bands. According to the results of 
\citet{reines2010}, a nebular emission contribution which is greater than 
$80\%$ of the total flux (Nos.~7b, 23) indicates an extremely young object 
($t<1$~Myr). From the other side, the true colours of objects Nos.~7b, 23 
correspond to an age $t>10$~Myr.

A contradiction between the extremely high contributions of nebular emission 
to the broadband fluxes, which correspond to very young ages ($t<4$~Myr) 
\citep{reines2010}, and considerable red $B-V$ indices, 
which correspond to older ages ($t>10$~Myr), needs an explanation.

A possible cause of such a disagreement is a displacement between the 
photometric centres of the stars and those of the gas emissions, i.e. a 
displacement between a centre of photometric aperture and a centre of slit, 
which leads to a different light extinction for the stars than for the gas 
emissions. This results in wrong estimations of the colours of the star 
clusters, because the assumption that the stars and gas suffer from the same 
amount of extinction is not valid in such cases.

Fig.~\ref{figure:map2336} shows that the stellar photometric centres of 
objects No.~9, No.~22 and No.~23 in NGC~2336, visible through the 
$B$ band, are outside the slit. The centre of the slit lies in the gas 
emission photometric centre of the SF~region, observed through the 
H$\alpha$ filter. The map of object No.~7 in Fig.~\ref{figure:map2336} 
shows that slit `b' crosses the edge of SF~region 
No.~7, whereas slit `a' crosses the centre of the SF~region. As a result, 
object No.~7b has an extremely red colour index $B-V$ (see 
Fig.~\ref{figure:sakh4}) and a high contribution from nebular emission 
($95\%$ in $B$). Object No.~7a has an ordinary colour index $B-V$ and 
a medium contribution from the nebular emission ($11\%$ in $B$).

Thus the spatial displacement between the photometric centres of the stars 
and of the gas emissions in SF~regions such as No.~7b, No.~9, No.~22, and 
No.~23 leads to wrongly estimated extinction and overestimated 
contribution of nebular emission in the broad bands. Both lead to wrong 
colour--magnitude values of the star clusters.

SF~regions such as No.~7b, No.~9, No.~22, and No.~23 in NGC~2336, discussed 
here, and similar objects in other galaxies, are excluded from the further 
analysis for the determination of the physical parameters of the star 
clusters, age and masses.

Objects Nos.~7b, 22, and 23 in NGC~2336 have unusual locations 
on another diagram, too (see Fig.~\ref{figure:ew_rha2}). Object No.~9 in 
NGC~2336 shows an unreasonable value of its equivalent width 
${\rm EW(H}\alpha)>1500$\AA\, (one of the 
circles in Fig.~\ref{figure:ew_rha2}). Object No.~22 in NGC~2336 shows 
an unreasonable ratio of fluxes $F({\rm H}\alpha)/F(R_{\rm stars})>1$ 
(marked as a triangle in Fig.~\ref{figure:ew_rha2}). Another two SF~regions, 
No.~7b and No.~23 in NGC~2336, also have extremely high contributions 
in the $U$ and $B$ bands: $80\%-100\%$ (No.~7b is marked as a triangle in 
Fig.~\ref{figure:ew_rha2}, see also these objects in
Fig.~\ref{figure:sakh4}).

Further discussion of this problem will be continued below in 
Section~\ref{sect:EWha}.

\subsection{EW(H$\alpha$) and spectrophotometric fluxes}
\label{sect:EWha}

\begin{figure}
\vspace{0.3cm}
\resizebox{1.00\hsize}{!}{\includegraphics[angle=000]{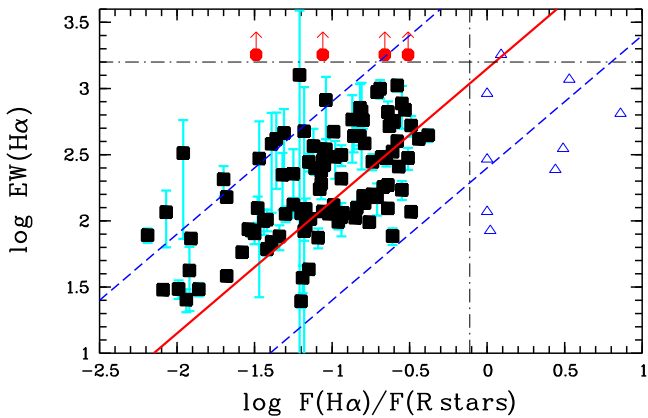}}
\caption{Diagram $\log F({\rm H}\alpha)/F(R_{\rm stars})$ versus 
$\log$~EW(H$\alpha$) for the SF~regions. Dot dashed lines are upper limits 
of the equivalent width EW(H$\alpha$) (horizontal line) and of the ratio of 
the H$\alpha$ flux to star emission flux in the $R$ band (vertical line), 
computed in \citet{reines2010}. The solid line is a linear fit, computed for 
SF~regions (black squares), located under the upper limits of the 
EW(H$\alpha$) and $F({\rm H}\alpha)/F(R_{\rm stars})$ (horizontal and 
vertical dot dashed lines). Dashed lines are upper and lower $95\%$ 
prediction limits of the linear fit. Circles show the SF~regions with an 
unreasonable ${\rm EW(H}\alpha)>1500$\AA. SF~regions with an unreasonable 
ratio $F({\rm H}\alpha)/F(R_{\rm stars})>1$ are marked as triangles. See the 
text for details.
}
\label{figure:ew_rha2}
\end{figure}

In Fig.~\ref{figure:ew_rha2}, we compare the equivalent width 
EW(H$\alpha$) with the ratio of the H$\alpha$ flux to the pure stellar 
emission flux in the $R$ band for the studied objects. We get the equivalent 
widths EW(H$\alpha$) from our spectroscopic observations presented in 
\citet{gusev2012} and \citet{gusev2013}. The stellar emission fluxes 
$F(R_{\rm stars})$ are derived from our $R$ band photometry data, reduced 
for the light absorption and nebula emission contribution. In the cases of 
the three galaxies (NGC~628, NGC~6946, and NGC~7331), we used the available 
spectrophotometric H$\alpha$ fluxes (Table~\ref{table:phot}). For 
the other four galaxies, with unknown H$\alpha$ photometry, we corrected 
the absolute fluxes received through the slit, by a factor that accounts 
for the difference between the area of the round aperture, used by the 
photometry, and the area of the slit (for details see 
Section~\ref{sect:sfrs}). Fig.~\ref{figure:ew_rha2} shows that most objects 
(black squares) are located in the left lower part of the diagram under the 
upper limits of the equivalent width EW(H$\alpha$) (horizontal dot dashed 
line) and of the ratio of the H$\alpha$ flux to star emission flux in the 
$R$ band (vertical dot dashed line), computed in \citet{reines2010}. These 
limits are appropriate for the youngest SF~regions, with ages $\sim1$~Myr. 
The distribution of these SF~regions on the diagram (black squares) can be 
described by the following linear regression (solid line):
\begin{eqnarray}
\log{\rm EW}({\rm H}\alpha)=\log\frac{F({\rm H}\alpha)}{F(R_{\rm stars})}+(3.15\pm0.05), \nonumber
\end{eqnarray}
where the value of the constant $3.15\pm0.05$ is in good agreement with the 
effective bandwidth ($1580$\AA) of the $R$ filter determined by 
\citet{bessell1990}. The uncertainties of the values of EW(H$\alpha$) are used as weights 
in the linear fitting. Dashed lines show the upper and lower $95\%$ prediction 
limits of the linear fit, with a standard deviation of $\sigma=0.37$ and a
correlation coefficient of $r = 0.67$.

Four objects with unreasonable equivalent widths 
${\rm EW(H}\alpha)>1500$\AA\, (marked as circles) and nine objects 
with unreasonable ratios $F({\rm H}\alpha)/F(R_{\rm stars})>1$ (marked as 
triangles) were not taken into account in the computation of the linear 
regression.

Most of these objects show either extremely blue or extremely red colours 
or even undeterminable true colours because of an unreasonably large 
contribution of the nebulosity emission, greater than the total flux in the 
used $UBVR$ broadbands (see Table~\ref{table:phot}). As noted above (see 
Section~\ref{sect:Nebular}), the extreme characteristics of these 
objects may indicate a spatial deviation between the photometric centres 
of the H\,{\sc ii}~region and the star clusters associated with it. In such 
cases, the slit of the spectrograph crosses the centre of the 
H\,{\sc ii}~region, but covers the edge of the star cluster.

\begin{figure*}
\resizebox{1.00\hsize}{!}{\includegraphics[angle=000]{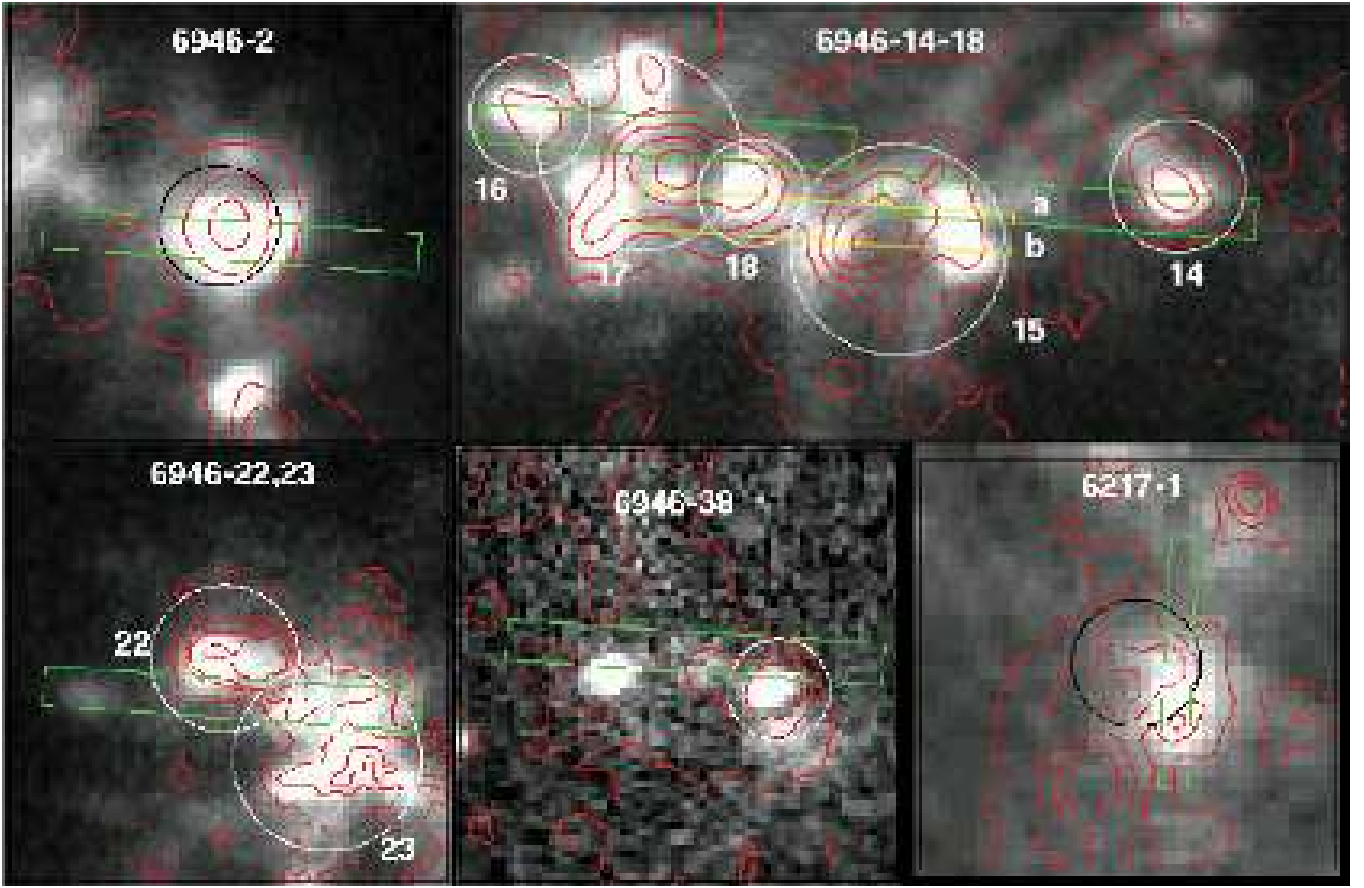}}
\caption{Same as Fig.~\ref{figure:map2336}, but for SF~regions Nos.~2, 
14--18, 22, 23, and 38 in NGC~6946 and No.~1 in NGC~6217. The letters `a' and 
`b' indicate the position of the slits for the measurement of the spectra of 
objects Nos. 14 and 15a and Nos. 15b and 18, respectively. The size of 
the images is $16.5\times16.5$~arcsec$^2$ or $33.1\times16.5$~arcsec$^2$ 
for the areas in NGC~6946 and $23.3\times23.3$~arcsec$^2$ for the area in 
NGC~6217. North is upward and East is to the left. See the text for details.
}
\label{figure:map6946}
\end{figure*}

Such spatial separations were observed earlier by \citet{maizapellaniz1998}. 
A two-dimensional spectrophotometric map of the central region of NGC~4214, 
obtained by \citet{maizapellaniz1998}, shows that the stars, gas, and dust 
clouds in the brightest SF~regions are spatially separated. The dust is 
concentrated at the edges of the region of ionisation and primarily 
influences the nebular emission lines, whereas the stellar continuum is 
located in a region that is relatively free from dust and gas. Thus, the 
assumption about $A({\rm stars})=A({\rm Balmer})$, adopted here, is not 
valid in such SF~regions. This leads to wrong colours for the star clusters 
embedded in such SF~regions. We excluded abnormal objects from further 
analysis when determining the physical parameters of the star clusters 
embedded in the SF~regions. In Fig.~\ref{figure:map2336} and 
Fig.~\ref{figure:map6946} we present maps of the SF~regions which are marked 
by the triangles and circles in Fig.~\ref{figure:ew_rha2}. 
Fig.~\ref{figure:map2336} shows that the 
centre of the ionised gas cloud lies at the edge of the star forming region 
(the slit is located in the photometric centre, visible through the H$\alpha$ 
filter) and there are no stars or a very small number of stars that can 
make a significant contribution to the continuum emission observed through 
the slit. The photometric centre, observed through the $B$ filter, lies 
outside the slit's area. 

A special case is object No.~38 in NGC~6946 (Fig.~\ref{figure:map6946}). It 
has an extremely high ${\rm EW(H}\alpha)>1500$\AA\, and an unreasonably low 
$A_V<A(V)_{\rm Gal}$. The object also has maximum deviation from the 
one-to-one line in Fig.~\ref{figure:spectr_photom}. Possibly, it was 
incorrectly identified.

\subsection{Twice observed objects}
\label{sect:double}

Among the studied SF~regions there is a set of objects that were observed 
twice. In these cases, we took weighted averages of the measured metallicities, 
absorptions, EW(H$\alpha$)s, and $U$, $B$, $V$, $R$, and $I$ fluxes. For 
each measurement, `a' and `b', of the twice observed SF~regions, we 
calculated the weights in the following way:
$p_a = F({\rm H}\beta)_a/(F({\rm H}\beta)_a+F({\rm H}\beta)_b)$ and 
$p_b = F({\rm H}\beta)_b/(F({\rm H}\beta)_a+F({\rm H}\beta)_b)$. Here 
$F({\rm H}\beta)_a$, $F({\rm H}\beta)_b$ are non-dereddened fluxes in 
the H$\beta$ line of the spectra `a', `b', taken from \citet{gusev2012} and 
\citet{gusev2013}. This procedure was followed for 10 out of the 12 
measurement pairs for which both measurements, `a' and `b', showed no 
extremely high nebular emission contribution ($<40\%$) and 
EW$({\rm H}\alpha)<1500$\AA.

Regions No.~7b in NGC~2336 and No.~8b in NGC~6946 have extremely high 
nebular contributions $>50\%$ in every passband, and unreasonable ratios 
$F({\rm H}\alpha)/F(R_{\rm stars})>1$ (see 
Sections~\ref{sect:Nebular} and \ref{sect:EWha}). We excluded these 
objects from further consideration. In the case of objects  
No.~7 in NGC~2336 and No.~8 in NGC~6946, we assume the photometric and 
spectroscopic quantities obtained for objects No.~7a in NGC~2336 and 
No.~8a in NGC~6946.

\section{Comparison with models}
\label{sect:n2336}

Here we compare the observed photometric properties of the SF~regions with a 
numerical SSP-model, with an underlying Salpeter IMF, valid within a 
stellar mass range below $m_{\rm up}$ and above $m_{\rm low}$.

Since the range of the SF~region's masses varies from $10^4 M_{\odot}$ up to 
$10^7 M_{\odot}$, we constructed models in {\it Standard} modes according to
the technique described in \citet{piskunov2011}. The standard mode reproduces 
the features of standard SSP models with a continuously populated IMF, 
whereas the extended mode takes into account the effect of a randomly 
populated IMF. As shown by \citet{piskunov2009}, the effect of the 
randomly populated discrete IMF plays an important role in comparing 
colours of low mass clusters ($M_{\rm cl} < 10^4 M_{\odot}$) with 
the synthetic colours of models.

We used a grid of isochrones provided by the Padova group 
\citep{bertelli1994,girardi2000,marigogirardi2007,marigo2008} via the 
online server CMD\footnote{http://stev.oapd.inaf.it/cgi-bin/cmd}. We used 
the prior sets of stellar evolutionary tracks (version 2.3), described in 
\citet{marigo2008}, instead of the latest Padova models (version 2.5), 
described in \citet{bressan2012}, because the latest version 2.5 is computed 
for a narrower interval of initial masses, ranging from $0.1 M_{\odot}$ to 
$12 M_{\odot}$, while for our purposes we need the interval of initial masses 
to reach up to $100 M_{\odot}$.

As we discussed previously in \citet*{gusev2014}, the multiple structure of 
unresolved star forming complexes does not affect their integrated $B-V$ 
colour indices and thereby the results of the estimations of the age of a 
YMC. Since we use the integrated $B$ luminosities and integrated $B-V$ colours 
of unresolved star complexes in the CMD, we can assign the parameters of the 
model of a single massive cluster to the unresolved multiple star clusters.

\begin{figure*}
\vspace{10.0mm}
\resizebox{0.85\hsize}{!}{\includegraphics[angle=000]{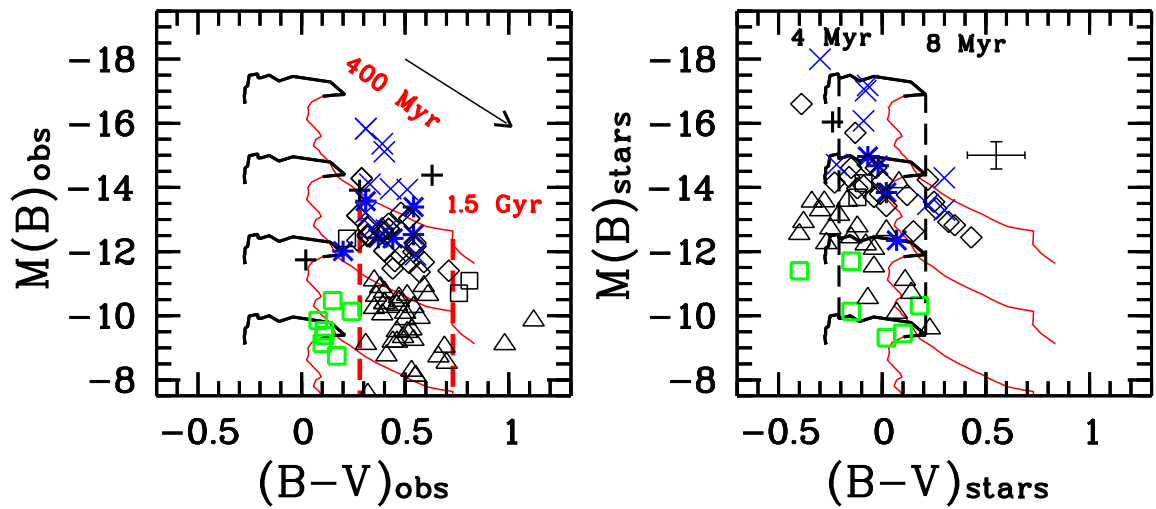}}
\caption{Observed (left) and true (right) colours and luminosities of 
SF~regions in studied galaxies compared to SSP models. Grey 
squares: NGC~628, stars: NGC~783, rhombuses: NGC~2336, crosses (+): 
NGC~6217, triangles: NGC~6946, black squares: NGC~7331, crosses 
($\times$): NGC~7678. The big cross in the right panel indicates the mean 
of the absolute error of true colour--magnitude values. Evolutionary tracks 
for the models with Salpeter's IMF and $Z=0.012$ are shown. See the text 
for details.
}
\label{figure:sakh6}
\end{figure*}

In Fig.~\ref{figure:sakh6} (left panel) we show the observed (not 
deredenned, not corrected for the impact of nebular emission) 
$B-V$ colours versus $M_B$ luminosities of the star forming regions in 
the studied galaxies. Fig.~\ref{figure:sakh6} also shows the evolutionary 
tracks of the SSP models. The models younger than 10~Myr are shown as black 
thick lines, the models between 10~Myr and 1.5~Gyr are shown as grey 
lines. We do not show SF~regions with extremely high nebular emission 
contributions $>40\%$ in the $B$ and/or $V$ passbands (see 
Section~\ref{sect:Nebular}), regions with ${\rm EW(H}\alpha)>1500$\AA, and 
objects with unreasonable ratios $F({\rm H}\alpha)/F(R_{\rm stars})>1$ (see 
Section~\ref{sect:EWha}) in this figure.

Because all the objects in the left panel of Fig.~\ref{figure:sakh6} are 
located in the unreal age interval between 400~Myr and 1.5~Gyr, a light 
absorption correction is needed. Remember that all our objects 
contain giant H\,{\sc ii} regions ionised by massive star clusters 
younger than 10 Myr. Fig.~\ref{figure:sakh6} shows that the extinction 
vector is parallel to the evolution tracks for 400~Myr and 1.5~Gyr. This is a 
clear demonstration of how `age--absorption' degeneracy works. To avoid the 
degeneracy problem, we use the deredenning method based on the independent 
measurements of the Balmer decrements (see Section~\ref{sect:obs_eff}).

In Fig.~\ref{figure:sakh6} (right panel) we compare the true colours 
and luminosities, which are corrected for light extinction and impact 
of nebular emission of star clusters (cluster complexes), with SSP models 
(black and grey lines) in the colour--magnitude diagram. As 
we noted above (Section~\ref{sect:intro}), we call the colours and 
luminosities which are uneffected by both the extinction and the nebula 
emission to broad bands the `true' colours and luminosities of the young 
massive cluster complexes.

We eliminated from Fig.~\ref{figure:sakh6} (right panel) the objects with 
large errors of $A_V$ estimations ($\Delta A_V>0.8$~mag), as well as the 
objects with underestimated or overestimated light extinction, with extremely 
red or blue $B-V$ indices. An example of one of the two objects with 
underestimated extinction, No.~1 in NGC~6217, is shown in 
Fig.~\ref{figure:map6946}. Most of objects with 
overestimated light extinction are SF~regions in NGC~6946 (see 
Table~\ref{table:phys}). For example, we discuss three 
`abnormal' objects, Nos.~2, 17, and 22 in NGC 6946 (triangles). 
These objects have extremely blue $B-V$ colour indices, which 
cannot be fitted by the youngest models of star clusters. At the same time, 
the low relative contributions of the nebular emission to the $U$, $B$, 
$V$ fluxes indicate older ages $t>5$~Myr, according to Starburst99 model 
evolutionary isochrones for the nebular continuum presented in 
\citet{reines2010}. Note that the light absorption estimates for all 
three objects are extremely high: $A_V$=3.84, 3.86, and 4.59 mag, 
respectively, while the characteristic value of the light absorption for 
SF~regions in NGC~6946 is $A_V=2.2\pm0.9$ mag. This could indicate a 
possible overestimation of the light absorption of the stellar continuum 
observed through $UBVRI$ filters. Note that the estimations of light 
absorption are based on the observed Balmer decrement values under the 
assumption that the emission from stars embedded in the SF~region is absorbed 
in the same way as the emission in the lines of ionised hydrogen surrounding 
the star clusters in the SF~region. This is valid when the clumping of the 
gas and dust distribution in the space of the star forming complex 
is not high, otherwise the light extinction of the stars is not equal to
the light extinction of the emission of ionised gas. It seems to us 
that this is exactly what happens in the three abnormal objects, Nos.~2, 17, 
and 22. Fig.~\ref{figure:map6946} shows the displacements between the 
photometric centres in these objects, visible through the $B$ and H$\alpha$ 
filters. The case of object No.~22 can be explained as a displacement between 
the gas cloud and the star clusters along the radial direction.

Fig.~\ref{figure:sakh6} (right panel) shows that most of the objects are 
in a narrow age interval ($t<8$~Myr). This result agrees with the time 
interval determined by the life time of the giant H\,{\sc ii} regions. 
The big cross indicates the mean of the absolute error of true 
colour--magnitude values. The displacements of most objects from the 
model's area lie within the error interval.

There are 5 objects in distant galaxies NGC~2336 (three rhombuses), NGC~7678 
(two crosses `$\times$'), which are located outside of the area of the young 
models ($t<10$~Myr), and can be fitted with older models between 300~Myr 
and 700~Myr. As the sizes of all 5 objects belong to the range from 300~pc to 
600~pc, they are cluster complexes. As the evolutionary tracks for greater 
ages ($t>50$~Myr) are parallel to a `reddening vector', we meet the well known 
age--absorption degeneracy in the case of these 5 cluster complexes. There are 
two possible interpretations.

\begin{figure*}
\vspace{8.0mm}
\resizebox{1.00\hsize}{!}{\includegraphics[angle=000]{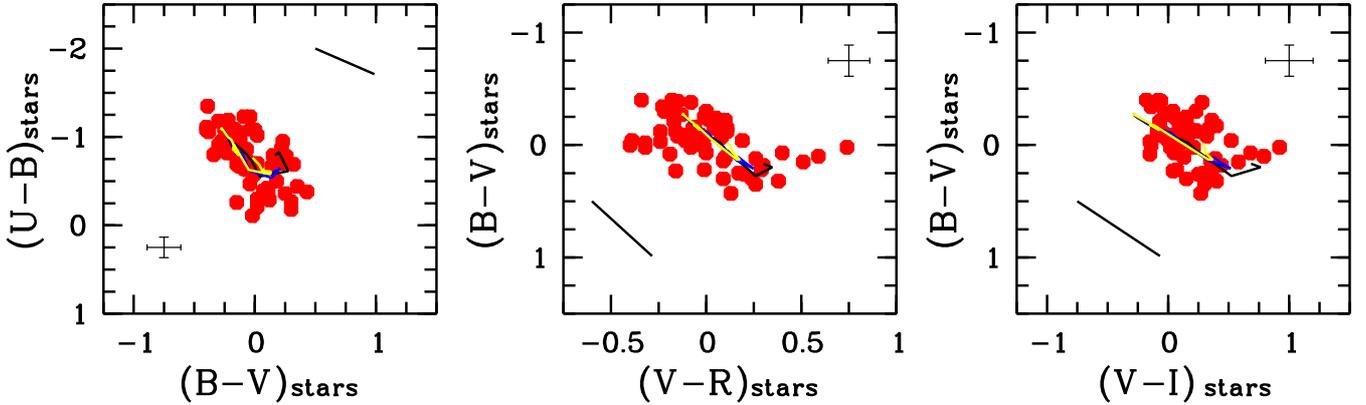}}
\caption{Colour--colour diagrams $(U-B)-(B-V)$, $(B-V)-(V-R)$, 
and $(B-V)-(V-I)$ for the true colours of cluster complexes in 
the studied galaxies. Black lines show SSP models with continuously populated 
IMF with Salpeter's slope $\alpha=-2.35$ and $Z=0.019$, grey lines show models 
with $Z=0.012$, and light lines show models with $Z=0.008$. The SF~regions 
with estimated ages and masses are shown as grey circles. The error crosses in 
the diagrams shows the mean accuracy of the colours of these objects. The 
black straight line in the corner of the diagrams is parallel to the 
extinction vector.
}
\label{figure:ccd1}
\end{figure*}

First, a possible spatial displacement between the photometric centres of 
the nebular emission and stellar continuum provides a spatially inhomogeneous 
extinction, so our assumption that the stars and gas suffer from the same 
amount of extinction is invalid here. In the case of these 5 cluster 
complexes, it could be evidence for an underestimation of the light 
extinction and $A_V({\rm stars})>A_V({\rm Balmer})$.

Second, near young clusters, which ionize the gas in the SF~region,
there are clusters with older stellar populations. The large sizes of these 
5 objects can also indicate a possible coexistence of old and 
extremely young star clusters in the same SF~region.

On the other hand, the deviations of these 5 objects from the area of the 
young models ($t<10$~Myr) do not exceed the errors of the photometric 
properties. In Section~\ref{sect:phys}, these objects are fitted with young 
models (see Table~\ref{table:phys}).

Fig.~\ref{figure:ccd1} plots SSP model tracks for the two-colour diagrams. 
The true colours, derived for our star forming regions, are superimposed on 
these diagrams. The offsets between the positions of the SF~regions and  
the theoretical tracks are comparable with the mean accuracy of the 
derivations of the true colours, shown as big crosses on the diagrams.

The true magnitudes and colours of the SF~regions, which are shown in 
Figs.~\ref{figure:sakh6} (right panel) and \ref{figure:ccd1}, are given 
in Table~\ref{table:true}.

\section{Physical parameters of the star forming regions}
\label{sect:phys}

Multicolour photometry provides a useful tool for constraining the physical 
parameters of YMC complexes in star forming regions. Here we use the 
method of the minimisation of $O-C$ parameters (observed minus computed):
\begin{eqnarray}
O-C=[[(U-B)_{\rm obs}-(U-B)_{\rm model}]^2+ \nonumber \\
+[(B-V)_{\rm obs}-(B-V)_{\rm model}]^2+[M_{\rm obs}(B)-M(B)_{\rm model}]^2]^{1/2} \nonumber
\end{eqnarray}
to constrain the ages and masses of the YMC complexes in the star forming 
regions. Under the concept of `observed parameters' we place the true colours 
and luminosities, which are those corrected for the light extinction and the 
impact of nebular emission. The procedure for finding the values of the 
physical parameters, ages ($t$) and YMC's masses ($m_{\rm cl}$) can be 
subdivided into the following steps. First, the models of stellar population, 
computed for the derived (from spectroscopy) chemical abundance, are presented 
in the form of a grid of colour indices and luminosities for a broad 
range of variation of searching parameters $t(i)$ and $m_{\rm cl}(j)$. 
Here, the indices $i$, $j$ are the numbers of rows and columns 
in a two dimensional grid of photometric parameters. The table step of the 
$\log(t)$ parameter variation is 0.05~dex. The table step of the 
$\log(m_{\rm cl})$ parameter variation depends on the interval of luminosity 
variations of YMC complexes in a given galaxy: 
\begin{eqnarray}
h_{\log(m_{\rm cl})}=\frac{\log(m_{\rm cl}(max))-\log(m_{\rm cl}(min))}{N}, \nonumber
\end{eqnarray}
where $N$ is the number of the evolutionary tracks. For 
every node $(i, j)$, we calculated the value of the $(O-C)_{i,j}$ 
parameter. 

The second step is the search for the grid node in which the 
$(O-C)_{i,j}$ parameter has a minimum value. We accept corresponding 
values $t=t(i)$, $m_{\rm cl}=m_{\rm cl}(j)$ as a solution of the problem for 
a given individual young massive cluster or cluster complex. The ages and 
masses of YMC complexes, computed through the minimization of the $O-C$ 
parameters (observed minus computed), are presented in columns (7) and (8) of 
Table~\ref{table:phys}. The uncertainty introduced through the light 
absorption correction and accounting for a gas emission contribution is the 
source of the rather low accuracies of the age estimations, especially for 
very small ages ($t<3$~Myr), where the colour gradients of age are very steep. 
In Table~\ref{table:phys} we present the errors of the ages 
and masses computed according to Gauss's law of error propagation.

In the assessment of the ages and masses, we did not use the $V-R$ and $V-I$ 
colour indices because, in the case of YMCs, the $R$ and $I$ fluxes are weakly 
sensitive to changes in age, and the actual observational errors lead to large 
uncertainties.

\begin{table*}
\caption[]{\label{table:phys}
The physical parameters of the star forming regions.
}
\begin{center}
\begin{tabular}{cccccccccccc} \hline \hline
H\,{\sc ii} & $r$   & $r$/$R_{25}$ & $A_V$ & EW(H$\alpha$) & $Z$ & 
$I_B$(gas)/  & $t$   & 
$m/M_{\odot}$      & $d$  & Structure$^a$ & Notes$^b$ \\
region      & (kpc) &              & (mag) & (\AA) &     & 
$I_B$(total) & (Myr) & 
($10^4 M_{\odot}$)  & (pc) &               & \\
1 & 2 & 3 & 4 & 5 & 6 & 7 & 8 & 9 & 10 & 11 & 12 \\
\hline
\multicolumn{11}{c}{NGC~628} \\
 1 & 4.55 & 0.415 & $1.58\pm0.67$ & $476\pm^{550}_{476}$ & 
$-$               & 0.05 & 
 $2.0\pm^{6.1}_{2.0}$ & $6.50\pm1.02$ &  65 & dbl & \\
 2 & 3.11 & 0.283 & $1.24\pm0.51$ & $102\pm19$ & 
$-$               & 0.05 & 
 $6.3\pm2.6$          & $5.00\pm1.17$ & 130 & dbl, pt & \\
 3 & 2.58 & 0.235 & $0.70\pm0.45$ & $583\pm308$ & 
$-$               & 0.28 & 
 $-$                  & $1.21\pm0.06$ &  70 & st & \\
 4 & 2.77 & 0.253 & $0.79\pm0.67$ & $296\pm269$ & 
$0.0043\pm0.0017$ & 0.05 & 
 $6.3\pm0.4$          & $1.25\pm0.23$ & 110 & st & \\
 5 & 1.34 & 0.122 & $0.64\pm0.54$ & $>1500$ & 
$-$               & 0.12 & 
 $-$                  & $-$           & 115 & dbl & 2 \\
 6 & 1.51 & 0.138 & $1.81\pm2.05$ & $25\pm5$ & 
$-$               & 0.01 & 
 $-$                  & $-$           &  90 & dbl & 5 \\
 7 & 4.01 & 0.366 & $0.00\pm0.28$ & $674\pm430$ & 
$0.0106\pm0.0014$ & 0.43 & 
 $-$                  & $-$           & 130 & ring & 1 \\
 8 & 5.31 & 0.484 & $0.26\pm0.26$ & $716\pm365$ & 
$0.0091\pm0.0011$ & 0.32 & 
 $7.9\pm1.4$          & $1.00\pm0.06$ &  75 & st, pt & \\
 9 & 5.42 & 0.494 & $1.86\pm0.28$ & $>1500$ & 
$0.0082\pm0.0013$ & 0.58 & 
 $-$                  & $-$           &  45 & st & 1, 3 \\
10 & 6.25 & 0.571 & $0.06\pm0.47$ & $305\pm97$ & 
$0.0087\pm0.0018$ & 0.20 & 
 $6.3\pm0.9$          & $1.45\pm0.20$ & 220 & compl & \\
\multicolumn{11}{c}{NGC~783} \\
 1  & 10.60 & 0.729 & $0.60\pm0.32$ & $151\pm41$ & 
$0.0068\pm0.0010$ & 0.34 & 
 $6.3\pm1.9$          &  $10.0\pm2.2$  &  750 & dif, pt & \\
 2  & 12.83 & 0.881 & $0.17\pm1.17$ & $42\pm21$ & 
$-$               & 0.03 & 
 $-$                  &  $-$           & 1000 & ring & 5 \\
 3  &  8.33 & 0.572 & $1.09\pm0.19$ & $299\pm57$ & 
$0.0120\pm0.0009$ & 0.26 & 
 $-$                  &  $91.2\pm16.9$ & 1300 & dif &  \\
 4  & 11.96 & 0.822 & $1.92\pm0.42$ & $43\pm4$ & 
$0.0075\pm0.0014$ & 0.09 & 
 $5.6\pm0.1$          &  $95.5\pm52.9$ &  550 & dif & \\
 5  & 10.17 & 0.699 & $1.30\pm0.21$ & $172\pm28$ & 
$-$               & 0.25 & 
 $6.3\pm2.9$          &  $38.0\pm6.9$  &  500 & st, pt & \\
 6  & 11.15 & 0.766 & $0.94\pm0.19$ & $244\pm39$ & 
$0.0088\pm0.0007$ & 0.76 & 
 $-$                  & $-$            &  550 & st, pt & 1, 3 \\
 7  & 12.20 & 0.838 & $1.63\pm0.96$ & $25\pm3$ & 
$0.0062\pm0.0025$ & 0.09 & 
 $-$                  & $-$            &  550 & st & 5 \\
 8  &  8.42 & 0.578 & $1.56\pm0.19$ & $1170\pm^{1238}_{1170}$ & 
$0.0122\pm0.0012$ & 0.37 & 
 $-$                  & $-$            &  650 & st & 1, 3 \\
\multicolumn{11}{c}{NGC~2336} \\
 1    & 20.79 & 0.885 & $0.43\pm0.22$ & $140\pm59$ & 
$0.0056\pm0.0006$ & 0.07 & 
$10.0\pm1.1$ &  $47.4\pm5.7$  & 630 & dbl & \\
 2    & 16.89 & 0.719 & $0.96\pm0.17$ & $348\pm66$ & 
$0.0098\pm0.0008$ & 0.17 & 
 $6.3\pm2.9$ &  $34.2\pm5.3$  & 730 & dif & \\
 3    & 11.30 & 0.481 & $1.47\pm0.34$ & $98\pm12$ & 
$0.0133\pm0.0021$ & 0.12 & 
 $5.0\pm1.1$ &  $47.4\pm17.7$ & 390 & dif & \\
 4    &  7.79 & 0.331 & $0.87\pm0.27$ & $114\pm18$ & 
$-$               & 0.13 & 
 $7.9\pm1.3$ &  $84.1\pm11.0$ & 520 & dif & \\
 5    &  4.86 & 0.207 & $0.73\pm0.68$ & $31\pm5$ & 
$-$               & 0.03 & 
 $7.9\pm0.8$ &  $24.6\pm6.8$  & 370 & dif & \\
 6    & 23.02 & 0.979 & $2.39\pm0.13$ & $437\pm66$ & 
$0.0047\pm0.0004$ & 0.13 & 
 $-$         & $-$            & 470 & st, pt & 6 \\
 7    & 15.19 & 0.646 & $0.98\pm0.32$ & $416\pm241$ & 
$0.0090\pm0.0014$ & 0.11 & 
 $8.9\pm0.7$ &  $19.3\pm2.5$  & 470 & dif & \\
 8    & 12.99 & 0.553 & $0.68\pm0.15$ & $113\pm8$ & 
$0.0124\pm0.0008$ & 0.18 & 
 $7.1\pm2.7$ &  $31.5\pm3.3$  & 390 & st & \\
 9    & 10.27 & 0.437 & $0.23\pm0.17$ & $>1500$ & 
$0.0190\pm0.0015$ & 0.43 & 
 $-$         &  $-$           & 460 & dif & 1, 2 \\
10    & 11.82 & 0.503 & $1.21\pm0.12$ & $135\pm8$ & 
$0.0129\pm0.0006$ & 0.13 & 
 $5.0\pm0.1$ &  $71.4\pm11.1$ & 420 & st & \\
11    & 10.59 & 0.451 & $1.92\pm0.30$ & $287\pm101$ & 
$0.0077\pm0.0010$ & 0.11 & 
 $-$         &  $60.6\pm12.7$ & 330 & dif & \\
12    &  8.22 & 0.349 & $1.99\pm0.55$ & $58\pm5$ & 
$-$               & 0.04 & 
 $5.6\pm0.6$ &  $92.2\pm51.2$ & 550 & dif, pt & \\
13    & 14.84 & 0.632 & $2.62\pm0.17$ & $279\pm43$ & 
$0.0218\pm0.0025$ & 0.07 & 
 $-$         &  $552\pm169$   & 810 & dif & \\
14    & 22.83 & 0.971 & $1.51\pm0.13$ & $141\pm9$ & 
$0.0058\pm0.0003$ & 0.17 & 
 $4.5\pm1.0$ &  $77.5\pm7.9$  & 480 & st & \\
15    & 18.49 & 0.787 & $0.20\pm0.27$ & $123\pm25$ & 
$0.0063\pm0.0008$ & 0.08 & 
 $6.3\pm2.4$ &  $15.1\pm3.0$  & 580 & dif, pt & \\
16    & 12.11 & 0.515 & $1.25\pm0.14$ & $122\pm8$ & 
$0.0101\pm0.0007$ & 0.22 & 
 $7.1\pm4.7$ &  $29.0\pm3.9$  & 730 & dbl & \\
17    & 14.05 & 0.598 & $1.23\pm0.11$ & $189\pm12$ & 
$0.0122\pm0.0006$ & 0.20 & 
 $6.3\pm0.6$ &  $190\pm36$    & 860 & st, pt & \\
18    & 17.18 & 0.731 & $2.82\pm0.81$ & $117\pm30$ & 
$-$               & $>1$ & 
 $-$         &  $-$           & 430 & st & 1, 3, 5 \\
19    & 13.13 & 0.559 & $1.52\pm0.30$ & $84\pm8$ & 
$0.0093\pm0.0012$ & 0.43 & 
 $-$         &  $-$           & 450 & dif & 1, 3 \\
20    & 14.49 & 0.616 & $2.26\pm0.28$ & $98\pm2$ & 
$0.0107\pm0.0012$ & 0.16 & 
 $-$         &  $51.4\pm13.3$ & 320 & st, pt &  \\
21    & 13.50 & 0.574 & $0.75\pm0.26$ & $237\pm91$ & 
$0.0124\pm0.0014$ & 0.18 & 
 $7.1\pm3.5$ &  $11.8\pm1.7$  & 270 & dif, pt & \\
22    & 17.88 & 0.761 & $1.04\pm0.11$ & $916\pm207$ & 
$0.0091\pm0.0004$ & $>1$ & 
$-$          &  $-$           & 320 & st, pt & 1, 3 \\
23    & 17.47 & 0.743 & $0.19\pm0.08$ & $443\pm48$ & 
$0.0086\pm0.0004$ & 0.83 & 
$-$          &  $-$           & 380 & st, pt & 1 \\
24    & 15.61 & 0.664 & $1.79\pm0.54$ & $75\pm13$ & 
$0.0067\pm0.0015$ & 0.09 & 
 $3.5\pm2.1$ &  $37.1\pm19.8$ & 560 & dif, pt &  \\
25    & 15.32 & 0.652 & $1.84\pm0.36$ & $1267\pm^{2610}_{1267}$ & 
$0.0074\pm0.0012$ & 0.07 & 
 $3.5\pm^{4.2}_{3.5}$ & $65.8\pm28.5$ & 520 & dif, pt &  \\
26    & 12.43 & 0.529 & $1.97\pm0.68$ & $77\pm11$ & 
$0.0082\pm0.0023$ & 0.21 & 
 $-$         &  $43.7\pm18.2$ & 260 & dif, pt &  \\
27    & 12.88 & 0.548 & $1.07\pm0.54$ & $76\pm16$ & 
$0.0071\pm0.0017$ & 0.07 & 
 $5.6\pm0.4$ &  $34.2\pm16.5$ & 510 & dif, pt & \\
28    & 20.17 & 0.858 & $0.17\pm0.11$ & $297\pm40$ & 
$0.0081\pm0.0005$ & 0.57 & 
 $-$         & $-$            & 420 & ring & 1 \\
\multicolumn{11}{c}{NGC~6217} \\
 1 & 3.30 & 0.479 & $0.21\pm0.21$ & $74\pm5$ & 
$0.0128\pm0.0012$ & 0.05 & 
$-$         & $-$                    & 495 & dbl, pt & 6 \\
 2 & 4.41 & 0.640 & $1.63\pm0.67$ & $78\pm10$ & 
$-$               & 0.01 & 
$2.5\pm2.4$ & $448\pm293$ & 300 & dbl, pt & \\
 3 & 3.24 & 0.471 & $1.71\pm0.54$ & $461\pm239$ & 
$0.0078\pm0.0018$ & 0.05 & 
$-$         & $-$                    & 185 & st & 6 \\
\multicolumn{11}{c}{NGC~6946} \\
 1 & 5.34 & 0.402 & $2.69\pm0.26$ & $437\pm108$ & 
$0.0123\pm0.0013$ & 0.12 & 
 $-$         & $-$           &  60 & st & 6 \\
\hline
\end{tabular}\\
\end{center}
\begin{flushleft}
$^a$ compl: complex structure (more than three separated objects); 
dbl: double object; dif: separated object with diffuse 
profile; pt: brighter part (core) of an extended star forming region; 
ring: ring structure; st: separated object with 
star-like profile; tr: triple object \\
$^b$ 1: objects with $I$(gas)/$I$(total) $>40\%$ in the $B$ and/or $V$; 
2: objects with ${\rm EW(H}\alpha)>1500$\AA; 
3: objects with $F({\rm H}\alpha)/F(R_{\rm stars})>1$; 
4: an object lacking any photometry; 
5: objects with $\Delta A_V>0.8$~mag; 
6: objects for which, apparently, the condition 
$A{\rm (stars)} = A{\rm (Balmer)}$ is not satisfied. \\ 
\end{flushleft}
\end{table*}

\setcounter{table}{3}
\begin{table*}
\caption[]{
Continued}
\begin{center}
\begin{tabular}{cccccccccccc} \hline \hline
H\,{\sc ii} & $r$   & $r$/$R_{25}$ & $A_V$ & EW(H$\alpha$) & $Z$ & 
$I_B$(gas)/  & $t$   & 
$m/M_{\odot}$      & $d$  & Structure$^a$ & Notes$^b$ \\
region      & (kpc) &              & (mag) & (\AA) &     & 
$I_B$(total) & (Myr) & 
($10^4 M_{\odot}$)  & (pc) &               & \\
1 & 2 & 3 & 4 & 5 & 6 & 7 & 8 & 9 & 10 & 11 & 12 \\
\hline
\multicolumn{11}{c}{NGC~6946} \\
 2 & 6.51 & 0.490 & $3.84\pm0.08$ & $519\pm25$ & 
$0.0088\pm0.0003$ & 0.14 & 
 $-$         & $-$          & 110 & dif & 6 \\
 3 & 3.99 & 0.300 & $1.16\pm0.73$ & $107\pm30$ & 
$-$               & 0.15 & 
 $7.9\pm1.7$ &  $1.24\pm0.14$ & 135 & ring & \\
 4    & 1.75 & 0.131 & $0.41\pm0.19$ & $106\pm8$ & 
$-$               & 0.41 & 
 $-$         & $-$          &  90 & dif & 1 \\
 5    & 1.91 & 0.144 & $1.54\pm0.36$ & $68\pm6$ & 
$-$               & 0.05 & 
 $7.1\pm2.1$ &  $1.24\pm0.13$ & 110 & dif & \\
 6    & 2.95 & 0.222 & $1.92\pm0.15$ & $92\pm4$ & 
$0.0118\pm0.0008$ & 0.07 & 
 $4.5\pm0.5$ & $18.4\pm2.3$ & 180 & compl & \\
 7    & 2.14 & 0.161 & $3.35\pm1.53$ & $37\pm9$ & 
$-$               & 0.04 & 
 $-$         & $-$          &  45 & dif, pt & 5 \\
 8    & 2.21 & 0.166 & $1.93\pm0.09$ & $151\pm11$ & 
$0.0108\pm0.0005$ & 0.04 & 
 $7.1\pm2.2$ & $3.28\pm0.19$ & 115 & dif, pt &  \\
 9    & 2.28 & 0.171 & $2.37\pm0.11$ & $939\pm77$ & 
$0.0047\pm0.0004$ & 0.19 & 
 $3.2\pm2.6$ & $22.8\pm2.3$ & 150 & dbl, pt &  \\
10    & 2.75 & 0.207 & $2.05\pm0.15$ & $118\pm6$ & 
$0.0325\pm0.0022$ & 0.10 & 
 $5.6\pm0.3$ & $14.8\pm1.4$ &  80 & dbl, pt & \\
11    & 3.32 & 0.250 & $1.97\pm0.15$ & $577\pm80$ & 
$0.0295\pm0.0011$ & 0.21 & 
 $5.6\pm0.2$ & $8.65\pm0.65$ &  65 & st, pt &  \\
12    & 3.13 & 0.236 & $2.39\pm0.13$ & $174\pm9$ & 
$-$               & 0.08 & 
 $4.5\pm2.0$ & $28.3\pm2.9$ & 135 & dif, pt & \\
13    & 3.12 & 0.235 & $2.95\pm0.15$ & $264\pm28$ & 
$0.0160\pm0.0010$ & 0.10 & 
 $2.5\pm0.5$ & $14.8\pm1.2$ &  40 & st, pt & \\
14    & 3.87 & 0.291 & $3.35\pm0.11$ & $998\pm160$ & 
$0.0147\pm0.0007$ & 0.16 & 
 $-$         & $-$          &  65 & dif & 6 \\
15    & 3.98 & 0.299 & $2.15\pm0.15$ & $622\pm122$ & 
$0.0151\pm0.0012$ & 0.25 & 
 $1.3\pm^{5.2}_{1.3}$ & $31.6\pm1.1$ & 145 & dbl, pt & \\
16    & 4.34 & 0.326 & $3.76\pm0.15$ & $664\pm135$ & 
$0.0161\pm0.0010$ & 0.20 & 
 $-$         & $-$          &  65 & dbl & 6 \\
17    & 4.22 & 0.318 & $3.86\pm0.13$ & $385\pm27$ & 
$0.0175\pm0.0009$ & 0.11 & 
 $-$         & $-$          &  70 & st, pt & 6 \\
18    & 4.12 & 0.310 & $1.35\pm0.13$ & $180\pm12$ & 
$0.0041\pm0.0003$ & 0.29 & 
$10.0\pm0.5$ & $4.06\pm0.09$ &  55 & st, pt & \\
19    & 5.00 & 0.376 & $2.99\pm0.11$ & $1055\pm107$ & 
$0.0110\pm0.0005$ & 0.18 & 
 $-$         & $-$          & 125 & dbl & 6 \\
20    & 4.81 & 0.362 & $2.28\pm0.15$ & $>1500$ & 
$0.0063\pm0.0004$ & 0.20 & 
 $-$         & $-$          &  90 & dif & 1 \\
21    & 5.14 & 0.387 & $2.26\pm0.11$ & $570\pm55$ & 
$0.0109\pm0.0005$ & 0.22 & 
 $-$         & $-$          & 125 & dif, pt & 6 \\
22    & 6.44 & 0.485 & $4.59\pm0.13$ & $304\pm14$ & 
$0.0090\pm0.0008$ & 0.09 & 
 $-$         & $-$          &  75 & dbl, pt & 6 \\
23    & 6.41 & 0.483 & $2.71\pm0.13$ & $334\pm29$ & 
$0.0110\pm0.0006$ & 0.26 & 
 $-$         & $-$          &  85 & dif, pt & 6 \\
24    & 2.35 & 0.177 & $0.39\pm0.96$ & $30\pm4$ & 
$-$               & 0.06 & 
 $-$         & $-$          &  65 & st, pt & 5 \\
25    & 3.38 & 0.254 & $2.56\pm0.13$ & $417\pm35$ & 
$0.0116\pm0.0006$ & 0.21 & 
 $-$         & $-$          & 200 & tr & 6 \\
26    & 4.92 & 0.371 & $1.60\pm0.11$ & $159\pm6$ & 
$0.0072\pm0.0003$ & 0.22 & 
 $4.5\pm0.7$ & $9.63\pm0.86$ & 120 & dbl, pt & \\
27    & 4.72 & 0.356 & $2.31\pm0.08$ & $691\pm47$ & 
$0.0079\pm0.0003$ & 0.20 & 
 $-$         & $-$          & 145 & st, pt & 6 \\
28    & 5.10 & 0.384 & $2.54\pm0.41$ & $113\pm15$ & 
$0.0064\pm0.0011$ & 0.05 & 
 $3.5\pm3.1$ & $28.3\pm9.3$ & 100 & st, pt & \\
29    & 4.95 & 0.373 & $1.81\pm0.32$ & $115\pm12$ & 
$-$               & 0.13 & 
 $5.6\pm0.3$ & $6.97\pm1.24$ & 225 & compl & \\
30    & 6.00 & 0.452 & $2.37\pm0.08$ & $560\pm15$ & 
$0.0057\pm0.0003$ & 0.30 & 
 $1.1\pm^{4.4}_{1.1}$ & $83.3\pm2.0$ & 215 & tr, pt &  \\
31    & 6.47 & 0.487 & $1.26\pm0.17$ & $316\pm57$ & 
$0.0059\pm0.0005$ & 0.15 & 
 $6.3\pm0.3$ & $1.91\pm0.13$ &  80 & ring, pt & \\
32    & 6.35 & 0.478 & $2.67\pm0.17$ & $132\pm8$ & 
$0.0063\pm0.0005$ & 0.08 & 
 $2.0\pm^{6.1}_{2.0}$ & $22.8\pm1.1$ &  80 & dif, pt & \\
33    & 2.47 & 0.186 & $1.66\pm0.41$ & $292\pm96$ & 
$0.0113\pm0.0020$ & $>1$ & 
 $-$         &  $-$          &  65 & st, pt & 1, 3 \\
34    & 2.16 & 0.162 & $2.22\pm0.13$ & $209\pm15$ & 
$0.0177\pm0.0009$ & 0.11 & 
 $3.5\pm1.5$ &  $6.97\pm0.63$ & 100 & dbl, pt & \\
35    & 0.43 & 0.032 & $3.35\pm0.36$ & $38\pm2$ & 
$-$               & 0.03 & 
 $7.1\pm2.1$ & $54.1\pm18.0$ &  80 & dbl & \\
36    & 3.10 & 0.234 & $1.39\pm0.15$ & $117\pm6$ & 
$0.0148\pm0.0009$ & 0.38 & 
 $6.3\pm1.0$ & $4.06\pm0.33$  & 155 & tr, pt &  \\
37    & 2.88 & 0.217 & $1.64\pm0.62$ & $84\pm13$ & 
$-$               & 0.08 & 
 $6.3\pm0.7$ & $12.0\pm4.3$ & 225 & ring, pt & \\
38    & 5.29 & 0.398 & $0.54\pm0.17$ & $>1500$ & 
$0.0162\pm0.0013$ & 0.13 & 
 $-$         & $-$          &  55 & st & 2 \\
39    & 4.30 & 0.323 & $2.67\pm0.11$ & $400\pm21$ & 
$0.0126\pm0.0005$ & 0.17 & 
 $-$         & $-$          &  80 & st, pt & 6 \\
\multicolumn{11}{c}{NGC~7331} \\
 1    & 9.78 & 0.488 & $0.19\pm0.60$ & $55\pm13$ & 
$0.0070\pm0.0019$ & --   & 
 $-$                 & $-$           &  90 & dif, pt & 4 \\
 2    & 5.45 & 0.272 & $3.27\pm1.45$ & $30\pm2$ & 
$-$               & 0.01 & 
 $-$                 & $-$           & 225 & tr & 5 \\
 3    & 3.48 & 0.173 & $2.65\pm0.36$ & $382\pm168$ & 
$0.0057\pm0.0013$ & 0.05 & 
$4.0\pm0.5$ &  $50.2\pm16.4$ & 110 & st & \\
 4    & 9.06 & 0.451 & $1.07\pm0.17$ & $251\pm28$ & 
$0.0090\pm0.0006$ & 0.11 & 
$5.0\pm1.2$ &  $35.2\pm4.5$ & 410 & compl &  \\
\multicolumn{11}{c}{NGC~7678} \\
 1 & 13.02 & 0.901 & $2.01\pm1.09$ & $227\pm121$ & 
$0.0059\pm0.0025$ & 0.08 & 
 $-$                 & $-$            & 300 & st & 5 \\
 2 &  8.57 & 0.593 & $0.68\pm0.45$ & $223\pm103$ & 
$0.0084\pm0.0016$ & 0.08 & 
$7.1\pm1.0$          &   $32.9\pm8.6$ & 590 & st & \\
 3 &  8.10 & 0.560 & $0.30\pm0.21$ & $367\pm128$ & 
$0.0040\pm0.0006$ & 0.16 & 
$8.9\pm2.8$          &   $90.6\pm13.8$ & 620 & dif & \\ 
 4 &  8.11 & 0.561 & $0.64\pm0.17$ & $819\pm405$ & 
$0.0090\pm0.0008$ & 0.24 & 
$-$                  & $-$            & 520 & st & 6 \\
 5 &  8.15 & 0.564 & $0.13\pm0.41$ & $100\pm22$ & 
$0.0119\pm0.0021$ & 0.08 & 
$7.1\pm4.0$          &   $30.0\pm10.4$ & 430 & st & \\
 6 &  7.02 & 0.486 & $1.64\pm0.30$ & $69\pm5$ & 
$0.0091\pm0.0011$ & 0.05 & 
$5.6\pm0.2$          &  $274\pm148$ & 450 & dif, pt & \\
 7 &  9.63 & 0.666 & $1.47\pm0.11$ & $191\pm9.0$ & 
$0.0074\pm0.0003$ & 0.11 & 
$4.0\pm0.8$          &  $627\pm190$ & 740 & dif & \\
 8 &  8.56 & 0.592 & $1.71\pm0.08$ & $330\pm10$ & 
$0.0086\pm0.0003$ & 0.17 & 
$2.2\pm2.2$ & $2740\pm340$ & 790 & dbl & \\
 9 & 11.87 & 0.821 & $1.69\pm0.15$ & $280\pm39$ & 
$0.0093\pm0.0006$ & 0.21 & 
$1.6\pm^{3.6}_{1.6}$ & $157\pm14$ & 460 & st &  \\
10 &  5.56 & 0.384 & $1.52\pm0.11$ & $104\pm5$ & 
$0.0080\pm0.0004$ & 0.09 & 
$5.6\pm0.4$          &  $627\pm188$ & 490 & st, pt & \\
\hline
\end{tabular}\\
\end{center}
\begin{flushleft}
$^a$ compl: complex structure (more than three separated objects);
dbl: double object; dif: separated object with diffuse
profile; pt: brighter part (core) of an extended star forming region; 
ring: ring structure; st: separated object with
star-like profile; tr: triple object \\
$^b$ 1: objects with $I$(gas)/$I$(total) $>40\%$ in the $B$ and/or $V$; 
2: objects with ${\rm EW(H}\alpha)>1500$\AA; 
3: objects with $F({\rm H}\alpha)/F(R_{\rm stars})>1$; 
4: an object lacking any photometry; 
5: objects with $\Delta A_V>0.8$~mag; 
6: objects for which, apparently, the condition 
$A{\rm (stars)} = A{\rm (Balmer)}$ is not satisfied \\ 
\end{flushleft}
\end{table*}

We estimated the ages for 57 and masses for 63 YMC complexes of 102 
observed SF~regions. We believe that the condition 
$A({\rm Balmer})=A({\rm stars})$ is valid for these regions. The estimates
of the ages for 6 objects have been obtained with low accuracies, the errors 
of the age estimation exceeding the range of ages of 1--8~Myr.
 
\begin{figure}
\vspace{10.0mm}
\resizebox{0.90\hsize}{!}{\includegraphics[angle=000]{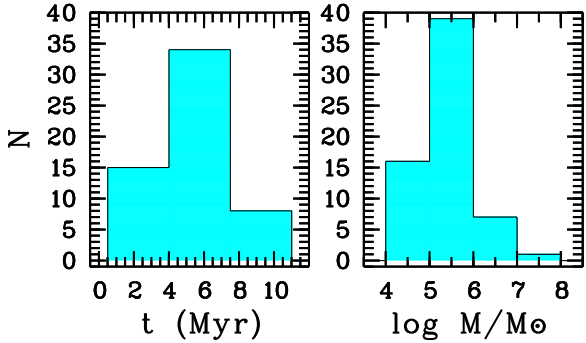}}
\caption{Frequency distribution of ages (left panel) and masses (right 
panel) of star complexes in all studied galaxies.
}
\label{figure:sakh10} 
\end{figure}

The left histogram in Fig.~\ref{figure:sakh10} shows the age frequency 
distribution of 57 SF~regions in studied galaxies. The mean value of the ages 
for our sample is $5.5\pm2.1$~Myr. The drastic decay of the number of YMC 
complexes between 8 and 10~Myr is in agreement with the lifetime estimates of 
H\,{\sc ii}~regions. The relatively small number of YMC clusters with 
ages $<4$~Myr can be explained by the effect of selection: very young 
clusters/complexes have a high extinction. During the first $3-4$~Myr of life 
of a YMC, its absorption decreases from $A_V\sim10$~mag to $\approx1$~mag, 
while its luminosity increases insignificantly 
\citep[see][and references therein; see also the evolutionary tracks in 
Fig.~\ref{figure:sakh6}]{whitmore2011}. As a result, the apparent magnitude 
of a YMC of the same mass will be maximal for ages $4-7$~Myr.

The distances to the studied galaxies range from 7~Mpc up to 70~Mpc. As a 
result, we observe different absolute fluxes and different mass intervals for 
the YMC complexes in the different galaxies. The right histogram in 
Fig.~\ref{figure:sakh10} and Table~\ref{table:phys} show that the masses of 
the YMC complexes range from $10^4 M_{\odot}$ in the nearby galaxies NGC~628 
and NGC~6946 to $10^7 M_{\odot}$ in the distant galaxy NGC~7678; however, 
most of them have masses $10^5-10^6 M_\odot$.

Table~\ref{table:phys} shows that the extinction in the studied star forming 
regions varies greatly, from 0 up to 3.5~mag. This result is in 
good agreement with the reddening measurements for the sample of 49 disc, halo 
and nuclear star clusters in M82 \citep{konstantopoulos2009}.

The following physical parameters of the SF~regions are presented in 
Table~\ref{table:phys}. Deprojected galactocentric distances in units of kpc 
and in units of isophotal radius, corrected for the Galactic extinction and 
inclination effects, are given in columns (2) and (3), respectively. 
Column~(4) presents the estimated values of the light absorption $A_V$ 
computed from the observed Balmer decrements. These values include Galactic 
extinctions $A(V)_{\rm Gal}$. The equivalent widths, EW(H$\alpha$), obtained 
in \citet{gusev2012} and \citet{gusev2013} are shown in column~(5). The 
chemical abundances $Z$, estimated from spectroscopic observations 
\citep{gusev2012,gusev2013} using the equations of \citet{pilyugin2011}, 
are given in column~(6). The estimated values of the nebular emission 
contribution in the $B$ band are given in column~(7). The ages, $t$, 
and masses, $m$, of the YMC complexes are shown in columns~(8) 
and (9), respectively. The estimated diameters are presented in column~(10). 
The structure of the SF~regions is described in column~(11). Lastly, the 
notes are given in column~(12).

Note that the diameters of the SF~regions in NGC~2336 presented in 
Table~\ref{table:phys} are larger than those published in \citet{gusev2003}. 
In the present paper we estimate the diameter as the geometric mean of 
$d_{max}$ and $d_{min}$, instead of $d_{min}$ as in \citet{gusev2003}.

\section{Comparison of extragalactic SF~regions and Milky Way open clusters 
in the integrated colour--magnitude diagram}
\label{sect:ocs}

The currently accepted paradigm for star formation assumes that open 
clusters are born in a super cluster. They are not formed in isolation, but in 
stellar complexes born out of giant molecular clouds. As the giant 
molecular clouds disappear after the star formation is complete, they may seed 
the galactic disc with families of young clusters (cluster complexes). These 
families gradually disperse to become individual clusters and, eventually, 
field populations. We investigate the existence of the evolutionary relation 
between YMC complexes embedded in giant H\,{\sc ii} regions and open 
clusters using a comparative analysis of the evolution of the integrated 
photometric parameters for the studied YMC complexes and the open star 
clusters in our Galaxy in a colour--magnitude diagram.

\begin{figure}
\vspace{0.6cm}
\hspace{0.6cm}
\resizebox{0.92\hsize}{!}{\includegraphics[angle=000]{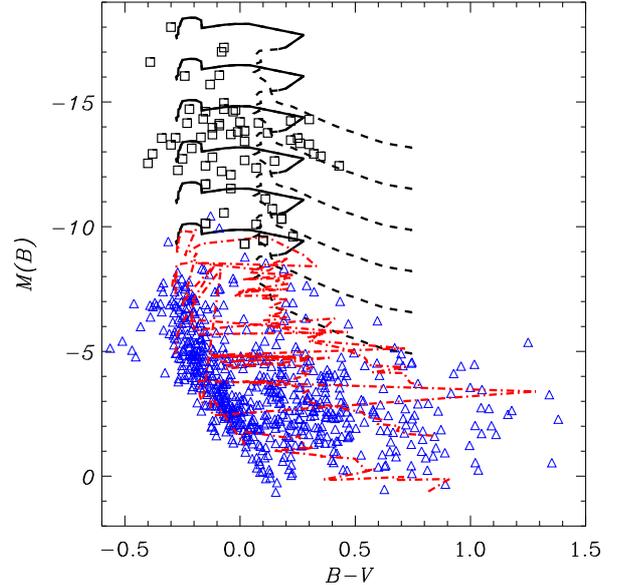}}
\caption{True colours and luminosities of YMC complexes (squares), compared 
with the {\it Standard} mode of SSP models (continuously populated IMF), and 
of OCs in the Milky Way (triangles), compared with the {\it Extended} SSP 
mode (randomly populated IMF). Only YMC complexes with estimated ages and 
masses are shown. In black we show the tracks of the {\it Standard} mode with 
an adopted characteristic metallicity of the studied YMCs of $Z = 0.012$, 
drawn in the age interval from 1 to 10~Myr as solid, and in the age 
interval from 10~Myr to 1.5~Gyr as dashed lines. The tracks were 
computed for the following masses of star clusters: $2\cdot 10^7 M_{\odot}$, 
$4.4\cdot 10^6 M_{\odot}$, $9.6\cdot 10^5 M_{\odot}$, 
$2.1\cdot 10^5 M_{\odot}$, $4.6\cdot 10^4 M_{\odot}$, and 
$1\cdot 10^4 M_{\odot}$. Grey dot-dashed lines show the {\it Extended} mode 
tracks for characteristic metallicity of OCs ($Z = 0.019$) computed for the 
$1\cdot 10^4 M_{\odot}$, $9\cdot 10^3 M_{\odot}$, $1.2\cdot 10^3 M_{\odot}$, 
and $150 M_{\odot}$ in the age interval from 1~Myr to 9.8~Gyr.
}
\label{figure:sakh9}
\end{figure}

Fig.~\ref{figure:sakh9} shows the colour--magnitude diagram for the YMC 
complexes in the studied galaxies and for the Galactic open clusters (OCs). 
We use a catalogue of 650 OCs and compact associations in the Milky Way 
\citep{kharchenko2005,piskunov2005} that were identified in the All-Sky 
Compiled Catalogue of 2.5 million stars (ASCC-2.5) complete to $V\approx11.5$ 
\citep{kharchenko2001}. The integrated absolute magnitudes and intrinsic 
colours of the OCs were estimated by \citet{kharchenko2009}. The true 
photometric quantities of the YMC complexes, unaffected by extinction and 
nebula emission, were estimated from the observed ones taking into account 
the interstellar reddening and nebula emission contribution in photometric 
bands (see Sections~\ref{sect:observ} and \ref{sect:obs_eff}).

Fig.~\ref{figure:sakh9} shows that the YMC complexes in the galaxies and the 
OCs in the Milky Way form a continuous sequence of luminosities/masses and 
colour/ages. This diagram shows a hypothetical colour--magnitude diagram that 
will be observed in remote galaxies when the depth of present day 
extragalactic surveys is increased approximately by 10~mag. 

It is worth noting that the observed continuity exhibits the hierarchical 
properties of star formation but it is not connected with the properties of 
the YMCs themselves. The SF~region can consist, in its brightest part, of 
several young massive stellar clusters (a YMC complex), which can be observed 
as a single object because of the weak spatial resolution at large distances. 
As we noted in \citet{gusev2014}, the multiple structure of the unresolved 
complex of YMCs does not affect their integrated brightness $M_B$ and colour 
index $B-V$. One can estimate the number of the brightest Milky Way clusters 
(BMWC) required to reproduce the positions of the studied SF~regions in the 
CMD. It is enough to move the BMWC up to $M_B \approx -15$ mag. A value of 
$\Delta M_B = -5$~mag corresponds to a luminosity excess of 
$L_{\rm YMC}(B)/L_{\rm BMWC}(B)\approx 100$, where $L_{\rm YMC}(B)$ is the 
typical luminosity of our cluster complexes, and $L_{\rm BMWC}(B)$ is the 
luminosity of the BMWC. The corresponding mass can be found with the help of 
the mass--luminosity ratio. The SSP models show that for young clusters 
($\log t\le7$), the ratio $m_{\rm cl}/L_V<0.1$, both for standard and extended 
modes \citep[see Fig.~12 in][]{piskunov2011}. Since young clusters are 
brighter in $B$ than they are in $V$, the ratio $m_{\rm cl}/L_B$ is even 
smaller than 0.1. Then the typical mass of a complex that emits the observed 
flux will be on the order of $10\,m_{\rm BMWC}$, i.e. our complexes may 
not contain more than 10 BMWC. Note that in the Solar Neighbourhood, one can 
identify many young complexes with sizes from 150--700~pc, containing tens 
of associations, star clusters, and H\,{\sc ii} regions \citep{efremov1988}. 
So we can conclude that the observed YMCs are comparable to what we know 
in the Solar Neighbourhood.

In Fig.~\ref{figure:sakh9}, we show the 
{\it Standard} mode of SSP models with continuously populated IMF computed 
for high luminosity ($M_B<-9$, $m_{\rm cl}>10^4 M_{\odot}$ and age 
$t < 10$~Myr) YMC complexes in galaxies and the {\it Extended} mode of 
SSP models with randomly populated IMF, computed for relatively low 
luminosity Galactic OCs ($M_B>-9$, $50 M_{\odot}<m_{\rm cl}<10^4 M_{\odot}$ 
and 1~Myr~$<t<1$~Gyr). The dispersion of the integrated colours of the YMC 
complexes in the galaxies is due to both colour measurement errors and the 
natural dispersion of their physical parameters (ages). A comparison with 
the models shows that the YMC complexes belong to the young area of models, 
with $t < 10$~Myr, while the OCs cover the area of the models with a wide age 
interval, from 1~Myr up to 1~Gyr. We excluded from Fig.~\ref{figure:sakh9} 
those SF~regions for which the ages and masses were not obtained.

\section{Discussion}
\label{sect:discus}

Our estimates of the physical parameters of the YMC clusters are based on the 
assumption $A({\rm gas})=A({\rm stars})$. In Section~\ref{sect:obs_eff} we 
discussed obvious cases where this condition is not valid. 
\citet{whitmore2011} proposed an evolutionary classification scheme of YMCs 
based on {\it HST} observations of M83. They showed that in clusters with 
ages $1-4$~Myr, the ionised gas is spatially coincident with the cluster stars. 
For these clusters, the condition $A({\rm gas})=A({\rm stars})$ is satisfied. 
YMCs with ages $\approx5$~Myr and older have a large ionised gas bubble 
surrounding the cluster. The radius of the bubble is larger than 20~pc. In 
the nearby galaxies, such as NGC~628 and NGC~6946, the resolution of our 
observations $\approx40$~pc allows us to separate the gas and star clusters 
in space. Sometimes we can not use our spectroscopic data for the estimation 
of $A({\rm stars})$ for such objects (see Sections~\ref{sect:obs_eff}, 
\ref{sect:n2336}, and Figs.~\ref{figure:map2336}, \ref{figure:map6946}). 
However, for the majority of 
the SF~regions older than 5~Myr, we can adequately estimate $A({\rm stars})$. 
YMCs with ages $\approx4-5$~Myr have a small H\,{\sc ii} bubble surrounding 
the cluster. The radii of the bubbles are $7-20$~pc \citep{whitmore2011}. 
The stage of the partially embedded cluster when the cluster blows 
a gas bubble is quite short, about 1~Myr \citep{hollyhead2015}. Our 
observations do not allow us to separate gas bubbles from star clusters. As 
a result, our values of $A({\rm stars})$ will be overestimated for these 
YMCs. This is indicated by a significant number of star clusters with 
an extremely blue $B-V$ colour index in NGC~6946 (Table~\ref{table:phys}, 
see the objects marked `6' in the notes). Note that the ratio of the 
number of extremely blue YMCs to the number of YMCs with obtained ages and 
masses in the nearby galaxies NGC~628, NGC~6946, and NGC~7331 (12/29) is 
close to the ratio of the number of partially embedded YMCs to the number of 
embedded and exposed YMCs (16/50) according to \citet{hollyhead2015}.

YMC complexes in more distant galaxies are conglomerates of star clusters, 
associations, and ionised gas bubbles. The effects of the spatial distribution 
of star groupings and H\,{\sc ii} gas bubbles are smoothed for the distant 
SF~regions. The averaged gas and star extinctions are approximately equal to 
each other. Only a few objects in the distant galaxies have colour indices 
outside the limits of the model tracks (Tables~\ref{table:phys}, 
\ref{table:true}).

\begin{figure}
\vspace{0.5cm}
\resizebox{0.90\hsize}{!}{\includegraphics[angle=000]{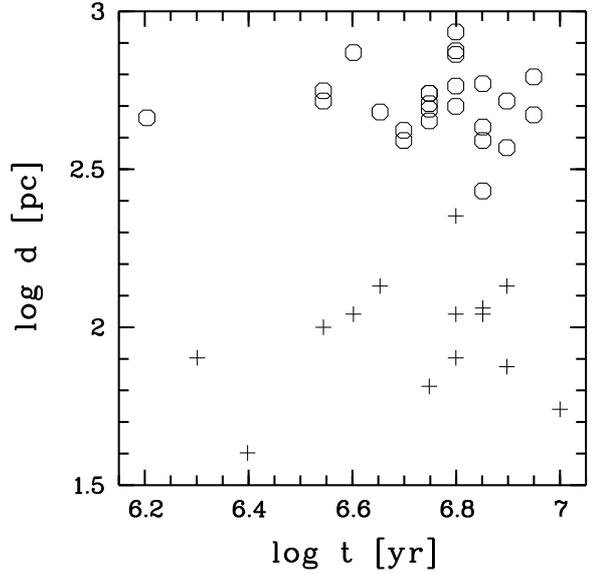}}
\caption{`Age--size' diagram for SF~regions in nearby (crosses) and 
distant (circles) galaxies. See the text for details.
}
\label{figure:t_d}
\end{figure}

As seen from the evolutionary tracks in Figs.~\ref{figure:sakh6} and 
\ref{figure:ccd1}, the colours of YMC clusters vary slightly during the first 
$3-4$~Myr of a SF~region's life. The ages of SF~regions younger than 4~Myr can 
be estimated with low accuracy using our technique. The colours of YMC 
clusters with ages from 4 to 8~Myr stringently depend on the age 
(Fig.~\ref{figure:sakh6}). The age of these SF~regions can be estimated 
with considerable accuracy. The errors of age depend only on the errors of 
the true colour indices.

The photometric ages of the SF~regions' population younger than 10~Myr 
are consistent with the maximum lifetime of giant H\,{\sc ii}~regions 
and the age distribution of 45 star forming complexes, fitted with the 
simultaneous star formation models (SSF model), with a mean age 
$4.5\pm^{4.5}_{2.5}$~Myr \citep{sakhibov2000}. Note that \cite{sakhibov2000} 
used grids for both models: the simultaneous star formation model (SSF) 
and the infinite continuous star formation model (CSF) for the confrontation 
of the intrinsic colour indices of 113 star forming complexes in 22 spiral 
and irregular galaxies with the predicted colour indices. The star forming 
complexes, fitted with infinite continuous star formation models (CSF model), 
show a mean age $10\pm^{30}_{8}$~Myr. The spread in the age of 14 intense 
starbursts in irregular and compact galaxies, estimated by 
\cite{mashesse1999}, is narrower ($2.5-6.5$~Myr).

The combination of broadband multicolour photometry and emission spectra 
of SF~regions, used in current research, provides age estimations only 
for young massive cluster complexes with ages less than 10~Myr. 
\cite{konstantopoulos2009} derived the ages of 44 bright isolated star 
clusters in the range 30~Myr to 270~Myr, using absorption spectra combined 
with photometric data.

\begin{figure}
\vspace{0.5cm}
\resizebox{0.90\hsize}{!}{\includegraphics[angle=000]{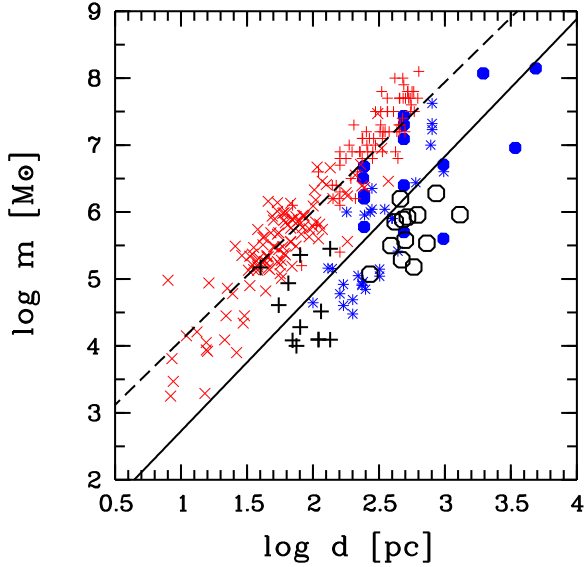}}
\caption{The `size--mass' diagram for SF~regions of our sample 
(symbols are the same as in Fig.~\ref{figure:t_d}), YMC complexes in 
very distant ($z\sim1.5$) galaxy Sp~1149 (dark filled circles) and cluster 
complexes in the local galaxies (dark stars) by \citet{adamo2013}, GMCs from 
\citet{bolatto2008} (grey crosses `$\times$') and \citet*{wei2012} (grey 
crosses `+'). The solid line is a linear fit, computed for local and 
high-$z$ YMC complexes in \citet{adamo2013}. The dashed line is a linear fit, 
computed for GMCs from the \citet{bolatto2008} sample. See the text for 
details.
}
\label{figure:d_m}
\end{figure}

Modern high resolution studies suggest that SF~regions present an 
age spread comparable to the size of the regions 
\citep{bastian2005,whitmore2011,kim2012}. Thus, the size of 
an H\,{\sc ii}~region is a function of the age of the stellar population. 
The resolution of our photometric observations does not allow us to explore 
the `age--size' dependence for individual star clusters. The sizes of 
the H\,{\sc ii}~regions beginning from 40~pc depend weakly on the age of the 
stellar population \citep[see Fig. 4 (top) in][for instance]{whitmore2011}. 
A different situation is observed for star associations. The `age--size' 
relation for star associations was first established by \citet{efremov1998} 
for star groupings in LMC and was later confirmed in \citet{sakhibov2001} for 
SF~regions in other galaxies. Associations in the Milky Way and in other 
galaxies inside and outside the Local Group also show correlations 
between age and size \citep[see Fig. 8 in][]{zwart2010}.

Fig.~\ref{figure:t_d} shows the dependence between the age and the size of 
the studied SF~regions. In order to make a homogeneous sample, we excluded 
multiple (double, triple, and complex) objects from the graph (see column 
(11) in Table~\ref{table:phys}). As seen from the figure, the SF~regions 
(stellar associations) in the nearby galaxies NGC~628, NGC~6946, and NGC~7331 
satisfy the `age--size' relation. Larger star associations are older. 
Single YMCs have smaller sizes. A possible example of such a YMC is the 
relatively old and compact SF~region No.~18 in NGC~6946, with an age of 
10~Myr and diameter of 55~pc.

The sizes of the large SF~regions in the distant galaxies (large YMC complexes) 
do not correlate with age (Fig.~\ref{figure:t_d}). Obviously, all of them
have a complex structure and contain a conglomerate of numerous YMCs and 
associations. Each cluster/association expands with age inside the complex, 
but the size of the complex depends basically on the physical parameters of the
surrounding interstellar matter and the magnetic field 
\citep{elmegreen2003a,elmegreen2003b,gusev2013b}.

\begin{figure}
\vspace{0.5cm}
\resizebox{0.90\hsize}{!}{\includegraphics[angle=000]{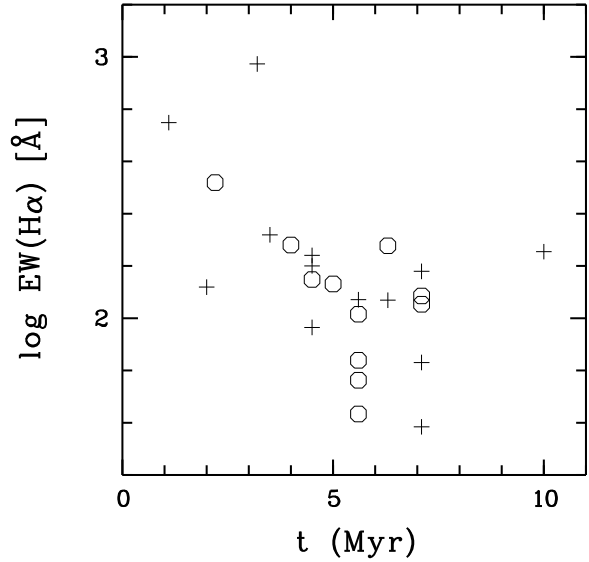}}
\caption{The `age--EW(H$\alpha$)' diagram for the studied SF~regions. Symbols 
are the same as in Fig.~\ref{figure:t_d}. See the text for details.
}
\label{figure:t_ew}
\end{figure}

\citet{larson1981} found a correlation between the sizes and the masses of 
giant molecular clouds (GMCs), $m\sim d^2$. This correlation has been 
repeatedly confirmed \citep[see, e.g.,][]{hopkins2012}. As is known, YMC 
complexes are the direct descendants of GMCs. As a result, the 
mass--size relation for YMC complexes was found to be close to that of GMCs 
\citep[see][and references therein]{adamo2013}. \citet{adamo2013} found a 
relation $m\sim d^{2.0\pm0.3}$ for YMC complexes and $m\sim d^{1.9\pm0.1}$ 
for GMCs from the sample of \citet{bolatto2008}. These relations are plotted 
in Fig.~\ref{figure:d_m}.

We examined the dependence between the sizes and the masses for the studied 
SF~regions (Fig.~\ref{figure:d_m}). As in Fig.~\ref{figure:t_d}, we excluded 
multiple objects from the graph. In addition, we excluded SF~regions with 
errors of mass estimation larger than $\pm20\%$. As can be seen in the figure, 
there is no correlation between the masses and the ages for the SF~regions in 
the nearby galaxies. Among the objects in NGC~628 
and NGC~6946 which have diameters $\sim60-130$~pc, we observe both 
compact regions with masses $\sim10^5 M_{\odot}$ and doughy regions with 
masses $\sim10^4 M_{\odot}$. Nevertheless, they follow the general
relation for YMCs (Fig.~\ref{figure:d_m}). One of the YMCs (No.~13 in 
NGC~6946) is located in the area occupied by GMCs. Note that this cluster is 
the core of a larger complex and it is the youngest object of our sample among 
the regions with confident estimates of ages ($2.5\pm0.5$~Myr).

We did not find a very strong correlation between the sizes and the masses for 
the YMC clusters in the distant galaxies (Fig.~\ref{figure:d_m}). 
Nevertheless, they 
follow the relation $m\sim d^2$ for YMC complexes, as well as data 
from \citet{adamo2013}. Obviously, the correlation which is observed for 
YMC complexes is a relic of the `size--mass' dependence for their ancestors,
the GMCs. The vertical offset between the GMCs and the YMC complexes in 
the diagram could be due to the efficiency of star formation: only a fraction 
of the gas in the GMCs will form stars \citep{bastian2005,adamo2013}.

The equivalent widths of the H$\alpha$ and H$\beta$ emission lines can be 
used to infer the ages of YMC clusters 
\citep*[e.g.,][]{copetti1986,alonso1996,reines2008,reines2010}. 
Starburst99 evolutionary models show that EW(H$\alpha$) in 
H\,{\sc ii}~regions falls from $>1000$\AA\, in the youngest YMCs to 
$\approx30-40$\AA\, in the regions with an age of 10~Myr \citep{reines2010}. 
The dependence between EW(H$\alpha$) and the ages of the studied SF regions 
is shown in Fig.~\ref{figure:t_ew}. In contrast to the samples in 
Fig.~\ref{figure:t_d}, \ref{figure:d_m}, we included objects 
of any structure in the figure, but excluded objects with errors of 
EW(H$\alpha$) larger than $\pm10\%$. Fig.~\ref{figure:t_ew} shows that 
the SF~regions follow the Starburst99 evolutionary models, however several 
objects of intermediate age ($5-7$~Myr) have lower EW(H$\alpha$) than 
predicted by the models. There are no differences between the YMC/star 
associations in the nearby galaxies and those of the YMC/star complexes in the 
distant galaxies in the distribution in the `$t$--EW(H$\alpha$)' diagram. 
Note only a larger dispersion in the distribution of clusters/associations 
(crosses) compared to complexes (circles) in Fig.~\ref{figure:t_ew}.

\section{Conclusions}
\label{sect:concl}

In this paper, we have presented a combination of spectroscopic and 
photometric studies of the disc cluster population in seven spiral galaxies. 
Our primary goal was to derive spectroscopic information on star forming 
regions (extinction, chemical abundance, relative contributions of nebular 
continuum and emission lines to the total observed flux) and photometric 
information on young massive clusters (true colour, luminosity, 
mass, age) embedded in SF~regions. Combining the information from 
spectroscopy and imaging, we found that disc clusters in external galaxies 
and open clusters in the Milky Way show a uniform sequence (evolutionary/mass).

We summarise the following conclusions.

1. Cluster reddening ranges from 0 to 3~mag in the studied spirals, which 
is in good agreement with the spectroscopic reddening estimations by 
\citet{konstantopoulos2009} for 44 isolated disc clusters in the M82. 
Accounting for the light extinction, based on independent spectroscopic 
observations, provides photometric ages of young massive clusters 
(cluster complexes) embedded in SF~regions within 10~Myr.

2. The relative contributions of the nebular continuum and the emission 
lines to the total observed flux for most of the star forming regions are 
lower than $40\%$ of the total fluxes of the SF~regions, with a mean value of 
$\approx(20-22)\%$ in the $B$ and $V$ bands (Fig.~\ref{figure:sakh5}). This 
means that the impact of nebular emission on integrated broadband 
photometry is not greater than the observational (photometric $+$ 
spectroscopic) errors.

3. Estimates of ages for 57 and masses for 63 of 102 YMC 
complexes embedded in H\,{\sc ii}~regions were obtained. The photometric 
ages of the SF~regions population younger than 10~Myr are consistent with 
the maximum lifetime of giant H\,{\sc ii}~regions. 

4. The derived masses of YMCs range from $10^4 M_{\odot}$ in the nearby 
galaxy NGC~628 to $10^7 M_{\odot}$ in the most distant NGC~7678. 
More than $80\%$ of the YMC complexes have masses between $10^5 M_{\odot}$ 
and $10^6 M_{\odot}$. The lowest mass estimate of $10^4 M_{\odot}$ for the 
objects in NGC~628 and NGC~6946 belongs to the mass interval of the youngest 
Galactic open clusters. This is an argument for a uniform evolutionary 
sequence of extragalactic star forming regions and Galactic OCs.

5. Extragalactic young massive clusters and open star clusters in the 
Milky Way represent a single evolutionary sequence of objects at different 
stages of their evolution. The observed YMCs are comparable to
what we know in the Solar Neighbourhood.

\section*{Acknowledgments}

We are grateful to the referee for constructive comments. 
The authors acknowledge the use of the HyperLeda database 
(http://leda.univ-lyon1.fr), the NASA/IPAC Extragalactic Database 
(http://ned.ipac.caltech.edu), and the Padova group online server CMD 
(http://stev.oapd.inaf.it). This study was supported by the Russian 
Science Foundation (project no. 14--22--00041).

\appendix

\section{}

\begin{table*}
\caption[]{\label{table:true}
True magnitudes and colours of SF~regions.
}
\begin{center}
\begin{tabular}{cccccccccccc} \hline \hline
NGC & No. & $M_B$ & $\Delta M_B$ & $U-B$ & $\Delta (U-B)$ & 
$B-V$ & $\Delta (B-V)$ & $V-R$ & $\Delta (V-R)$ & $V-I$ & $\Delta (V-I)$ \\
\hline
 628 &   1 & $-11.40$ & 0.912 & $-1.11$ & 0.180 & $-0.40$ & 0.229 & $-0.18$ & 0.153 & $-0.08$ & 0.318 \\
 628 &   2 & $-11.70$ & 0.715 & $-0.26$ & 0.153 & $-0.15$ & 0.181 &   0.05  & 0.122 &   0.12 & 0.251 \\
 628 &   3 &  $-9.31$ & 0.640 & $-1.02$ & 0.137 &   0.02  & 0.190 &   0.74  & 0.138 &   0.92 & 0.272 \\
 628 &   4 & $-10.13$ & 0.932 & $-0.69$ & 0.200 & $-0.15$ & 0.249 &   0.04  & 0.183 &   0.09 & 0.368 \\
 628 &   8 &  $-9.45$ & 0.348 & $-0.42$ & 0.074 &   0.10  & 0.092 &   0.59  & 0.064 &   0.79 & 0.136 \\
 628 &  10 & $-10.31$ & 0.658 & $-0.50$ & 0.132 &   0.18  & 0.167 &   0.30  & 0.113 &   0.44 & 0.231 \\
 783 &   1 & $-12.35$ & 0.543 & $-0.39$ & 0.240 &   0.07  & 0.282 &   0.40  & 0.227 &   0.68 & 0.344 \\
 783 &   3 & $-14.68$ & 0.274 & $-0.11$ & 0.068 & $-0.02$ & 0.101 & $-0.32$ & 0.100 &   0.11 & 0.167 \\
 783 &   4 & $-14.96$ & 0.632 & $-0.63$ & 0.162 & $-0.07$ & 0.213 & $-0.05$ & 0.164 &   0.08 & 0.272 \\
 783 &   5 & $-13.84$ & 0.302 & $-0.30$ & 0.083 &   0.02  & 0.098 & $-0.32$ & 0.095 &   0.02 & 0.156 \\
2336 &   1 & $-13.34$ & 0.454 & $-0.78$ & 0.140 &   0.26  & 0.145 &   0.20  & 0.115 &   0.25 & 0.203 \\
2336 &   2 & $-13.76$ & 0.255 & $-0.29$ & 0.093 &   0.12  & 0.085 &   0.26  & 0.056 &   0.39 & 0.107 \\
2336 &   3 & $-14.15$ & 0.501 & $-0.82$ & 0.145 & $-0.23$ & 0.179 &   0.01  & 0.151 &   0.15 & 0.234 \\
2336 &   4 & $-14.24$ & 0.295 & $-0.87$ & 0.141 &   0.22  & 0.152 &   0.29  & 0.130 &   0.36 & 0.210 \\
2336 &   5 & $-12.92$ & 0.960 & $-0.69$ & 0.245 &   0.32  & 0.275 &   0.38  & 0.197 &   0.40 & 0.388 \\
2336 &   7 & $-12.63$ & 0.481 & $-0.79$ & 0.165 &   0.15  & 0.149 &   0.51  & 0.111 &   0.58 & 0.204 \\
2336 &   8 & $-13.55$ & 0.207 & $-0.36$ & 0.067 &   0.25  & 0.068 &   0.17  & 0.061 &   0.29 & 0.107 \\
2336 &  10 & $-14.61$ & 0.175 & $-0.81$ & 0.071 & $-0.15$ & 0.073 & $-0.07$ & 0.064 &   0.10 & 0.114 \\
2336 &  11 & $-14.05$ & 0.445 & $-1.23$ & 0.164 & $-0.09$ & 0.156 & $-0.03$ & 0.112 &   0.12 & 0.205 \\
2336 &  12 & $-14.71$ & 0.811 & $-0.74$ & 0.244 & $-0.07$ & 0.244 &   0.08  & 0.180 &   0.12 & 0.340 \\
2336 &  13 & $-16.60$ & 0.255 & $-1.35$ & 0.083 & $-0.39$ & 0.085 & $-0.15$ & 0.066 & $-0.06$ & 0.107 \\
2336 &  14 & $-14.32$ & 0.192 & $-1.00$ & 0.085 & $-0.16$ & 0.087 & $-0.06$ & 0.078 &   0.16 & 0.131 \\
2336 &  15 & $-12.82$ & 0.445 & $-0.44$ & 0.150 &   0.35  & 0.159 &   0.26  & 0.114 &   0.30 & 0.222 \\
2336 &  16 & $-13.42$ & 0.267 & $-0.21$ & 0.183 &   0.02  & 0.187 & $-0.01$ & 0.174 &   0.21 & 0.225 \\
2336 &  17 & $-15.70$ & 0.143 & $-0.67$ & 0.045 & $-0.13$ & 0.054 & $-0.06$ & 0.039 &   0.23 & 0.083 \\
2336 &  20 & $-14.20$ & 0.396 & $-1.08$ & 0.119 &   0.00  & 0.129 & $-0.40$ & 0.108 & $-0.14$ & 0.185 \\
2336 &  21 & $-12.44$ & 0.388 & $-0.38$ & 0.164 &   0.43  & 0.172 &   0.13  & 0.174 &   0.27 & 0.286 \\
2336 &  24 & $-13.97$ & 0.783 & $-1.05$ & 0.228 & $-0.12$ & 0.228 & $-0.24$ & 0.166 & $-0.09$ & 0.310 \\
2336 &  25 & $-14.61$ & 0.509 & $-1.09$ & 0.140 & $-0.03$ & 0.156 & $-0.24$ & 0.116 & $-0.15$ & 0.224 \\
2336 &  26 & $-13.70$ & 0.970 & $-1.23$ & 0.235 & $-0.04$ & 0.285 & $-0.39$ & 0.217 &   0.00 & 0.388 \\
2336 &  27 & $-13.80$ & 0.725 & $-0.57$ & 0.183 & $-0.01$ & 0.221 & $-0.03$ & 0.172 &   0.05 & 0.311 \\
6217 &   2 & $-16.04$ & 0.942 & $-1.16$ & 0.210 & $-0.24$ & 0.279 &   0.05  & 0.213 &   0.06 & 0.378 \\
6946 &   3 &  $-9.60$ & 1.028 & $-0.95$ & 0.231 &   0.23  & 0.302 & $-0.16$ & 0.222 &   0.04 & 0.437 \\
6946 &   5 & $-10.09$ & 0.522 & $-0.38$ & 0.225 &   0.07  & 0.309 &   0.00  & 0.285 &   0.05 & 0.435 \\
6946 &   6 & $-13.14$ & 0.207 & $-0.90$ & 0.047 & $-0.21$ & 0.058 & $-0.17$ & 0.041 & $-0.07$ & 0.077 \\
6946 &   8 & $-11.11$ & 0.217 & $-0.38$ & 0.057 &   0.11  & 0.078 &   0.10  & 0.081 &   0.13 & 0.127 \\
6946 &   9 & $-13.28$ & 0.151 & $-0.91$ & 0.037 & $-0.30$ & 0.054 &   0.00  & 0.042 &   0.23 & 0.078 \\
6946 &  10 & $-12.78$ & 0.217 & $-0.84$ & 0.057 & $-0.12$ & 0.068 & $-0.16$ & 0.051 & $-0.05$ & 0.087 \\
6946 &  11 & $-12.22$ & 0.207 & $-0.85$ & 0.047 & $-0.08$ & 0.068 &   0.08  & 0.051 &   0.32 & 0.117 \\
6946 &  12 & $-13.56$ & 0.179 & $-0.90$ & 0.042 & $-0.28$ & 0.051 & $-0.16$ & 0.037 & $-0.03$ & 0.078 \\
6946 &  13 & $-12.72$ & 0.217 & $-1.09$ & 0.067 & $-0.25$ & 0.088 & $-0.01$ & 0.071 &   0.15 & 0.117 \\
6946 &  15 & $-12.92$ & 0.148 & $-1.06$ & 0.047 & $-0.38$ & 0.058 & $-0.08$ & 0.051 &   0.28 & 0.087 \\
6946 &  18 & $-10.71$ & 0.179 & $-0.67$ & 0.042 &   0.14  & 0.051 &   0.09  & 0.037 &   0.35 & 0.078 \\
6946 &  26 & $-12.44$ & 0.151 & $-0.85$ & 0.037 & $-0.15$ & 0.054 &   0.11  & 0.042 &   0.34 & 0.088 \\
6946 &  28 & $-13.55$ & 0.545 & $-0.80$ & 0.111 & $-0.34$ & 0.140 & $-0.23$ & 0.095 & $-0.14$ & 0.193 \\
6946 &  29 & $-12.08$ & 0.433 & $-0.67$ & 0.100 & $-0.04$ & 0.122 & $-0.11$ & 0.087 &   0.08 & 0.174 \\
6946 &  30 & $-13.58$ & 0.123 & $-1.09$ & 0.031 & $-0.17$ & 0.047 &   0.08  & 0.028 &   0.40 & 0.059 \\
6946 &  31 & $-10.55$ & 0.245 & $-0.74$ & 0.063 & $-0.07$ & 0.085 & $-0.01$ & 0.076 &   0.16 & 0.127 \\
6946 &  32 & $-12.54$ & 0.235 & $-1.08$ & 0.063 & $-0.40$ & 0.075 & $-0.34$ & 0.056 & $-0.18$ & 0.107 \\
6946 &  34 & $-12.26$ & 0.179 & $-0.96$ & 0.052 & $-0.27$ & 0.071 & $-0.13$ & 0.057 &   0.02 & 0.108 \\
6946 &  35 & $-14.16$ & 0.489 & $-0.38$ & 0.110 &   0.08  & 0.136 & $-0.19$ & 0.096 & $-0.15$ & 0.184 \\
6946 &  36 & $-11.53$ & 0.217 & $-0.47$ & 0.067 & $-0.04$ & 0.088 &   0.19  & 0.071 &   0.52 & 0.127 \\
6946 &  37 & $-12.66$ & 0.837 & $-0.70$ & 0.174 &   0.02  & 0.228 & $-0.08$ & 0.169 &   0.02 & 0.319 \\
7331 &   3 & $-14.13$ & 0.519 & $-0.76$ & 0.150 & $-0.09$ & 0.186 & $-0.11$ & 0.176 & $-0.07$ & 0.284 \\
7331 &   4 & $-13.69$ & 0.245 & $-0.97$ & 0.063 & $-0.12$ & 0.075 &   0.11  & 0.056 &   0.23 & 0.107 \\
7678 &   2 & $-13.47$ & 0.622 & $-0.70$ & 0.162 &   0.22  & 0.203 & $-0.01$ & 0.164 &   0.07 & 0.302 \\
7678 &   3 & $-14.30$ & 0.330 & $-0.18$ & 0.088 &   0.30  & 0.115 &   0.24  & 0.089 &   0.34 & 0.166 \\
7678 &   5 & $-13.30$ & 0.565 & $-0.27$ & 0.151 &   0.30  & 0.190 &   0.09  & 0.165 &   0.15 & 0.273 \\
7678 &   6 & $-16.07$ & 0.415 & $-0.64$ & 0.104 & $-0.09$ & 0.136 & $-0.05$ & 0.112 &   0.01 & 0.195 \\
7678 &   7 & $-17.18$ & 0.151 & $-0.86$ & 0.037 & $-0.07$ & 0.044 & $-0.08$ & 0.032 &   0.05 & 0.058 \\
7678 &   8 & $-18.00$ & 0.088 & $-1.18$ & 0.031 & $-0.30$ & 0.037 & $-0.22$ & 0.028 &   0.04 & 0.059 \\
7678 &   9 & $-14.71$ & 0.227 & $-1.19$ & 0.077 & $-0.22$ & 0.108 &   0.10  & 0.101 &   0.36 & 0.187 \\
7678 &  10 & $-17.01$ & 0.179 & $-0.68$ & 0.052 & $-0.08$ & 0.061 & $-0.15$ & 0.047 & $-0.04$ & 0.088 \\
\hline
\end{tabular}\\
\end{center}
\end{table*}

\begin{thebibliography}{}

\bibitem [Adamo et al. (2013)]{adamo2013}
          Adamo A., \"{O}stlin G., Bastian N., Zackrisson E., 
          Livermore R.~C., Guaita L., 2013, ApJ, 766, 105

\bibitem [Alonso-Herrero et al. (1996)]{alonso1996}
          Alonso-Herrero A., Arag\'{o}n-Salamanca A., Zamorano J., Rego M., 
          1996, MNRAS, 278, 417

\bibitem [\protect\citeauthoryear{Artamonov, Bruevich \& Gusev}{Artamonov et al.}{1997}]{artamonov1997}
          Artamonov B.~P., Bruevich V.~V., Gusev A.~S., 1997, 
          Astron. Rep., 41, 577

\bibitem [Artamonov et al. (1999)]{artamonov1999}
          Artamonov B.~P., Badan Y.~Y., Bruyevich V.~V., 
          Gusev~A.S., 1999, Astron. Rep., 43, 377

\bibitem [Artamonov et al. (2010)]{artamonov2010}
          Artamonov B.~P. et al., 2010, Astron. Rep., 54, 1019

\bibitem [Bastian et al. (2005)]{bastian2005}
          Bastian N., Gieles M., Efremov Y.~N., Lamers H.~J.~G.~L.~M., 
          2005, A\&A, 443, 79

\bibitem [Belley \& Roy (1992)]{belley1992}
          Belley J., Roy J.-R., 1992, ApJS, 78, 61

\bibitem [Bertelli et al. (1994)]{bertelli1994}
          Bertelli G., Bressan A., Chiosi C., Fagotto F., Nasi E., 
          1994, A\&AS, 106, 275

\bibitem [Bessell (1990)]{bessell1990}
          Bessell M.~S., 1990, PASP, 102, 1181

\bibitem [Bolatto et al. (2008)]{bolatto2008}
          Bolatto A.~D., Leroy A.~K., Rosolowsky E., Walter F., Blitz L., 
          2008, ApJ, 686, 948

\bibitem [Bressan et al. (2012)]{bressan2012}
          Bressan A., Marigo P., Girardi L., Salasnich B., Dal Cero C.,
          Rubele S., Nanni A., 2012, MNRAS, 427, 127

\bibitem [Brown \& Mathews (1970)]{brown1970}
          Brown R.~L., Mathews W.~G., 1970, ApJ, 160, 939

\bibitem [Bruevich et al. (2007)]{bruevich2007}
          Bruevich V.~V., Gusev A.~S., Ezhkova O.~V., Sakhibov F.~K., 
          Smirnov M.~A., 2007, Astron. Rep., 51, 222

\bibitem [\protect\citeauthoryear{Copetti, Pastoriza \& Dottori}{Copetti et al.}{1986}]{copetti1986}
          Copetti M.~V.~F. Pastoriza M.~G., Dottori H.~A., 1986, A\&A, 
          156, 111

\bibitem [Dinerstein (1990)]{dinerstein1990}
          Dinerstein H.~L., 1990, in \textit{The Interstellar Medium in 
          Galaxies}, eds. H.A. Thronson (Jr.) and J.M. Shull (Dordrecht: 
          Kluwer), 257

\bibitem [\protect\citeauthoryear{Epinat, Amram \& Marcelin}{Epinat et al.}{2008}]{epinat2008}
          Epinat B., Amram P., Marcelin M., 2008, MNRAS, 390, 466

\bibitem [Efremov \& Elmegreen (1998)]{efremov1998}
          Efremov Y.~N., Elmegreen B., 1998, MNRAS, 299, 588

\bibitem [Efremov \& Sitnik (1988)]{efremov1988}
          Efremov Y.~N., Sitnik T.~G., 1988, Soviet Astronomy Letters, 14, 347

\bibitem [Elmegreen \& Efremov (1996)]{elmegreen1996}
          Elmegreen B.~G., Efremov Y.~N., 1996, ApJ, 466, 802

\bibitem [Elmegreen, Elmegreen \& Leitner (2003)]{elmegreen2003a}
          Elmegreen B.~G., Elmegreen D.~M., Leitner S.~N., 2003, ApJ, 
          590, 271

\bibitem [Elmegreen et al. (1996)]{elmegreen1996b}
          Elmegreen B.~G., Elmegreen D.~M., Salzer J.~J., Mann H., 1996, 
          ApJ, 467, 579

\bibitem [Elmegreen et al. (2003)]{elmegreen2003b}
          Elmegreen B.~G., Leitner S.~N., Elmegreen D.~M., Cuillandre J.-C., 
          2003b, ApJ, 593, 333

\bibitem [Gieles \& Portegies Zwart (2011)]{gieles2011}
          Gieles M., Portegies Zwart S.~F., 2011, MNRAS, 410, L6

\bibitem [Girardi et al. (2000)]{girardi2000}
          Girardi L., Bressan A., Bertelli G., Chiosi C., 2000, A\&AS,
          141, 371

\bibitem [Gusev (2006a)]{gusev2006a}
          Gusev A.~S., 2006a, Astron. Rep., 50, 167

\bibitem [Gusev (2006b)]{gusev2006b}
          Gusev A.~S., 2006b, Astron. Rep., 50, 182

\bibitem [Gusev \& Efremov (2013)]{gusev2013b}
          Gusev A.~S., Efremov Y.~N., 2013, MNRAS, 434, 313

\bibitem [Gusev \& Park (2003)]{gusev2003}
          Gusev A.~S., Park M.-G., 2003, A\&A, 410, 117

\bibitem [\protect\citeauthoryear{Gusev, Egorov \& Sakhibov}{Gusev et al.}{2014}]{gusev2014}
          Gusev A.~S., Egorov O.~V., Sakhibov F., 2014, MNRAS, 437, 1337

\bibitem [Gusev et al. (2015)]{gusev2015}
          Gusev A.~S., Guslyakova S.~A., Novikova A.~P., Khramtsova M.~S., 
          Bruevich V.~V., Ezhkova O.~V., 2015, Astron. Rep., 59, 899

\bibitem [Gusev et al. (2012)]{gusev2012}
          Gusev A.~S., Pilyugin L.~S., Sakhibov F., Dodonov S.~N.,
          Ezhkova O.~V., Khramtsova M.~S., 2012, MNRAS, 424, 1930

\bibitem [Gusev et al. (2009)]{gusev2009}
          Gusev A.~S., Piskunov A.~E., Sakhibov F.~K., Kharchenko N.~V., 
          2009, Astron. Lett., 35, 679

\bibitem [\protect\citeauthoryear{Gusev, Sakhibov \& Dodonov}{Gusev et al.}{2013}]{gusev2013}
          Gusev A.~S., Sakhibov F.~H., Dodonov S.~N., 2013, Astr. Bull., 
          68, 40

\bibitem [Hollyhead et al. (2015)]{hollyhead2015}
          Hollyhead K., Bastian N., Adamo A., Silva-Villa E., Dale J., 
          Ryon J.~E., Gazak Z., 2015, MNRAS, 449, 1106

\bibitem [Holtzman et al. (1995)]{holtzman1995}
	  Holtzman J.~A., Burrows C.~J., Casertano S., Hester J.~J., 
          Trauger J.~T., Watson A.~M., Worthey G., 1995, PASP, 107, 1065

\bibitem [Hopkins (2012)]{hopkins2012}
          Hopkins P.~F., 2012, MNRAS, 423, 2016

\bibitem [\protect\citeauthoryear{Izotov, Thuan \& Lipovetsky}{Izotov et al.}{1994}]{izotov1994}
          Izotov Y.~I., Thuan T.~X., Lipovetsky V.~A., 1994, ApJ, 435, 64

\bibitem [Kaplan \& Pikelner (1979)]{kaplan1979}
          Kaplan S.~A., Pikelner S.~B., 1979, The physics of the 
          interstellar medium. Nauka, Moscow, p. 592 (In Russian)

\bibitem [Kennicutt \& Hodge (1980)]{kennicutt1980}
          Kennicutt R.~C., Hodge P.~W., 1980, ApJ, 241, 573

\bibitem [Kharchenko (2001)]{kharchenko2001}
          Kharchenko N.~V., 2001, Kinematika Fizika Nebesnykh 
          Tel, 17, 409

\bibitem [Kharchenko et al. (2005a)]{kharchenko2005}
          Kharchenko N.~V., Piskunov A.~E., R\"{o}ser S., 
          Schilbach E., Scholz R.-D., 2005a, A\&A, 438, 1163

\bibitem [Kharchenko et al. (2005b)]{piskunov2005}
          Kharchenko N.~V., Piskunov A.~E., R\"{o}ser S., 
          Schilbach, E., Scholz R.-D., 2005b, A\&A, 440, 403

\bibitem [Kharchenko et al. (2009)]{kharchenko2009}
          Kharchenko N.~V., Piskunov A.~E., R\"{o}ser S., 
          Schilbach E., Scholz R.-D., Zinnecker, H.,
          2009, A\&A, 504, 681

\bibitem [Kim et al. (2012)]{kim2012}
          Kim et al., 2012, ApJ, 753, 26

\bibitem [Konstantopoulos et al. (2009)]{konstantopoulos2009}
          Konstantopoulos I.~S., Bastian N., Smith L.~J., 
          Westmoquette M.~S., Trancho G., Gallagher J.~S., 
          2009, ApJ, 701, 1015

\bibitem [Landolt (1992)]{landolt1992}
          Landolt A.~U., 1992, AJ, 104, 340

\bibitem [Lang (1978)]{lang1978}
          Lang K.~R., 1978, Astrophysical Formulae. A Compendium for 
          the Physicist and Astrophysicist. Springer-Verlag, Berlin, 
          Heidelberg, New York, p. 783

\bibitem [Larsen (1999)]{larsen1999}
          Larsen S.~S., 1999, A\&AS, 139, 393

\bibitem [Larsen (2004)]{larsen2004}
          Larsen S.~S., 2004, A\&A, 416, 537

\bibitem [Larson (1981)]{larson1981}
          Larson R.~B., 1981, MNRAS, 194, 809

\bibitem [Ma\'{i}z-Apell\'{a}niz et al. (1998)]{maizapellaniz1998}
          Ma\'{i}z-Apell\'{a}niz J., Mas-Hesse J.~M., 
          Mu\~{n}oz-Tu\~{n}\'{o}n C., V\'{i}lchez J.~M., Casta\~{n}eda H.~O., 
          1998, A\&A, 329, 409

\bibitem [Marigo \& Girardi (2007)]{marigogirardi2007}
          Marigo P., Girardi L., 2007, A\&A, 469, 239

\bibitem [Marigo et al. (2008)]{marigo2008}
          Marigo P., Girardi L., Bressan A., Groenewegen 
          M.~A.~T., Silva L., Granato G.~L., 2008, A\&A, 482, 883

\bibitem [Mas-Hesse \&  Kunth (1999)]{mashesse1999}
          Mas-Hesse J.~M., Kunth D., 1999, A\&A 349, 765

\bibitem [Osterbrock (1989)]{osterbrock1989}
          Osterbrock D.~E., 1989, Astrophysics of gaseous nebulae 
          and active galactic nuclei. University Science Books, 
          Mill Valley, CA, p. 422

\bibitem [Paturel at al. (2003)]{paturel2003}
          Paturel G., Petit C., Prugniel Ph., Theureau G., Rousseau J.,
          Brouty M., Dubois P., Cambresy L., 2003, A\&A, 412, 45

\bibitem [Pilyugin \& Mattsson (2011)]{pilyugin2011}
          Pilyugin L.S., Mattsson L., 2011, MNRAS, 412, 1145

\bibitem [Piskunov at al. (2009)]{piskunov2009}
          Piskunov A.~E., Kharchenko N.~V., Schilbach E., R\"{o}ser S., 
          Scholz R.-D., Zinnecker H., 2009, A\&A, 507, L5

\bibitem [Piskunov at al. (2011)]{piskunov2011}
          Piskunov A.~E., Kharchenko N.~V., Schilbach E., R\"{o}ser S., 
          Scholz R.-D., Zinnecker H., 2011, A\&A, 525, A122

\bibitem [\protect\citeauthoryear{Portegies Zwart, McMillan \& Gieles}{Portegies Zwart et al.}{2010}]{zwart2010}
          Portegies Zwart S.~F., McMillan S.~L.~W., Gieles~M., 2010, ARA\&A, 48, 431

\bibitem [\protect\citeauthoryear{Reines, Johnson \& Hunt}{Reines et al.}{2008}]{reines2008}
          Reines A.~E., Johnson K.~E., Hunt L.~K., 2008, AJ, 136, 1415

\bibitem [Reines et al. (2010)]{reines2010}
          Reines A.~E., Nidever D.~L., Whelan D.~G., Johnson K.~E., 2010, 
          ApJ, 708, 26

\bibitem [Sakhibov \& Smirnov (1990)]{sakhibov1990}
          Sakhibov F., Smirnov M.~A., 1990, Soviet 
          Astronomy, 34, 236

\bibitem [Sakhibov \& Smirnov (2000)]{sakhibov2000}
          Sakhibov F., Smirnov M.~A., 2000, A\&A, 354, 802

\bibitem [Sakhibov \& Smirnov (2001)]{sakhibov2001}
          Sakhibov F., Smirnov M.~A., 2001, Astron. Rep., 45, 1

\bibitem [Sakhibov et al. (2010)]{sakhibov2010}
          Sakhibov F., Gusev A.~S., Kharchenko N.~V., Piskunov A.~E., 
          2010, Proc. IAU Symp. 266, 522

\bibitem [Scalo (1986)]{scalo1986}
          Scalo J.~M., 1986, Fundamentals of Cosmic Physics, 11, 1

\bibitem [\protect\citeauthoryear{Wei, Keto \& Ho}{Wei et al.}{2012}]{wei2012}
          Wei L.~H., Keto E., Ho L.~C., 2012, ApJ, 750, 136

\bibitem [Whitmore et al. (2011)]{whitmore2011}
          Whitmore B.~C. et al., 2011, ApJ, 729, 78

\end{thebibliography}
\end{document}